\documentclass{aa}  

\usepackage{graphicx}
\usepackage{txfonts}
\usepackage{soul}  %used to strikethrough text
\usepackage{placeins} % to force the images to appear after title
\usepackage{hyperref}
\hypersetup{
    colorlinks=true,
    linkcolor=blue,
    filecolor=magenta,      
    urlcolor=cyan,
    citecolor=blue,
    pdfpagemode=UseThumbs, % UseThumbs (Thumbnails), UseOutlines (Bookmarks) and FullScreen
    }
\usepackage{makecell}
\usepackage{float}
\usepackage{natbib}
\usepackage{xcolor}
\usepackage[normalem]{ulem}
\bibpunct{(}{)}{;}{a}{}{,} % to follow the A&A style

\begin{document} 
   %\title{}
   %\subtitle{}
   \title{Opacity for realistic 3D MHD simulations of cool stellar atmospheres} 

   \author{A. Perdomo García
          \inst{1}\fnmsep\inst{2}%fnmsep\inst{4}
          \and
          N. Vitas\inst{1}\fnmsep\inst{2}%\fnmsep\inst{4}
          \and 
          E. Khomenko
          \inst{1}\fnmsep\inst{2}%\fnmsep\inst{5}
          \and 
          M. Collados
          \inst{1}\fnmsep\inst{2}%\fnmsep\inst{6}
          \and
          C. Allende Prieto\inst{1}\fnmsep\inst{2}
          \and
          I. Hubeny\inst{4}
          \and
          Y. Osorio
          \inst{1}\fnmsep\inst{2}\fnmsep\inst{3}
          }

   \institute{Instituto de Astrofísica de Canarias,
              38200 La Laguna, Tenerife, Spain 
              \\
              \email{andrperdomo@gmail.com}
         \and
             Departamento de Astrofísica de la Universidad de La Laguna,
             38200 La Laguna, Tenerife, Spain
         \and
            Isaac Newton Group of Telescopes, Apartado de Correos 321, 38700 Santa Cruz de La Palma, Canary Islands, Spain
        \and 
            Steward Observatory, University of Arizona, Tucson, USA
             }

   \date{Received Month day, year; accepted Month day, year}

% \abstract{}{}{}{}{} 
% 5 {} token are mandatory
 
  \abstract
   {Realistic 3D time-dependent simulations of stellar near-surface convection employ the opacity binning method for efficient and accurate computation of the radiative energy exchange. 
   The method provides several orders of magnitude of speed-up, but its implementation includes a number of free parameters.}
%    As the realistic simulations are increasingly used to interpred high-resolution observations, it is essential to control their error budget.
   {Our aim is to evaluate the accuracy of the opacity binning method as a function of the choice of these free parameters.  }
   {The monochromatic opacities computed with the \texttt{SYNSPEC} code are used to construct opacity distribution function (ODF) that is then verified through detailed comparison with the results of the \texttt{ATLAS} code. The opacity binning method is implemented with the \texttt{SYNSPEC} opacities for four representative cool main-sequence stellar spectral types (F3V, G2V, K0V, and M2V).}
   {The ODFs from \texttt{SYNSPEC} and \texttt{ATLAS} show consistent results for the opacity and bolometric radiative energy exchange rate $Q$ in case of the F, G, and K -- type stars. 
   Significant differences, coming mainly from the molecular line lists, are found for the M -- type star.
   It is possible to optimise a small number of bins to reduce the deviation of the results coming from the opacity grouping with respect to the ODF for the F, G, and K -- type stars. 
   In the case of the M -- type star, the inclusion of splitting in wavelength is needed in the grouping to get similar results, with a subsequent increase in computing time.
   In the limit of a large number of bins, the deviation for all the binning configurations tested saturates and the results do not converge to the ODF solution. 
   Due to this saturation, the $Q$ rate cannot be improved by increasing the number of bins to more than about 20 bins. The more effective strategy is to select the optimal location of fewer bins.
   }
   {}


   \maketitle
%
%-------------------------------------------------------------------

\section{Introduction} \label{sec:introduction}
    
    Radiative transfer is one of the key factors governing the structure and dynamics of stellar atmospheres and it has to be taken into account with special care when constructing a model atmosphere \citep{Hubeny_Mihalas_2014}.
    Detailed interactions between radiation and matter are quantitatively described by the opacity: the absorption coefficients of all contributing (bound-bound, bound-free, and free-free) processes due to all relevant particle species present in the atmosphere \citep{book2003rutten}.
    The complexity of the opacity function is enormous as it is visible from the plethora of atomic and molecular spectral lines in the spectra of late-type stars \citep{2002AIPC..636..134K, 2017CaJPh..95..825K} and its evaluation is computationally expensive.
    In combination with the already computationally demanding solution of the radiative transfer equation (RTE), the problem of calculating detailed monochromatic radiative losses and gains in the atmosphere, accounting explicitly for $\approx$$10^9$ lines and $\approx$$10^6$ wavelength points, is difficult even in one-dimensional (1D) modelling, and virtually impossible in three-dimensional (3D) time-dependent numerical simulations.
    Two important tricks are available, however, that reduce the computational effort by many orders of magnitude. 
    
    First, under the assumption of local thermodynamic equilibrium (LTE),
    opacity is a function of chemical composition, the turbulent velocity, wavelength, and two independent thermodynamic variables (for example, temperature $T$ and mass density $\rho$); it can therefore be precomputed and stored into lookup tables. 
    Even in the situations where some atoms are treated
    out of LTE, the opacity of the LTE background can be precomputed
    without the contribution of the NLTE lines, which are computed explicitly. 
    The opacities are computed as by-products in the codes and projects for modelling stellar atmospheres: the \texttt{ATLAS} family of codes \citep{1970ATLAS, 2003IAUS..210P.A20C, 2004MSAIS...5...93S, 2021A&A...653A..65W}, \texttt{MARCS} \citep{1975A&A....42..407G, 2008A&A...486..951G}, \texttt{PHOENIX} \citep{1975A&A....42..407G, 2008A&A...486..951G}, \texttt{SYNSPEC}/\texttt{TLUSTY} \citep{SYNSPEC, synspec_tlusty_guide_IV} and others. 

    Second, in the energy conservation equation of the system of magnetohydrodynamic equations, the radiative energy exchange is accounted for by a wavelength-integrated source term $Q$. 
    The distribution of radiative heating and cooling with wavelength is irrelevant, at least as long as the one-fluid MHD approach is used.
    Therefore, instead of solving monochromatic RTE and then integrating
    the results, 
    an approximate solution may be obtained by first combining or integrating the opacities in wavelength ranges and then solving the RTE for a number of representative opacity values. 
    This idea is implemented in the \textit{opacity distribution function} (ODF) defined by \citet{ODF1966}. 
    In their approach the monochromatic opacities are divided into hundreds of wavelength segments. The opacities in each segment are then sorted by magnitude to obtain a continuous monotonic distribution that is then discretised by averaging over several ($\approx$$10$) substeps.
    In that way the RTE can be solved for several thousands ODF values instead for millions of monochromatic opacities. 
    
    An important underlying assumption in the ODF approach is that there are no 
    substantial wavelength shifts in the spectrum along the different heights in the atmosphere. There are situations in which this assumption fails \citep[see chapter 17.6 in][]{Hubeny_Mihalas_2014}, especially for plasmas with velocity fields. Nevertheless, even in moving media, if we are not interested in detailed spectra and the velocities are not larger than the thermal velocity, we can still solve the RTE using ODFs.  %
    
    Precomputed ODF tables produced by the \texttt{ATLAS} code are widely used in stellar modelling and for computing emergent spectra from the models \citep[see][]{2014dapb.book...39K}.
    The ODF method was recently optimised and generalised by \citet{cernetic2019}. 
    Nevertheless, not even the speedup enabled by ODF is sufficient when atmospheres are modelled as time-dependent 3D systems. 
    A solution, known in the literature as the \textit{opacity binning} (OB) method, was initially proposed by \citet{norlund1982}. 
    In this method the opacities are grouped according to the optical depth in a representative atmosphere at which the monochromatic optical depth reaches unity.
    Very few groups (or \textit{bins}) are sufficient to 
    approximate 
    fairly accurately  
    the correct solution for the radiative energy exchange rate, although the explicit wavelength dependence of the opacity is eliminated.
    The number of groups and the distribution of the optical depth separators between them are free parameters of the method, while
    four groups are typically used, roughly representing the continuum, weak, intermediate, and strong lines. 
    The OB method opened the door to the efficient implementation of realistic 3D simulations that have revolutionised our understanding of the physics of stellar atmospheres. 
    The method was further developed by 
    \citet{ludwig_thesis, 1994A&A...284..105L}, \citet{voegler_thesis}, and \citet{2013ApJ...769...18T}. \citet{scatt_lines_binning} developed a variant that includes the effects of scattering. 
    
    The wavelength dependence of the opacities is disregarded in the OB method. 
    \citet{voegler_opacity_binning} and \citet{2013MSAIS..24...53L} noted that this approximation does not converge to the correct solution in the limit of an infinite number of bins. 
    Several other groups experimented with variations to the OB method. 
    \citet{2008A&A...488.1031C} used 12 bins in their study of the oxygen abundance, while \citet{2013A&A...554A.118P} and \citet{magic2013} introduced a variant of the method whereby the opacities are sorted by wavelength into several groups before sorting by optical depths.
    However, there is no unique criterion for choosing the number and distribution of the bins.
    The distribution of the opacities varies significantly with the effective temperature and chemical composition of a star and it may thus be expected that different distributions of bins are optimal for different stars.  
    Our aim in this paper is to test the OB method for four cool main-sequence stars (F3V, G2V, K0V, and M2V) and to design a strategy for finding optimal setups. 
    Our analysis is based on monochromatic opacities computed with the \texttt{SYNSPEC} code.
    The code (and its modelling counterpart \texttt{TLUSTY}) was recently significantly upgraded and equipped with an up-to-date list of atomic and molecular spectral lines necessary for the modelling of cool stellar types \citep{synspec_tlusty_guide_IV}. 
    The code is publicly available as open source\footnote{\url{https://www.as.arizona.edu/ehubeny/tlusty208-package/tl208-s54.tar.gz}} and supplemented with a Python wrapper \citep{synple}. 
    As we intend to use the code to prepare customised opacities for 3D simulations of near-surface convection in cool stars (Perdomo et al, in preparation) with the \texttt{MANCHA} code \citep{2017khomenko, 2018khomenko}, flexibility in selecting opacity contributors is of particular importance for us.
    
    In Sect.\ \ref{sec:method} we briefly introduce a set of the 1D models that represent our stars and specify the radiative transfer details. Computation of the monochromatic opacities using the \texttt{SYNSPEC} code are described in Sect.\ \ref{sec:monochromatic_opacity}. In Sect.\ \ref{sec:odf} the essentials of the ODF method are summarised: ODFs are constructed from the monochromatic opacities computed with \texttt{SYNSPEC}, and the results are compared with the results of the \texttt{ATLAS} code. The same data set is used to study strategies for OB in Sect.\ \ref{sec:opacity_binning}: we first test the importance of the location of the bin separators in optical depth (Sect.\ \ref{subsec:tau_bin_location}), then we explore the influence of the number of bins in optical depth (Sect.\ \ref{subsec:tau_bin_number}), and, finally, we study two different strategies for combined binning in optical depth and wavelength (Sect.\ \ref{subsec:tau_lambda_binning}). Our conclusions are presented in Sect.\ \ref{sec:conclusions}.

\section{Method} \label{sec:method}
    
    \subsection{Model atmospheres} \label{subsec:model_atms}

        \begin{figure}
        \includegraphics[width=8.8cm]{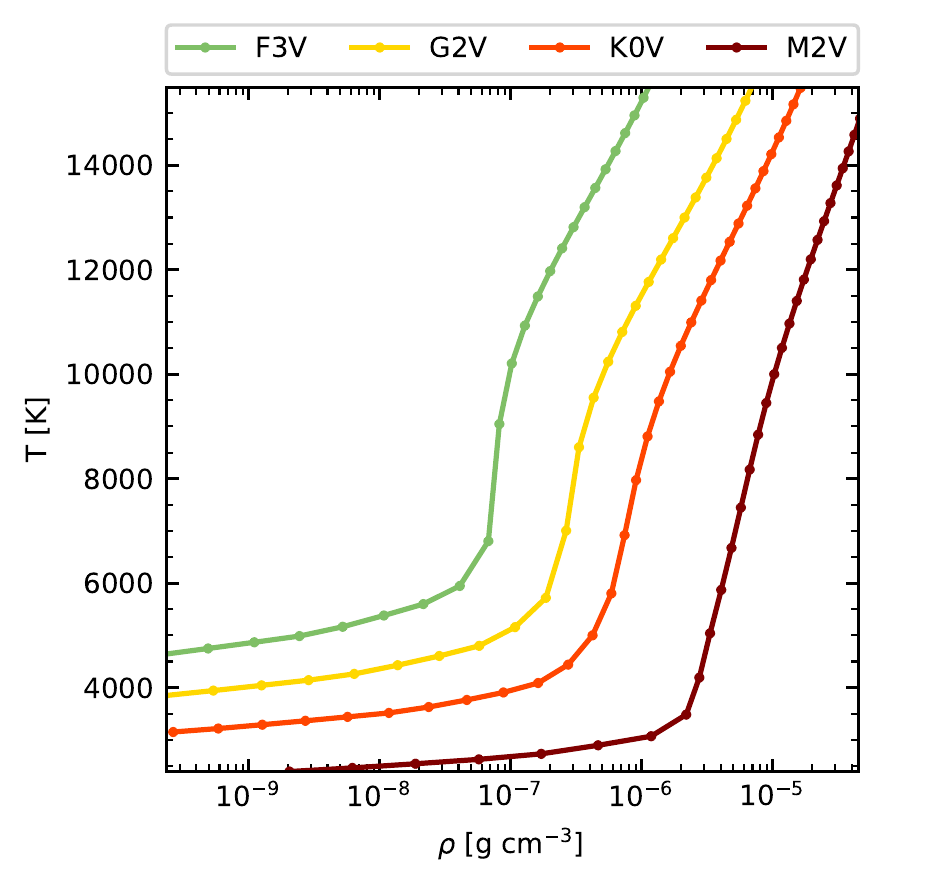}
        \caption{Temperature stratification of the F3V (\textit{green}), G2V (\textit{yellow}), K0V (\textit{orange}), and M2V (\textit{dark red}) 1D hydrostatic models used in this work. The same colour coding is used throughout the paper to label results for the 4 stellar types. The grid used to solve the RTE is marked by dots.}
         \label{fig:1d_mod_atms}
        \end{figure}
        
        To test the OB for a series of main sequence spectral types, a set of 1D stellar atmospheric models is used. These atmospheres are computed assuming hydrostatic equilibrium and using the mixing length theory to account for the convective energy transfer. 
        The models are initiated with radiation in the grey approximation and corrected to have representative temperature gradient at the top of the atmosphere by scaling the Hopf function $q(\tau)$ \citep[see chapter 17 from][and references 507--510 there]{Hubeny_Mihalas_2014} of the the Harvard-Smithsonian Reference Atmosphere \citep{1971HSRA} with the effective temperature of each star, following the expression $T^4(\tau) = \frac{3}{4}T_{\mathrm{eff}}^4 \times \left[\tau + q(\tau) \right]$. The models are shown in Fig.\ \ref{fig:1d_mod_atms} and are computed for spectral types F3V, G2V, K0V, and M2V, with effective temperature (K) and logarithm of gravity (cm s$^{-2}$) of [6890, 4.3], [5780, 4.4], [4855, 4.6], [3690, 4.8], respectively. 
        In the computation of the models we used the Rosseland opacities consistent with the opacities used in the present paper (see Sect.\ \ref{sec:monochromatic_opacity}). 
        More details about the models will be published elsewhere.

    \subsection{Radiative transfer computation} \label{subsec:rte}
         
         The RTE in a plane-parallel static atmosphere in LTE without considering line scattering is
         \begin{equation}
             \mu \frac{dI_\nu(z, \nu)}{dz} = \rho(z) \varkappa_\nu(z) \left[ B_\nu(z) - I_\nu(z) \right],
         \end{equation}
         where $I_\nu(z, \nu)$ is the specific intensity (erg s$^{-1}$ cm$^{-2}$ Hz$^{-1}$ ster$^{-1}$) at frequency $\nu$ in the direction $\mu$ and at the geometrical height $z$ in the atmosphere; $B_\nu$ is the Planck function; $\rho(z)$ is the mass density; and $\varkappa_\nu(z)$ is the monochromatic absorption coefficient per unit mass (cm$^2$ g$^{-1}$; referred to as \textit{opacity} throughout the text). The opacity is computed as the sum of the continuum opacity (mainly contributions from the free-free and bound-free processes) and the line opacity (bound-bound individual transitions included in the line lists for atoms and molecules),
         \begin{equation}
             \varkappa_\nu = \varkappa_\nu^{\mathrm{cont}} + \varkappa_\nu^{\mathrm{lines}}
         \end{equation}
         We restrict ourselves to the LTE case that is valid only in deep layers where $J \approx B$. Therefore our treatment of the coherent scattering terms (see App.\ \ref{app:opac_sources}) in the total opacity is only approximate \citep[e.g.\ ][]{hayek2010}.
         
         We solve the RTE in the vertical direction ($\mu = \pm 1$). 
         At each frequency $\nu$, the formal solution for outward and inward intensity ($I^+$, $I^-$) in LTE is:
         \begin{equation}
         \begin{split}
            I^{\pm}_\nu(z_i) & = I^{\pm}_\nu(z_{i\mp1}) \exp\left(-\Delta \tau_\nu^{i\mp1}\right) + \int_0^{\Delta \tau_\nu^{i\mp1}} B_\nu (t) \exp\left(t-\Delta \tau_\nu^{i\mp1}\right)dt \equiv \\
                             & \equiv I^{\pm}_\nu(z_{i\mp1}) \exp\left(-\Delta \tau_\nu^{i\mp1}\right) + \Delta I^{\pm}_\nu(z_i),
         \end{split}
         \end{equation}
         where $i$ is the height index for each point in the atmosphere, being zero at the bottom and increasing upwards; $\tau_\nu^i$ is the optical depth at the height with index $i$ ($d\tau_\nu(z) = \rho(z) \varkappa_\nu(z) dz$); and $\Delta \tau_\nu^{i+1} = \tau_\nu^{i+1}-\tau_\nu^i$ and $\Delta \tau_\nu^{i-1} = \tau_\nu^i-\tau_\nu^{i-1}$. To solve the RTE, we apply the short characteristics method \citep{shortcharactNLTE}, using the linear approximation of the Planck function:
         \begin{equation}
         \Delta  I_\nu^{\pm}(z_i) = \psi_0 B_\nu(z_i) + \psi_1 B_\nu(z_{i\mp1}).
         \end{equation}
         For each frequency $\nu$ we compute the coefficients
         for the local and the upwind points 
         \begin{equation}\label{General}
         \begin{split}
            \psi_0 &= 1 - \frac{1}{\Delta \tau_\nu} + \frac{1}{\Delta \tau_\nu} \exp\left(-\Delta \tau_\nu\right) , \\
            \psi_1 &= \frac{1}{\Delta \tau_\nu} - \frac{\Delta \tau_\nu+1}{\Delta \tau_\nu} \exp\left(-\Delta \tau_\nu\right). \\
         \end{split}
         \end{equation} 
         Since the method is formulated in a way that it is symmetrical along a ray direction, the coefficients are the same for the upward and inward intensities, and the only change needed is the specific $\Delta \tau_\nu$ in each case ($\Delta \tau_\nu=\Delta \tau_\nu^{i-1}$ for the upward intensities; $\Delta \tau_\nu=\Delta \tau_\nu^{i+1}$ for the inward ones). This first order short-characteristics scheme would not be sufficiently accurate if scattering was taken into account (i.e.\ if the formal solver was used in an iterative solution).
        
        The mean intensity $J_\nu$ (erg s$^{-1}$ cm$^{-2}$ Hz$^{-1}$ ster$^{-1}$) and the flux $\vec{\mathcal{F}}_\nu$ (erg s$^{-1}$ cm$^{-2}$ Hz$^{-1}$) can then be calculated for a vertical ray from the known $I_\nu$:
        \begin{equation}
            J_\nu (z_i) = \frac{1}{2} \int_{-1}^1 I_\nu (z_i) d\mu \equiv \frac{1}{2} \left( I^{+}_\nu(z_i) + I^{-}_\nu(z_i) \right),
        \end{equation}
        \begin{equation}
            \mathcal{F}_\nu (z_i)  = 2 \pi \int_{-1}^1 I_\nu (z_i) \mu d\mu \equiv 2 \pi \left( I^{+}_\nu(z_i) - I^{-}_\nu(z_i) \right).
        \end{equation}

        Finally, the monochromatic radiative energy exchange rate $Q_\nu$ (erg cm$^{-3}$ s$^{-1}$ Hz$^{-1}$), which accounts for the radiation sources and sinks in the energy conservation equation, can be computed either as the divergence of the radiative flux,
        \begin{equation}
            Q^{F}_\nu (z_i) = - \vec{\nabla} \vec{\mathcal{F}}_\nu \equiv -(\mathcal{F}_\nu (z_{i+1}) - \mathcal{F}_\nu (z_{i-1})) / (z_{i+1} - z_{i-1}),
        \end{equation}
        or directly from the mean intensity,
        \begin{equation}
            Q^{J}_\nu (z_i) = 4 \pi \varkappa_\nu (z_i) \rho (z_i) \left( J_\nu - B_\nu \right).
        \end{equation}
        
        The monochromatic rate is calculated as a weighted combination of the two, using equation 4.13 from \citet{voegler_thesis} \citep[following the suggestion from][]{bruls1999}:
        \begin{equation} 
        \label{eq:bruls_Q}
            Q_\nu (z_i) = \exp\left(-\tau^i_\nu/\tau_{\mathrm{h}}\right) Q^{J}_\nu (z_i) + \left[1 - \exp\left(-\tau^i_\nu/\tau_{\mathrm{h}}\right) \right] Q^{F}_\nu (z_i),
        \end{equation} 
        where $\tau_{\mathrm{h}}=0.1$. As it is explained in \citet{voegler_thesis}, this smooth transition between $Q^{J}_\nu$ and $Q^{F}_\nu$ avoids the accuracy problem that $Q^{J}_\nu$ has in the optically thick regime (where $J_\nu$ approaches $B_\nu$) and the errors coming from small variations of the orientation of the flux in the optically thin part of the atmosphere (where flux should be nearly constant), that are enhanced by the gradient in $Q^{F}_\nu$.

\section{Monochromatic opacity} \label{sec:monochromatic_opacity}
    
    \begin{figure*}
    \centering
    \includegraphics[width=17.6cm]{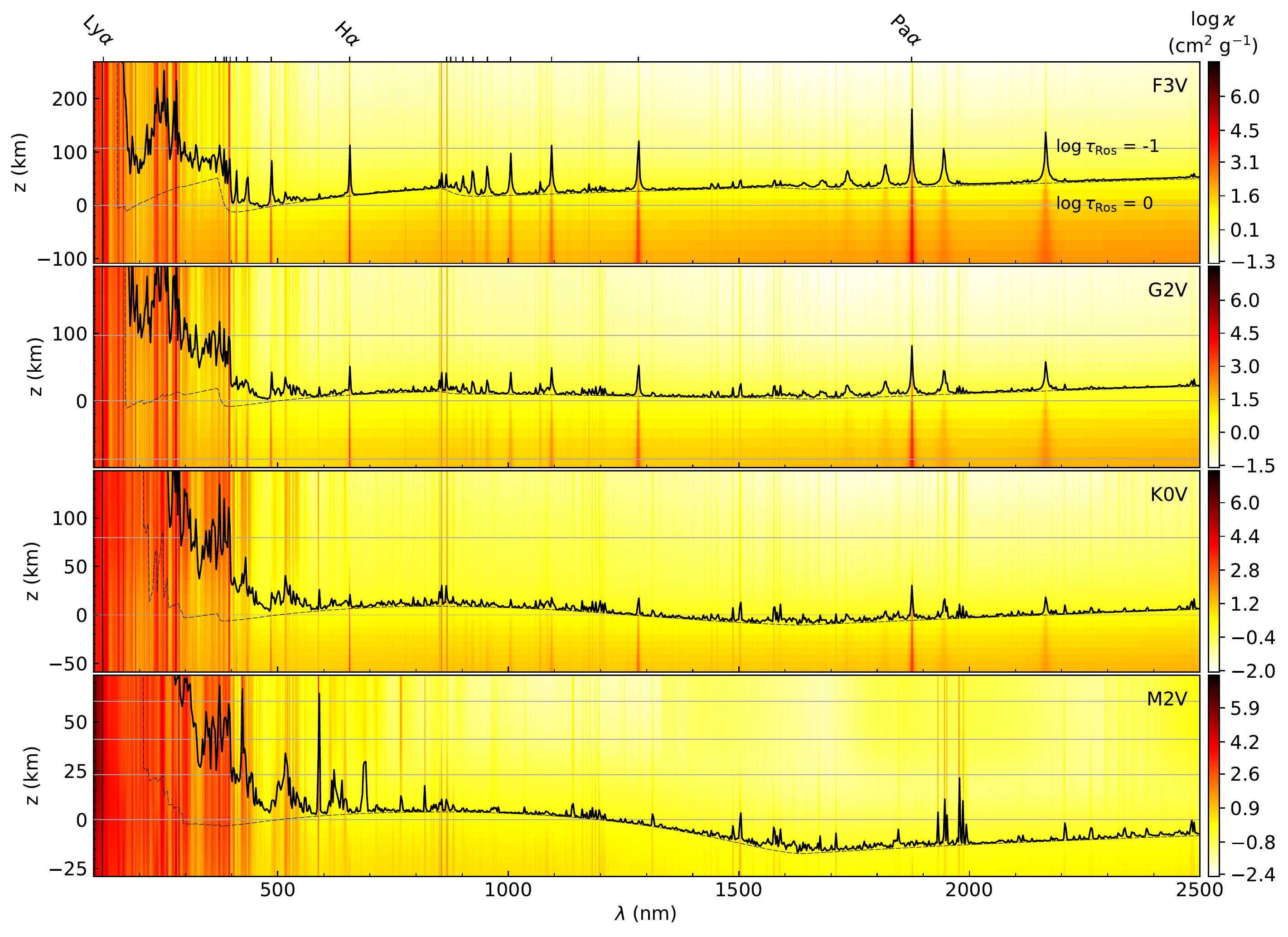}
    \caption{Monochromatic opacity (values shown in the \textit{colour--bars}) for the F3V (\textit{1st row}), G2V (\textit{2nd row}), K0V (\textit{3rd row}), and M2V stars (\textit{4th row}) versus wavelength (horizontal axis) and geometrical height (vertical axis). 
    \textit{Horizontal grey lines}: Rosseland optical depths. The zero geometrical height is taken where $\log \tau_{\mathrm{Ros}}=0$. Other lines are shifted by $\Delta$log$\tau_{\mathrm{Ros}}=1$. 
    \textit{Black dashed thin line}: height at which the continuum optical depth at each wavelength is unity ($\tau^{\mathrm{cont}}_{\lambda}=1$). \textit{Black solid thick line}: heights at which the total optical depth $\tau_{\lambda}$ is unity. Values are averaged over 25 nm intervals.
    \textit{Tick marks on the top} indicate wavelengths of the hydrogen series. Hydrogen lines gradually vanish with decreasing effective temperature. 
    The colour-bar values of each star range from $\min \log \varkappa $ to $\max \log \varkappa $.}
             \label{fig:monoch_opac_stars}
    \end{figure*}
    
    \begin{figure*}
    \centering
    \includegraphics[width=17.6cm]{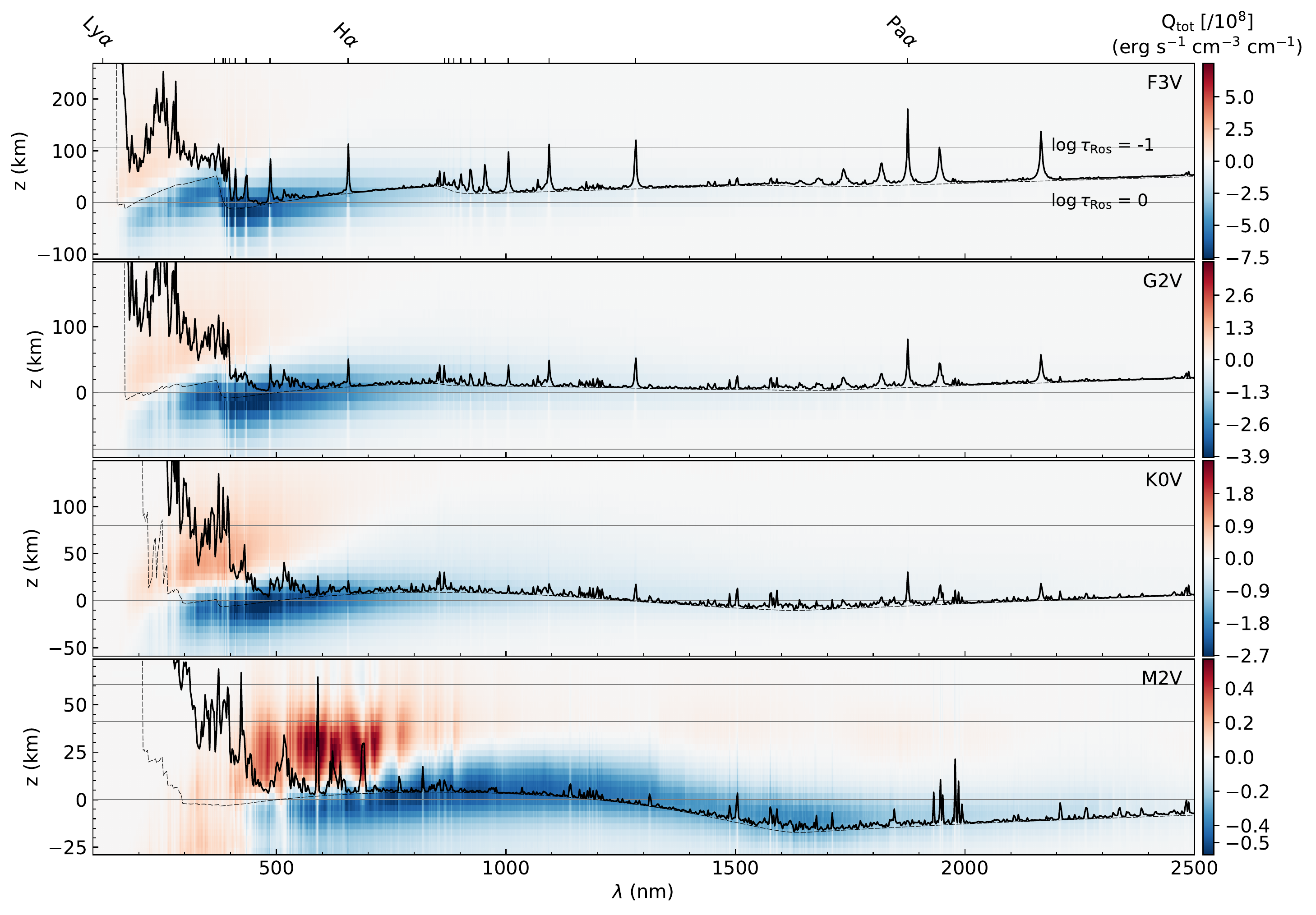}
    \caption{Monochromatic radiative energy exchange rate $Q_\nu$ (values shown in the \textit{colour--bars}) computed using the opacities from Fig.\  \ref{fig:monoch_opac_stars} for the models of the same four stellar types versus wavelength (horizontal axis) and geometrical height (vertical axis). Labels and overploted lines are same as in  Fig.\ \ref{fig:monoch_opac_stars}. The colour-bar values of each star range from $-\max |Q_\nu|$ to $\max |Q_\nu|$. \textit{Blue} tones in the colour-map correspond to negative $Q_\nu$, i.e.\ cooling in the atmosphere, and \textit{red} tones correspond to positive $Q_\nu$, i.e.\ heating.
    }
    \label{fig:monoch_Qr_stars}
    \end{figure*}

    For the computation of the monochromatic opacity $\varkappa_\nu$ we use \texttt{synple}
    , a Python wrapper for \texttt{SYNSPEC}
    . \texttt{SYNSPEC} is a general spectrum synthesis code that has been used to solve the RTE in different astrophysical scenarios, including cool stars (see, for example, \citealp{2003allende_NLTElatetype_I, 2003allende_NLTElatetype_II} and the \citealp{2010POLLUX, 2020apogee}). The main purpose of \texttt{SYNSPEC} is to synthesise spectra for a given model atmosphere, but it can also be used to compute lookup tables of monochromatic opacities for a given grid of thermodynamic quantities. The code solves the equation of state (EOS), and uses the number densities to compute the continuum and line opacities. 
    The EOS from \texttt{SYNSPEC} is fully consistent with the results of our own EOS solver (Vitas et al.\, in preparation), which is used to produce the EOS tables for the 3D simulations of stellar atmospheres with the \texttt{MANCHA3D} code \citep[][and Modestov et al.\ in preparation]{2010felipe,2017khomenko,2018khomenko}.

    The EOS in \texttt{SYNSPEC} is solved for a specified grid of temperature $T$ and mass density $\rho$. The number of nuclei per species (fixed by the abundances) and charge are conserved. 
    The code solves the chemical equilibrium equations to determine the populations for 38 atomic species. For temperatures lower than a certain value (in this work chosen to be 8000 K), neutral atoms and singly ionised atoms are included in the EOS (not taking into account higher ionisations) and the code also considers molecular formation (including 503 molecular species). 
    For higher temperatures, molecules are not included in the EOS, and higher ionisations are computed for the atoms. 
    For more details on the solution of the EOS in \texttt{SYNSPEC} (and other details and the general use of the code), we refer the reader to section 2.6 from \citet{synspec_tlusty_guide_IV}, and \citet{tlusty_guide_II, tlusty_guide_III}.

    Once the EOS is solved, the opacity is computed in the same ($T,\rho$) grid, using the atomic and molecular data from several sources (see App.\ \ref{app:opac_sources}). The data used by the code can be customised by the user by changing, for example, the line lists for the bound-bound transitions. The control over opacity contributors is important for the cooler stars, where omission of certain molecules (e.g.\ TiO and VO) or incomplete line lists may strongly affect both detailed spectral synthesis and the bulk opacities used for modelling. 
    
    For the ingredients listed in App.\ \ref{app:opac_sources}, we computed a lookup table using \texttt{synple}, with a microturbulence velocity of $2$ km s$^{-1}$ and for the solar abundances from \citet{AG89}. These solar abundances are outdated, but preferred in our study since they were used in the computation of the reference ODFs we will be comparing to. The $T$-axis of the grid has 60 values in the range $T \in [1995,125900]$ K, with constant step $\Delta \log T \simeq 0.03$ K (in the present text, $\log \equiv \log_{10}$).
    The $\rho$-axis has 35 values in the range $\rho \in [10^{-13}, 3.2 \times 10^{-3}]$ g cm$^{-3}$ with $\Delta \log \rho \simeq 0.3$. The table is computed for wavelengths in the range $\lambda \in [20, 9.5\times 10^{4}]$ nm, with $\Delta \log \lambda = 10^{-6}$ ($\lambda$ in nm). In total there is  around $7.7\times10^9$ data points in the table, stored in an HDF5 file of $230$ GB. Another table is computed for the same wavelength grid but with finer temperature and density axes (same number of points, but reduced ranges, $T \in [1995,19900]$ K and $\rho \in [10^{-9}, 3.2 \times 10^{-3}]$ g cm$^{-3}$). This table is more suitable for the M2V star, thus reducing the interpolation errors in the RTE solver.
    
    The total opacity (continuum and lines) in the wavelength range from $100$ nm to $2500$ nm is shown in Fig.\ \ref{fig:monoch_opac_stars}. It is interpolated for the temperatures and mass densities of the four 1D stellar models. The opacity clearly changes with different heights in the stellar models and with spectral type. 
    To check where the radiation at each wavelength contributes to the emergent radiation, we mark the geometrical heights at which $\tau_{\nu}=1$ (\textit{thick line}). This curve includes the contribution from both the line and the continuum opacity. Since the wavelength range is wide, we divided it into sub-intervals of 2.4 nm and computed the average height in each of them. Similarly, we overplot the formation heights of the continuum (\textit{thin line}), at which $\tau^{\mathrm{cont}}_{\nu}=1$.
    The ionisation edges and the typical shape of the bound-bound and bound-free H$^{-}$ opacity coefficient are visible for the $\tau^{\mathrm{cont}}_{\nu}=1$ curve. 
    In the UV the line forest elevates the height of the $\tau_\lambda=1$ curve. The strong spectral lines, especially the hydrogen series, are clearly visible as well as their relative decrease in strength with decreasing effective temperature. In the M star, the $\tau_\lambda=1$ curve is shaped by millions of molecular transitions.
    
    Following the method from Sect.\  \ref{sec:method}, we solve the RTE to obtain $Q_\nu$. 
    Once $Q_\nu$ is computed, we can integrate it in frequency using the trapezoidal rule:
    \begin{equation} 
        Q = \int_\nu Q_\nu d\nu \equiv \frac{1}{2} \sum_{i} (Q_{\nu_i} + Q_{\nu_{i+1}}) \Delta \nu_i. 
    \end{equation}
    The computed $Q_\nu$ is shown in Fig.\  \ref{fig:monoch_Qr_stars}, for the wavelength range and set of stellar models as in Fig.\ \ref{fig:monoch_opac_stars}. The radiative energy exchange happens mostly around the stellar surface. For decreasing effective temperature, the absolute value of $Q_\nu$ is large for a wider wavelength range and the significance of the contribution to $Q_\nu$ shifts from the visible for the F star towards the IR for the M star. Figure \ref{fig:monoch_Qr_stars} also shows how the relative magnitude and area covered by significant $Q_\nu$ for the heating in the atmosphere ($Q_\nu>0$) and the cooling ($Q_\nu<0$) changes for the different spectral types. 
    For the F star the maximum heating is around an order of magnitude smaller than the magnitude of the most intense cooling, and 
    peaks in the near-UV and blue wavelengths. For the M star, the heating is comparable to the cooling in magnitude, and 
    ranges from the visible to the near-IR. As in the opacity plot (Fig.\  \ref{fig:monoch_opac_stars}), the individual contributions to the line opacity from strong transitions are obvious for the hotter stars in the sample. In contrast, they almost disappear for the M star, while small structures due to the presence of the molecules become more evident.

    Figure \ref{fig:monoch_Qr_stars} provides a qualitative overview of the monochromatic radiative energy exchange. One must keep in mind that the computations presented here are done in the LTE approximation which is fairly good for most of the photospheric lines, but fails badly for lines that contribute to radiative losses in the chromosphere. However, the method of precomputed opacities is feasible only in the LTE approximation when there is unique mapping between the pair thermodynamic quantities and the computed total opacity. Our primary focus is therefore on the stellar surface (i.e.\ the top of the convection zone and the lower photosphere), from where the bulk of the radiation escapes the star. This is also the reason why we present the results for the bolometric $Q$ (erg cm$^{-3}$ s$^{-1}$) and not the ratio $Q/\rho$ (erg g$^{-1}$ s$^{-1}$), which would put more emphasis on the radiation exchange at lower densities.

\section{Opacity distribution functions} \label{sec:odf}

    \subsection{Construction of the opacity distribution function} \label{subsec:odf-construction}
    \begin{figure}
    \includegraphics[width=8.8cm]{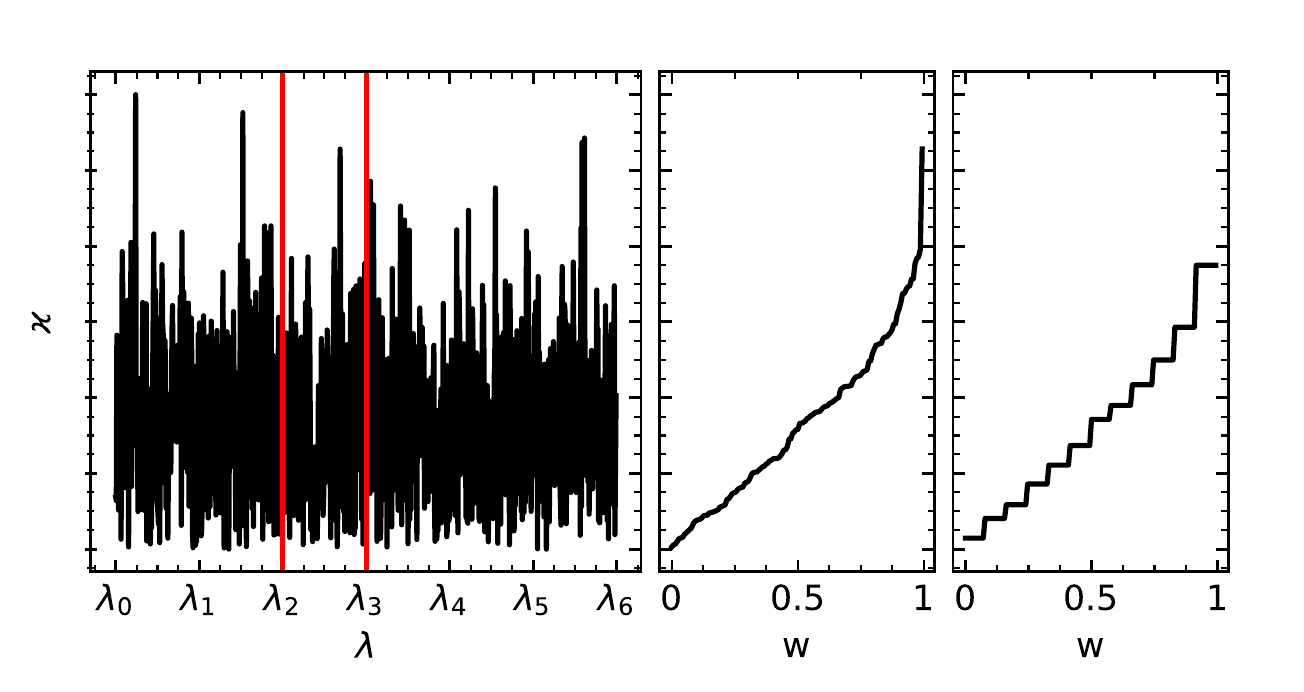}
    \caption{Three steps in construction of the ODF. \textit{Left}: schematic representation of monochromatic opacities in an interval of wavelengths divided into steps of $\lambda_i$. The \textit{red vertical lines} mark one ODF step with $\lambda \in [\lambda_2, \lambda_3]$. \textit{Middle}: opacities of that step sorted by magnitude. The distribution is mapped onto the unit interval $w \in [0, 1]$ and the wavelength dependence is lost within the step. \textit{Right}: sorted opacities discretised in ODF substeps.} 
             \label{fig:contruct_ODF}
    \end{figure}
    
    \begin{figure*}
    \centering
    \includegraphics[width=17.6cm]{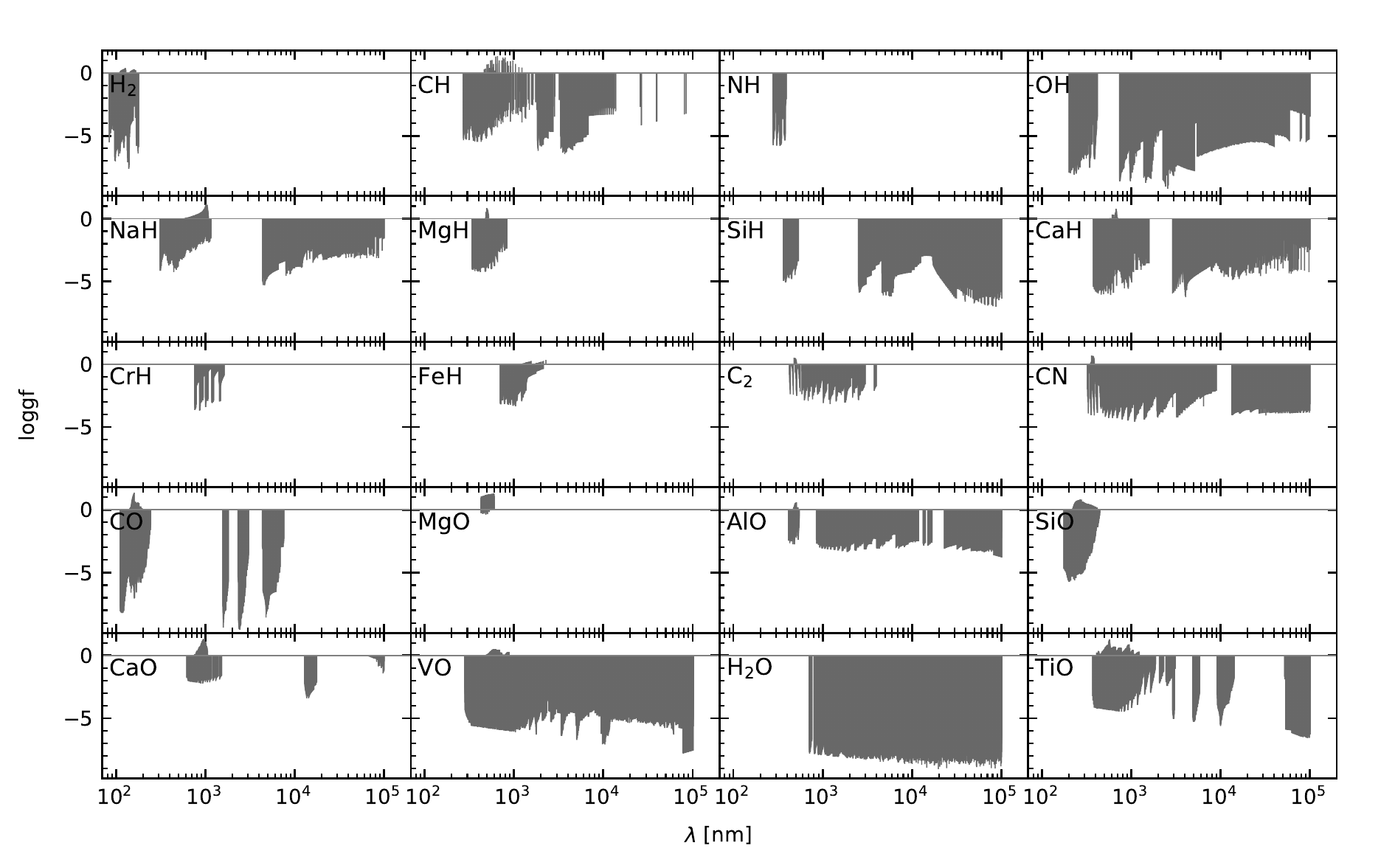}
    \caption{Oscillator strength $\log gf$ versus wavelength for all molecular spectral lines included in \texttt{SYNSPEC}. Notice that the transitions that mainly contribute to the opacity have typically values of $\log gf>-5$.}
             \label{fig:included_molecules}
    \end{figure*}

    \begin{figure}
    \includegraphics[width=8.8cm]{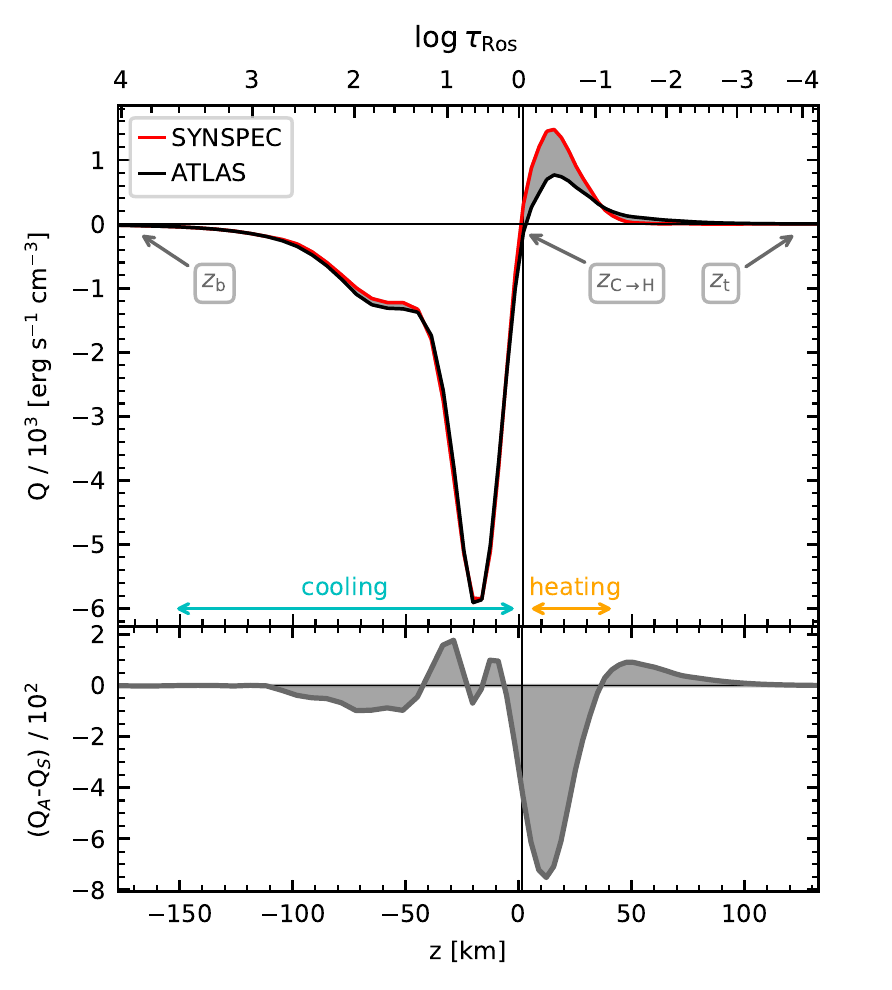}
    \caption{\textit{Top panel}: The radiative energy exchange rate $Q$ versus geometrical height $z$ for the M2V star computed using the ODF from \texttt{SYNSPEC} data (\textit{black solid line}) and the ODF from \texttt{ATLAS} (\textit{red solid line}). 
    The logarithm of the Rosseland optical depth is shown in the \textit{top axis}. 
    The area of the difference is shaded in \textit{grey}. The zero of the geometrical height $z$ is at the continuum optical depth at 500 nm $\tau_5=1$. The range of heights of the cooling ($Q<0$) and heating ($Q>0$) components are indicated in \textit{cyan} and \textit{orange} respectively. The heights $z_{\mathrm{b}}$, $z_{\mathrm{c \rightarrow h}}$, and $z_{\mathrm{t}}$ are used as integral bounds to calculate the deviation measures $\chi_{\mathrm{C}}$ and $\chi_{\mathrm{H}}$ (Eqs.\  \ref{eq:error_c}, \ref{eq:error_h}). The height $z_{\mathrm{c \rightarrow h}}$ is also marked with a \textit{vertical black line}. \textit{Bottom panel}: absolute difference between the $Q$ values.}
             \label{fig:error_dyagram}
    \end{figure}

    \begin{figure}
        \includegraphics[width=8.8cm]{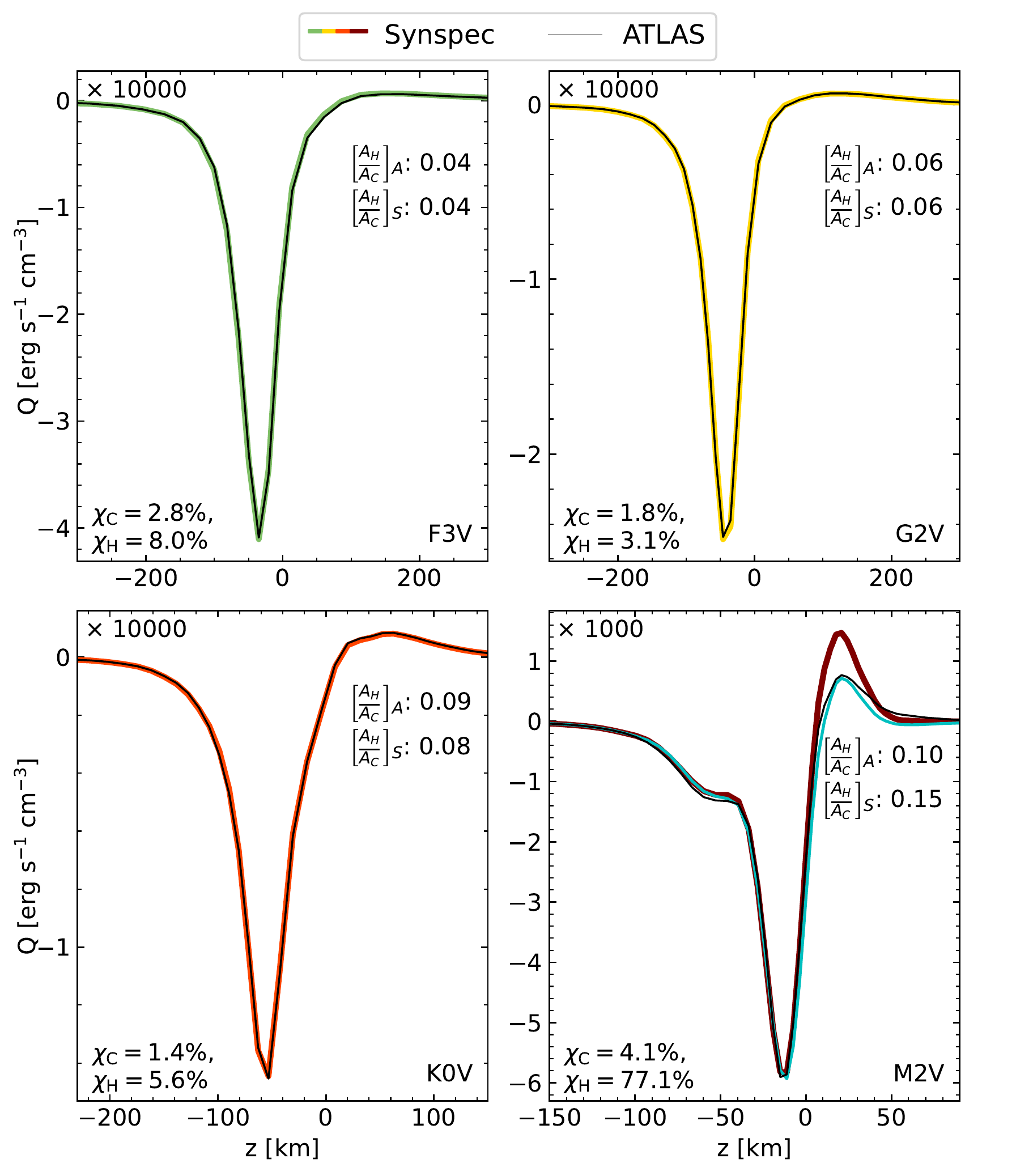}
        \caption{Bolometric radiative energy exchange rate $Q^{\mathrm{ODF}}$ against the geometrical height in the four stellar models. The results using ODF from \texttt{SYNSPEC} (\textit{coloured lines}) are compared to those using the ODF from \texttt{ATLAS} (\textit{thin black line}). \textit{Top left of each panel}: scaling factor of vertical axis. \textit{Bottom left of each panel}: difference between the two ODFs computed using Eqs.\  \ref{eq:error_c}, \ref{eq:error_h}. \textit{Top right of each panel}: ratio of the heating area with respect to the cooling area for the $Q$ for both data sets (\textit{A} for \texttt{ATLAS} and \textit{S} for \texttt{SYNSPEC}). In \textit{cyan in the bottom right panel} for the M2V star: ODF computed using monochromatic opacity from \texttt{SYNSPEC}, excluding the lines from TiO and VO. The corresponding deviations for that $Q$ curve are $\chi_{\mathrm{C}} \simeq 7$ and $\chi_{\mathrm{H}} \simeq 29$.}
        \label{fig:q_bolometric_atlas_vs_synspec}
    \end{figure}
    
    The ODF method \citep{1951labs_ODF, ODF1966} was invented to allow the accurate calculation of the radiative losses due to the millions of opacity contributors in a computationally efficient way. In the 1960s the solution for the RTE was prohibitive for more than a few wavelengths in 1D model atmospheres. Despite the exponential growth in computing power since then, the same problem of efficiently and accurately accounting for the millions spectral lines is still present today when the RTE is solved in 3D time-dependent simulations. 
    
    An ODF is constructed 
    to replace detailed monochromatic opacities by their statistical means in wavelength intervals, 
    following a procedure that preserves the accuracy of the computed bolometric radiative quantities while heavily reducing the amount of computation  
    \citep[see e.g.\ ][]{ODF1966,1970ATLAS,voegler_thesis}. First, one divides the full wavelength range into steps (an example for one step, defined by $\lambda \in [\lambda_2, \lambda_3]$, is shown at the left panel of Fig.\  \ref{fig:contruct_ODF}). Then, all the opacities within each of the steps are sorted by increasing magnitude (see the middle panel of Fig.\  \ref{fig:contruct_ODF}). 
    At this point the direct mapping between the monochromatic opacity values and their wavelengths is lost.
    To discretise this monotonic distribution within each step, a number of substeps are defined specifying a set of dimensionless weights $\omega_j$ ($\sum_j \omega_j=1$). The \textit{length} of a substep $\Delta \lambda_{i,j}$ is calculated in terms of the length of the step $\Delta \lambda_i$ and the weight $\omega_j$, so that $\Delta \lambda_{i,j} = \omega_j \Delta \lambda_i$\footnote{The length $\Delta \lambda_{i,j}$ is not an interval in the wavelength scale, but it counts a certain number of points in the distribution.}.
    Finally, the opacities are averaged computing the arithmetic mean within each substep (see the right panel of Fig.\  \ref{fig:contruct_ODF}).
    
    This procedure can be performed for any set of monochromatic opacities computed for a pair of thermodynamic quantities. One of them is usually the temperature $T$ and the other is either the density $\rho$ (as in \texttt{SYNSPEC}\footnote{\texttt{SYNSPEC} can also work with a grid of electron density or pressure.}) or the gas pressure $P$ \citep[as in the \texttt{ATLAS} code,][]{1970ATLAS}. \citet{kurucz1993_odf} created ODF tables\footnote{These can be found in \url{http://kurucz.harvard.edu/opacities.html}. 
    The particular table from \texttt{ATLAS} referred to in the present paper is in the section `Old Kurucz ODF files', computed for the solar abundances from \citet{AG89} and a turbulent velocity of $2$ km s$^{-1}$. The corresponding monochromatic opacities are not available.} in the described way for specific grids of ($T, P$). We apply the same strategy to the monochromatic opacities computed using \texttt{SYNSPEC} in a ($T, \rho$) grid, for the two monochromatic opacity tables described (see Sect.\ \ref{sec:monochromatic_opacity}). 
    The wavelength range of our \texttt{SYNSPEC} computation and derived ODF is shorter than the range covered by the \texttt{ATLAS} ODF, $\lambda \in [9,10^7]$ nm. When the \texttt{ATLAS} table is used to compute the bolometric RT quantities in the different model atmospheres, there is no difference between using it over its entire $\lambda$ range and in the reduced range of our \texttt{SYNSPEC} computations. This is not surprising since the emission of cool stars peaks in spectral regions far from the extremes that we left out from the full wavelength interval of the \texttt{ATLAS} tables. For our ODF calculation, the wavelength range is divided into 291 steps, which have exactly the same locations and the same 12 weights for the substeps as \texttt{ATLAS}. 
    The length of the steps is approximately proportional to the wavelength \citep[see the table from][to check the exact values]{kurucz1993_odf};  and the substeps are constructed using the weights $\omega_j = \left \{ 0.1, 0.1, 0.1, 0.1, 0.1, 0.1, 0.1, 0.1, 0.1, 0.05, 2/60, 1/60 \right \}$. This kind of non-uniform weighting is taken from the \texttt{ATLAS} code and had been demonstrated \citep{cernetic2019} to perform particularly well in the visible and infrared for the solar case, being less accurate than uniform weighting in the UV (only important to compute detailed spectra). Our ODF contains the total (continuum and lines) opacity in cm$^2$ g$^{-1}$, just as it is the case for \texttt{ATLAS}. 
    
    The tables from \citet{kurucz1993_odf} have been widely used in the context of MHD simulations of solar and stellar atmospheres \citep[see, for example,][and many others]{voegler_thesis, 2018khomenko,2013beeck}. The atomic and molecular line lists used to compute these tables \citep{kurucz1993_atoms, kurucz1993_molecules} are sufficiently complete and accurate to make these simulations closely resembling various observational diagnostics. However, apart from the molecules that are included (H$_2$, HD, MgH, NH, CH, SiH, OH, CH, C$_2$, CN, CO, SiO), many are missing (e.g.\ VO and TiO) which are dominant opacity sources in the spectra of M stars. \texttt{SYNSPEC}, however, allows for a more complete set of molecules (see App.\ \ref{app:opac_sources} and Fig.\  \ref{fig:included_molecules}) and includes an up-to-date collection of atomic and molecular data (partition functions and line lists). However, to evaluate the effect of the updated data, one has to compare the two data sets carefully. 

    \subsection{Comparison of opacity distribution function from \texttt{SYNSPEC} and \texttt{ATLAS}} \label{subsec:comparison_ODF_SYNS_vs_ATLAS}
    While there is overall good match between the two data sets, there are also significant differences. To understand better these differences, we compare the opacity of each step and substep of the ODFs interpolated for the values of $T$ and $\rho$ along the four 1D model of atmospheres. Four examples representing typical differences are shown in the App.\ \ref{app:details_A_vs_S}, in Figs.\  \ref{fig:ODF_AvsS_590-600}-\ref{fig:ODF_AvsS_193-197} 
    (where black lines represent the opacity from \texttt{ATLAS} along the models, and the coloured lines on top of them represent the the opacity from \texttt{SYNSPEC}).

    In parallel, we also compare the bolometric $Q$ rate computed using both ODFs. We apply the method presented in Sect.\  \ref{sec:method} to solve the RTE and compute $Q_{i,j}$ for every step $i$ and substep $j$; we then integrate in frequency following the ODF formalism:
    \begin{equation}
        Q^{\mathrm{ODF}} = \sum_i Q_{i} \equiv \sum_i \Delta \lambda_i \sum_j Q_{i,j} \omega_j,
    \end{equation}
    where $Q_{i}$ is the rate integrated in wavelength in the $i$th step.
    
    Comparing the values of $Q$ computed from different data sets is not straightforward. An example of $Q$ computed for the model of the M star using two opacity sets and their absolute difference is shown in Fig.\  \ref{fig:error_dyagram}. In the figure we identify the typical negative feature that corresponds to the strong cooling at the bottom of the photosphere and the relatively smaller positive one that shows the radiative heating (blanketing effect). Higher up and deeper down in the atmosphere, the values of $Q$ can be several orders of magnitude smaller. While the radiative losses at these heights may still be significant, we focus here on the dominant components around the optical depth unity. Wherever the values of $Q$ are small, the relative difference computed between them may be exaggerated with respect to the importance of these differences from the energy balance in the atmosphere. Moreover, as the sign of $Q$ flips in the region of interest (the interval of heights around $\tau = 1$) and the two $Q$ values computed from \texttt{ATLAS} and \texttt{SYNSPEC} do not necessary flip the sign at exactly the same height, the relative difference around that height might also be misleadingly high. Therefore, instead of computing the relative differences in the entire domain, we focus on the two dominant features and measure the deviation for the cooling and heating part separately. 

    For each of them we compare the area $A$ of the difference of $Q$ computed for each of the ODF data sets normalised to the area of one of them:  
    \begin{equation} \label{eq:error_c}
        \mathrm{\chi_{\mathrm{C}}} = \frac{ A \left( \left | Q^{(1)}  -  Q^{(2)} \right |  \right)_{\mathrm{C}} }{A \left( \left | Q^{(1)}  \right | \right)_{\mathrm{C}}},
    \end{equation} 
    \begin{equation} \label{eq:error_h}
        \mathrm{\chi_{\mathrm{H}}} = \frac{ A \left( \left | Q^{(1)}  -  Q^{(2)} \right |  \right)_{\mathrm{H}} }{A \left( \left | Q^{(1)}  \right | \right)_{\mathrm{H}}},
    \end{equation}
    where we choose as $Q^{(1)}$ and $Q^{(2)}$ the $Q$ computed with the ODF from \texttt{ATLAS} and \texttt{SYNSPEC}, respectively. In Sect.\  \ref{sec:opacity_binning} the same is used to measure discrepancy between the ODF and OB results. The areas are defined as $A \left( f(z) \right)_C = \int_{z_{\mathrm{b}}}^{z_{\mathrm{C \rightarrow H}}} f(z) dz$ and $A \left( f(z) \right)_H = \int_{z_{\mathrm{C \rightarrow H}}}^{z_{\mathrm{t}}} f(z) dz$. The height at $z_{\mathrm{C \rightarrow H}}$ is the height where $Q$ changes sign for the first time above the cooling component; $z_{\mathrm{b}}$ and $z_{\mathrm{t}}$ are, respectively, the highest point in the atmosphere under the surface ($z<0$ in the plots) and the lowest point over the surface ($z>0$) where $\left| Q \right|< 2 \times 10^{-4} \left| \min( Q ) \right|$ (see Fig.\  \ref{fig:error_dyagram}). If 
    $A \left( Q  \right)_{\mathrm{H}} < 1\% \; A \left( Q  \right)_{\mathrm{C}}$, $\chi_{\mathrm{H}}$ is not computed (these cases are marked as $\chi_{\mathrm{H}}=$ - - - $\%$ in the figures). 

    The radiative energy exchange terms $Q_i$ computed from the four models using the same steps as in Figs.\  \ref{fig:ODF_AvsS_590-600}-\ref{fig:ODF_AvsS_193-197} are shown in App.\ \ref{app:details_A_vs_S}, in Figs.\  \ref{fig:Q_AvsS_590-600}-\ref{fig:Q_AvsS_193-197}. The corresponding values of the deviation measures are indicated in each panel.
    
    When comparing the opacity from the ODF of the two data sets, we see a general match in behaviour and values in most of the visible and infrared intervals for the F, G, and K stars.
    An example for a step with a good match ($\lambda \in [590, 600]$ nm) is shown in Fig.\  \ref{fig:ODF_AvsS_590-600}. The $Q_i$ values computed in the same step closely overlap for these stars with all $\chi_{\mathrm{C}}$ and $\chi_{\mathrm{H}}$ between $1.4\%$ and $3.1\%$ (see, for example, Fig.\ \ref{fig:Q_AvsS_590-600}). 
    For the near-UV, visible, and IR, the steps that include strong lines show significant differences in the opacity. Some examples are the steps that include hydrogen H$\alpha$  
    (Fig.\  \ref{fig:ODF_AvsS_650-660}) or Na{\small I} D1 and D2 (Fig.\  \ref{fig:ODF_AvsS_580-590}). Similar discrepancies are found in the steps containing other strong lines like the Ca{\small II} H and K lines, the G band, H{\small I} $\beta$, Mg{\small I} b1, the Ca{\small II} triplet, and the most intense lines from the Paschen, Brackett, and Pfund series. In the case of all the hydrogen series the opacity from \texttt{ATLAS} is higher than the one from \text{SYNSPEC} in the deeper part of the atmospheres, but this reverses in the upper part (see Fig.\  \ref{fig:ODF_AvsS_650-660}). For these hydrogen lines, such opacity behaviour is present in all the substeps, although the reversal occurs deeper in the less opaque substeps. We notice a similar tendency in the G band. Another trend can be seen for the steps containing Ca{\small II} lines, Mg{\small I} b1 and Na{\small I} D1 and D2, for which the opacity from \text{SYNSPEC} is higher than the one from \text{ATLAS} at all the heights of the atmospheres for the most opaque substeps (see Fig.\  \ref{fig:ODF_AvsS_580-590}). The number of affected substeps varies between the 2 and 4 most opaque, depending on which intense line falls within the step. The mismatch is affected by the number of strong lines contained in the interval, the magnitude of their opacity with respect to the continuum, and their width relative to the length of the interval. 
    In the less opaque substeps, the two data sets give very close results.

    In the steps that contain strong spectral features, the radiative energy exchange rates generally show $\chi_{\mathrm{C}},\chi_{\mathrm{H}} \lesssim 5-10 \%$ for the F, G, and K stars (see, for example, Figs.\  \ref{fig:Q_AvsS_650-660},\ref{fig:Q_AvsS_580-590}). An exception happens for the G star for the step containing hydrogen H$\alpha$ %H{\small I} $\alpha$ 
    where $\chi_{\mathrm{H}} \simeq 25 \%$ (see Fig.\  \ref{fig:Q_AvsS_650-660}). However, in the latter example, the area of the heating feature is only 0.02 times the area of the cooling, making this discrepancy less important. 

    The differences between \texttt{SYNSPEC} and \texttt{ATLAS} for the strong atomic lines could be, for example, due to the wavelength resolution in the monochromatic opacity tables or the broadening parameters used in the line synthesis (e.g.\ Van der Waals broadening or damping constants). 
    In the case of the hydrogen lines, the difference is explained by the treatment of the Stark broadening. For this project, we used approximate Stark broadening profiles after \citet{hubeny1994}, while in \texttt{ATLAS} the tables from \citet{Vidal1973} are used. 
    Without having access to the monochromatic opacities from \texttt{ATLAS} it is difficult to pinpoint the exact reason of the discrepancies for the rest of the species. 
    Although both codes are open source, their complexity makes it very difficult to find the exact source of differences by comparing the actual source code.

    Apart from the spectral features discussed above, the comparison of the two opacity sets in the visible and IR for the F, G, and K stars reveals an excellent agreement. At the same time, there is a notable mismatch for the opacity in the UV between the two data sets (see an example in Fig.\  \ref{fig:ODF_AvsS_193-197}). These discrepancies are also evident in the corresponding values of $Q_i$ (see an example in Fig.\  \ref{fig:Q_AvsS_193-197}), for which we find differences of around $\chi_{\mathrm{C}}, \chi_{\mathrm{H}} \simeq 5-40\%$ for $\lambda \gtrsim 190$ nm (as in the example from Fig.\ \ref{fig:Q_AvsS_193-197}), and around $\chi_{\mathrm{C}}, \chi_{\mathrm{H}} \simeq 30-80\%$ for $\lambda < 190$ nm (in other steps in the ultraviolet).

     The differences between the two data sets in the UV are likely due to the use of photoionisation cross-sections from the Opacity Project \footnote{\url{http://cdsweb.u-strasbg.fr/topbase/publi.html}} and the Iron Project \citep{1995nahar, 1997bautista} in \texttt{SYNSPEC} \citep[see the discussion about the missing opacity in the Sun in][]{2003allende_NLTElatetype_II}, while in the \texttt{ATLAS} table from 1993 the cross-sections come mainly from \citet{Henry1970} and \citet{Peach1970}.

    With all the similarities found in the visible and infrared, and the nearly perfect match we see in the bolometric $Q$ for the F, G, and K stars (Fig.\  \ref{fig:q_bolometric_atlas_vs_synspec}), it may be concluded that both ODF tables are fairly similar and reproduce a close solution of the RTE. 
    
    However, both the comparison of the opacity and the computed values of $Q$ from \texttt{SYNSPEC} and \texttt{ATLAS} reveals much larger discrepancies for the M star. The opacities from \texttt{SYNSPEC} are, in general, larger than the ones from \texttt{ATLAS} for the points at the top of the M atmosphere. The corresponding values of $\chi_{\mathrm{C}}$ and $\chi_{\mathrm{H}}$ are also larger, for most of the steps in the visible and NIR they are in the range $\simeq$10--35\% ($\simeq$40--60\% in the worst cases). 

    As previously discussed, the discrepancies between the two tables for the later-type star are expected owing to different selection of molecules and line lists. The discrepancy is prominent in the heating component of the bolometric $Q$ (Fig.\  \ref{fig:q_bolometric_atlas_vs_synspec}), for which $\chi_{\mathrm{H}} \simeq 77$\%, in contrast to the values $\chi_{\mathrm{C}} < 3 \%$ and $\chi_{\mathrm{H}} < 10 \%$ found for the F, G, and K stars. To demonstrate the effect of certain molecules on the discrepancies between the data sets, we recompute the \texttt{SYNSPEC} table and $Q$, excluding TiO and VO from the line lists for the M star (blue curve in the bottom right panel of Fig.\  \ref{fig:q_bolometric_atlas_vs_synspec}). In this case, the value of $\chi_{\mathrm{H}}$ is considerably reduced, but it is still relatively high ($\simeq$29\%). Even if the selection of molecules in the two codes is identical, differences between the opacities are expected owing to different line lists and other atomic and molecular data. For example, the use in \texttt{SYNSPEC} of more up to date data for the opacities, in particular for H$_2$O and TiO, from EXOMOL \footnote{\url{http://www.exomol.com/data}}, or different partition functions and oscillator strengths.

\section{Opacity binning} \label{sec:opacity_binning}

   \begin{figure}
    \includegraphics[width=8.8cm]{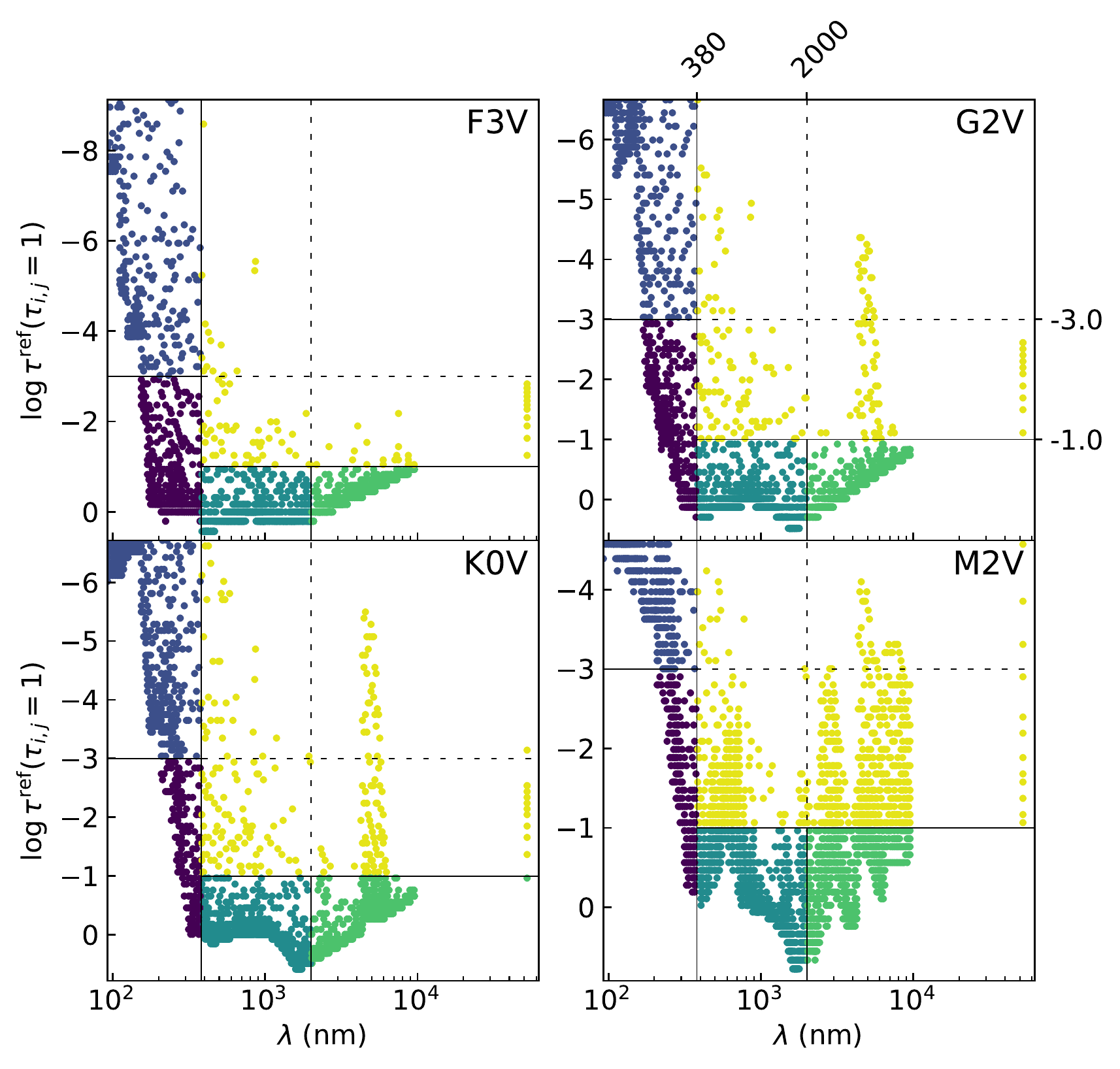}
    \caption{Example of the reference optical depth and wavelength of the ODF points organised into bins for the F3V, G2V, K0V, and M2V stars. As explained in the text, the Rosseland optical depth shown in the vertical axis is used as the reference $\tau^{\mathrm{ref}} (z_{i,j})$, where $z_{i,j}$ is the geometrical height at which $\tau_{i,j}=1$ for each step and substep of the ODF. We show the same $\{\tau, \lambda \}$ binning configuration for all the stars. The colours indicate different bins. Each point corresponds to one ODF substep, all centred in the middle wavelength of their corresponding step. The last column of dots at the right of each panel corresponds to the ODF step with $\lambda \in [9600, 9.5 \times 10^4]$ nm, and thus, appears separated from the rest of the dots.}
             \label{fig:binning_indexes_example}
    \end{figure}
    
    \begin{figure}
    \includegraphics[width=8.8cm]{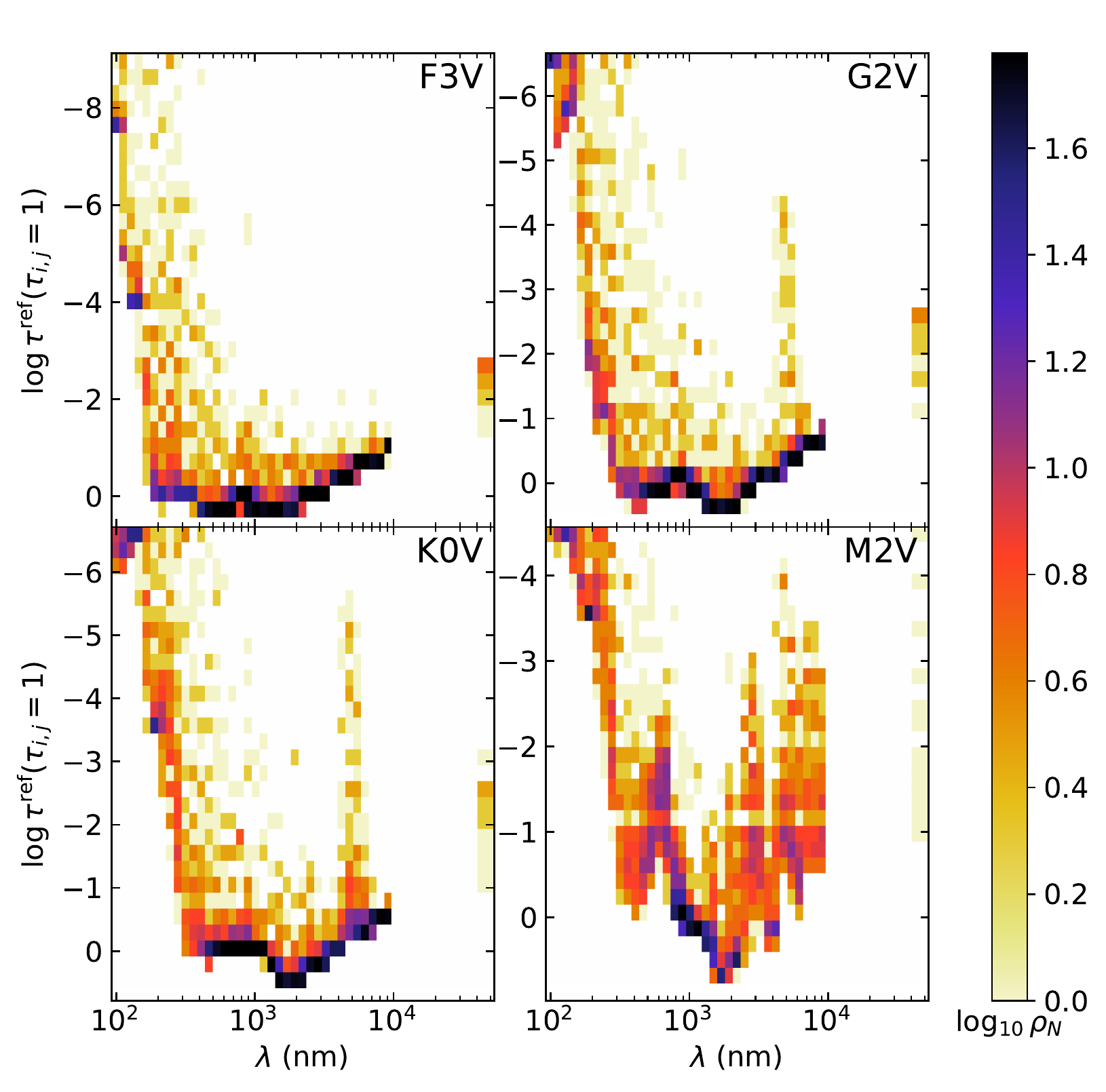}
    \caption{Histogram of the logarithm of the relative number of points from the ODF as a function of $\tau^{\mathrm{ref}} (z_{i,j})$ in the vertical axis and $\lambda$ in the horizontal, as in Fig.\  \ref{fig:binning_indexes_example}.  The density of the ODF points is the largest in the continuum (especially in the visible) and in the UV owing to the sampling of the ODF steps (see the description of the ODF tables in Sect.\  \ref{sec:odf}).} 
             \label{fig:histogram_indexes}
    \end{figure}

    A further approximation of the opacity computation is the OB method initially introduced by \citet{norlund1982}. Its implementation in the MURaM and CO$^5$BOLD %Co5bold 
    codes is described by \citet{voegler_thesis} and \cite{ludwig_thesis} respectively. In this method the opacities are sorted taking into account their distribution with the discretised optical depth $\tau^{\mathrm{ref}}$ of a representative one-dimensional model atmosphere. We use the Rosseland optical depth as reference. 
    An opacity bin $l$ is defined by selecting a pair of lower and upper $\tau$-separators ($\tau^{\mathrm{ref,l}}_0$, $\tau^{\mathrm{ref,l}}_1$).
    In a variation of the method \citep[see][]{2014MagicThesis} the wavelength dependence is partially preserved by grouping opacities by both optical depth and wavelength. In that case each bin is defined by two pairs of separators, one in optical depth ($\tau$-separators) and one in wavelength ($\lambda$-separators).
    
    We start from the ODF 
    constructed as it is described in Sect.\  \ref{sec:odf}.
    The optical depth is computed for each step $i$ and substep $j$ of the ODF  for all geometrical heights $z$ in a model atmosphere as
    \begin{equation}
        \tau_{i,j} (z) = \tau_{i,j} (z_{\mathrm{top}}) \, + \int_{z_{\mathrm{top}}}^z \varkappa_{i,j} (z)\, \rho\, \mathrm{d} z ,
    \end{equation}
    where $\rho$ refers to the mass density in the model atmosphere (and not the density in the grid of the opacity table) and the height dependence in $\varkappa_{i,j} (z)$ denotes that the opacity from the ODF is interpolated into the temperatures and densities from the model atmosphere. 
    The opacities are then grouped so that opacity of the ODF step $i$ and substep $j$ belong to bin $l$ if $\tau^{\mathrm{ref,l}}_0 \leq \tau^{\mathrm{ref}} (z_{i,j}) < \tau^{\mathrm{ref,l}}_1$, where $z_{i,j}$ is the height at which $\tau_{i,j}=1$. Additionally, when the $\lambda$-separators are used, every ODF value is sorted by wavelength so that the central wavelength of an ODF step falls between a pair of $\lambda$-separators. Figure \ref{fig:binning_indexes_example} shows examples of such grouping of the ODF values into five bins for our four stellar models. There are two $\lambda$-separators approximately dividing the wavelength axis into ultraviolet, visible, and infrared and there is one $\tau$-separator for the UV ($\log \tau = -3$) and a joint one for the visible and IR ($\log \tau = -1$). The corresponding 2D histograms of the ODF points are shown in Fig.\ \ref{fig:histogram_indexes}.  The minimal optical depth in the reference models vary with the stellar type and wavelength. We assign the top geometrical height of the atmosphere $z_{i,j}=z_{\mathrm{top}}$ to all ODF values with $\tau_{i,j}(z_{\mathrm{top}}) > 1$.
    
    After the opacities of the ODF have been distributed into bins, for every bin $l$ (that includes the ODF steps $i=i(l)$ and substeps $j=j(i,l)$ contained in the bin) and for every $T,\rho$ of the grid of the ODF we compute the Planck function, 
    \begin{equation}
        \mathrm{B}_l = \sum_{i(l)} \Delta \lambda_i \mathrm{B}_{i},
    \end{equation}    
    its derivative,
    \begin{equation}
        \left. \frac{\partial \mathrm{B}}{\partial \mathrm{T}}\right|_l = \sum_{i(l)} \Delta \lambda_i \frac{\partial \mathrm{B}_{i}}{\partial \mathrm{T}},
    \end{equation}   
    the Planck mean opacity,
    \begin{equation}
        \overline{\varkappa}^{\mathrm{Pl}}_l = \left( \sum_{i(l)} \Delta \lambda_i \mathrm{B}_{i} \sum_{j(i,l)} \omega_j \varkappa_{i,j} \right) \Bigg / \mathrm{B}_l,
    \end{equation}    
    and the Rosseland mean opacity,
    \begin{equation}
        \overline{\varkappa}^{\mathrm{Ro}}_l = \left. \frac{\partial \mathrm{B}}{\partial \mathrm{T}}\right|_l \Bigg / \left( \sum_{i(l)} \Delta \lambda_i \frac{\partial \mathrm{B}_{i}}{\partial \mathrm{T}} \sum_{j(i,l)} \frac{\omega_j}{\varkappa_{i,j}} \right), 
    \end{equation}    
    where $\Delta \lambda_i$, $\omega_j$, and $\varkappa_{i,j}$ are, respectively, the length of the wavelength step, the weight of the substep, and the opacity of the $i$th step and $j$th substep of the ODF. $\mathrm{B}_{i}$ is the Plank function in the middle of the wavelength step for each temperature $T$ in the ($T, \rho$) grid used to compute the ODF.
    
    Finally, following \citet{ludwig_thesis}, $\overline{\varkappa}^{\mathrm{Pl}}_l$ and $\overline{\varkappa}^{\mathrm{Ro}}_l$ are combined to get a mean value of the opacity for each bin:
    \begin{equation} \label{eq:mean_opac}
        \overline{\varkappa}_l = \left( 2^{-\tau_l /\tau_{\mathrm{thr}}} \right) \overline{\varkappa}^{\mathrm{Pl}}_l + \left( 1-2^{-\tau_l /\tau_{\mathrm{thr}}} \right) \overline{\varkappa}^{\mathrm{Ro}}_l,
    \end{equation}
    where the transition between thin and thick optical regimes is at the threshold optical depth $\tau_{\mathrm{thr}}=0.35$ and $\tau_l$ is computed 
    using pressure $p$ in the ODF grid and a stellar surface gravity $g$, approximately as $\tau_l = \overline{\varkappa}^{\mathrm{Ro}}_l p/g$. 
    With this definition, the opacity from each bin converges to the grey approximation for large enough optical depths (see the discussion about Fig.\ \ref{fig:OB_goes_grey} in App.\ \ref{app:not_convergence}).
    
    As in the ODF case, we compute the bolometric $Q$ rate, $Q^{\mathrm{OB}}$, by solving the RTE for each bin and summing over the bins:
    \begin{equation}
        Q^{\mathrm{OB}} = \sum_l Q_l,
    \end{equation}
    where $Q_l$ refers to partial rate of the bin $l$ computed using Eq.\ \ref{eq:bruls_Q}.
    
    We apply the described procedure to the ODF computed with \texttt{SYNSPEC} monochromatic opacities presented in Sect.\  \ref{sec:odf}. 
    There are several important free parameters in the procedure: the model atmosphere itself, the number of bins in $\tau$ and $\lambda$, and the locations of separators between the bins. 
    There is no obviously intuitive choice for these parameters. 
    We designed numerical experiments to test various alternatives.  
    As a deviation 
    measure of the radiative energy exchange rate $Q^{(2)}$ (computed using the binned opacity) with respect to the reference rate $Q^{(1)}$ (computed with ODF), \citet{voegler_opacity_binning} plotted the height-dependent absolute difference of $Q^{(1)}$ and $Q^{(2)}$ 
    divided by the density stratification $\rho$ of the atmosphere. The scaling with density is important to compare the radiative rates with other energy sources and sinks in the simulations. Alternatively, \citet{2014MagicThesis} opted for computing $\max|Q^{(1)}  -  Q^{(2)}|/ \max|Q^{(1)}|,$ to reduce the deviation measure to one value for a given choice of parameters. The latter choice is obviously more practical for 
    trial-and-error optimisation of the free parameters. However, in this approach the deviation is biased towards the dominant cooling component of $Q$.
    In our experiments, we take a compromise by using the deviation measures introduced in Eqs.\  \ref{eq:error_c} and \ref{eq:error_h}. The reference values of the $Q$ rate are those computed using ODF with \texttt{SYNSPEC} ($Q^{(1)}$) and the values that are tested ($Q^{(2)}$) are the ones calculated using the binned opacities. This choice of the deviation measures, allows us to automate the process of optimising the free-parameters, while it preserves separate information on both cooling and the heating component of $Q$. 
    As this is a multi-parameter study we break the problem down by first studying the influence of the location of the $\tau$-separators for a fixed number of bins (4) and with no bins in the wavelength direction. We then allow the number of $\tau$ bins to change. Finally, we test several cases including the $\lambda$ bins. 
    
    \subsection{$\tau$-bin location} \label{subsec:tau_bin_location}
    
    \begin{table}[]
    \caption{Percentage of the combinations of the three separators (4 bins) with certain $\chi_{\mathrm{C}}$ values for the four stellar types.}
    \begin{tabular}{ccccc}
    \hline
        & \multicolumn{4}{c}{Percentage of combinations}           \\ \hline
        & \multicolumn{1}{c}{$\chi_{\mathrm{C}}<20\%$} & \multicolumn{1}{c}{$\chi_{\mathrm{C}}<10\%$}   & \multicolumn{1}{c}{$\chi_{\mathrm{C}}<5\%$}& $\chi_{\mathrm{C}}<3\%$ \\ \hline
    F & \multicolumn{1}{c}{98}  & \multicolumn{1}{c}{74}   & \multicolumn{1}{c}{2.3} & 0  \\ \hline
    G & \multicolumn{1}{c}{100} & \multicolumn{1}{c}{96}   & \multicolumn{1}{c}{87}  & 43 \\ \hline
    K & \multicolumn{1}{c}{100} & \multicolumn{1}{c}{99}   & \multicolumn{1}{c}{93}   & 34 \\ \hline
    M & \multicolumn{1}{c}{100} & \multicolumn{1}{c}{28}   & \multicolumn{1}{c}{2.4}   & 0  \\ \hline
    \end{tabular}
    \label{tab:stats_cooling}
    \end{table}

    \begin{table}[]
    \caption{Percentage of the combinations of the three separators (4 bins) with certain $\chi_{\mathrm{H}}$ values for the four stellar types.}
    \begin{tabular}{ccccc}
    \hline
        & \multicolumn{4}{c}{Percentage of combinations}           \\ \hline
        & \multicolumn{1}{c}{$\chi_{\mathrm{H}}<50\%$} & \multicolumn{1}{c}{$\chi_{\mathrm{H}}<30\%$}   & \multicolumn{1}{c}{$\chi_{\mathrm{H}}<20\%$}& $\chi_{\mathrm{H}}<10\%$ \\ \hline
    F & \multicolumn{1}{c}{64} & \multicolumn{1}{c}{55}  & \multicolumn{1}{c}{50} & 16 \\ \hline
    G & \multicolumn{1}{c}{85} & \multicolumn{1}{c}{78}  & \multicolumn{1}{c}{55} & 8.9 \\ \hline
    K & \multicolumn{1}{c}{83} & \multicolumn{1}{c}{77}  & \multicolumn{1}{c}{62} & 6.8 \\ \hline
    M & \multicolumn{1}{c}{56} & \multicolumn{1}{c}{7.4} & \multicolumn{1}{c}{0.4} & 0 \\ \hline
    \end{tabular}
    \label{tab:stats_heating}
    \end{table}

    \begin{table}[]
    \caption{Percentage of the combinations of the three separators (4 bins) with certain simultaneous $\chi_{\mathrm{C}}$ and $\chi_{\mathrm{H}}$ values for the four stellar types.}
    \begin{tabular}{ccccc}
    \hline
        & \multicolumn{4}{c}{Percentage of combinations}           \\ \hline
        & \multicolumn{1}{c}{$\chi_{\mathrm{C}}<10\%$} & \multicolumn{1}{c}{$\chi_{\mathrm{C}}<10\%$} & \multicolumn{1}{c}{$\chi_{\mathrm{C}}<5\%$} & $\chi_{\mathrm{C}}<5\%$ \\ 
        & \multicolumn{1}{c}{$\chi_{\mathrm{H}}<50\%$} & \multicolumn{1}{c}{$\chi_{\mathrm{H}}<20\%$} & \multicolumn{1}{c}{$\chi_{\mathrm{H}}<20\%$} & $\chi_{\mathrm{H}}<15\%$ \\ \hline
    F & \multicolumn{1}{c}{64} & \multicolumn{1}{c}{50}  & \multicolumn{1}{c}{1.8} & 1.2 \\ \hline
    G & \multicolumn{1}{c}{85} & \multicolumn{1}{c}{55}  & \multicolumn{1}{c}{55}  & 32  \\ \hline
    K & \multicolumn{1}{c}{83} & \multicolumn{1}{c}{62}  & \multicolumn{1}{c}{62}  & 25  \\ \hline
    M & \multicolumn{1}{c}{14} & \multicolumn{1}{c}{0.4} & \multicolumn{1}{c}{0}   & 0   \\ \hline
    \end{tabular}
    \label{tab:stats_cooling_and_heating}
    \end{table}
    
    \begin{figure*}
        \centering
        \includegraphics[width=17.6cm]{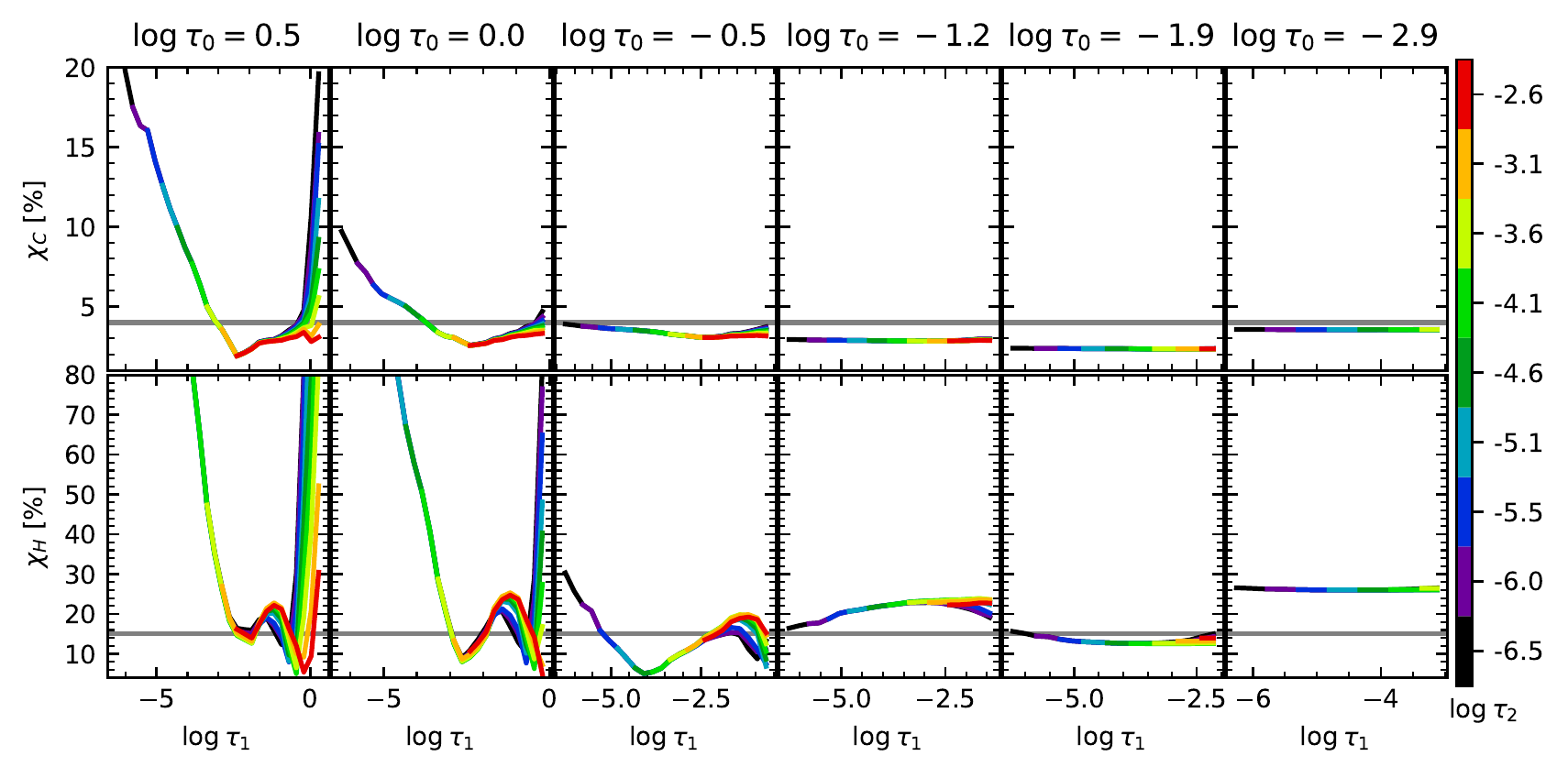}%.png}
        \caption{Deviation measures $\chi_{\mathrm{C}}$ (\textit{upper row}) and $\chi_{\mathrm{H}}$ (\textit{lower}) for the G2V star as function of the location of the separator $\tau_1$ when the other two separators are fixed. \textit{Columns} correspond to six different fixed values of $\tau_0$ (values indicated at the top of each column). Each curve corresponds to a different location of the top separator $\tau_2$. The colour of each curve indicates the location of $\tau_2$ used to compute that curve. We do not show all the possible locations for $\tau_0$ and $\tau_2$ to avoid overcrowding of the plots. \textit{The grey horizontal lines} show $\chi_{\mathrm{C}}=4\%$ and $\chi_{\mathrm{H}}=15\%$.} 
        \label{fig:chis_2x5_fixedTau0}
    \end{figure*}

    \begin{figure*}
        \centering
        \includegraphics[width=17.6cm]{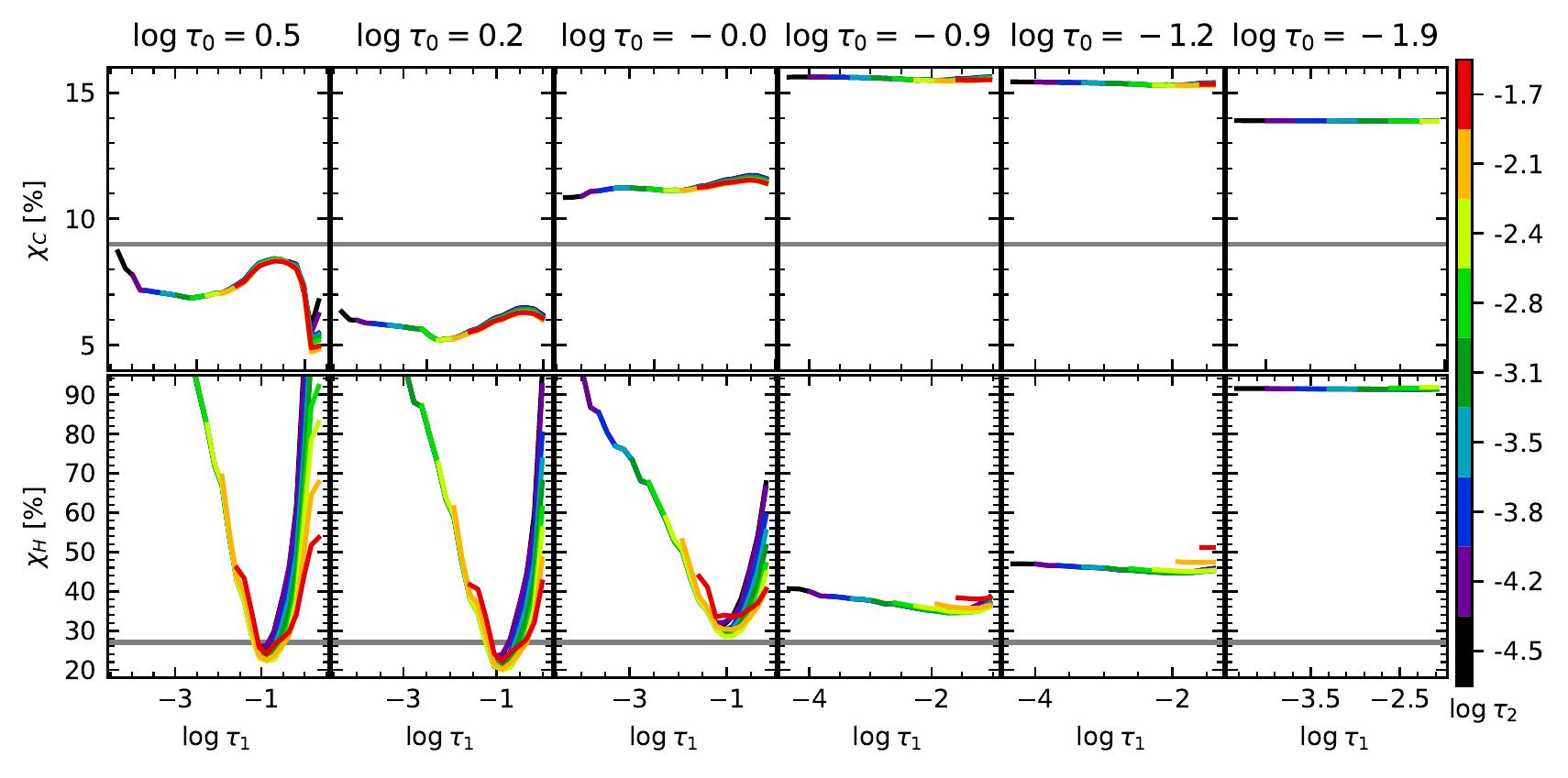}
        \caption{Same as in Fig.\  \ref{fig:chis_2x5_fixedTau0}, but for the M2V star. \textit{The grey horizontal lines} show $\chi_{\mathrm{C}}=9\%$ and $\chi_{\mathrm{H}}=27\%$.}
        \label{fig:chis_2x5_fixedTau0_M2V}
    \end{figure*}
    
    To understand the dependence of the $Q$ rate on the location of the $\tau$-separators for the four stars, we first fix the number of bins to 4 following \cite{norlund1982} and \citet{voegler_thesis}.
    Similarly to the experiment described in section  3.4 of \citet{cernetic2019}, we try all possible combinations of the $\tau$-separators from a discretised grid of optical depths for each of the stars. For the F3V star this gives us 9880 combinations selected from a grid of 40 equidistant values in $\log \tau^{\mathrm{ref}} \in [-9.0, 0.5]$; for K0V and G2V, 4060 combinations (grid of 30 equidistant values in $\log \tau^{\mathrm{ref}} \in [-6.5, 0.5]$); for M2V, 4060 combinations (grid of 30 equidistant values in $\tau^{\mathrm{ref}} \in [-4.5, 0.5]$). The grid is customised for each star to cover properly the relevant range in optical depths (see the vertical axis of Fig.\  \ref{fig:binning_indexes_example}). The $\tau$-separator with maximum optical depth is labelled as $\tau_0$. The indices of the separators increase with height in the atmosphere ($\tau_{i} > \tau_{i+1}$).
    
    The statistics for each star in our sample show different trends. In general, combinations of separators that give simultaneously small deviations $\chi_{\mathrm{C}}$ and $\chi_{\mathrm{H}}$ are easy to find for the G and K stars, while they become less frequent in the case of the F star and quite rare for the M star. The number of cases leading to the deviations below certain threshold is given in Tables \ref{tab:stats_cooling} and \ref{tab:stats_heating} separately for the cooling and the heating, and in Table \ref{tab:stats_cooling_and_heating} when both components are constrained together. 
    From Tables \ref{tab:stats_cooling} and \ref{tab:stats_heating} it is clear that the cooling part is much easier to replicate accurately than the heating one. Almost any combination gives deviations less than $20\%$ in the cooling for any of the stars, while only about half of them minimise $\chi_{\mathrm{H}}$ below $20\%$ for the F, G, and K and only a few for the M star. 
    Table \ref{tab:stats_cooling_and_heating} shows how many cases we find with both $\chi_{\mathrm{C}}$ and $\chi_{\mathrm{H}}$  smaller than certain thresholds indicated individually for each of the stars. For the M2V star the choice of 4 $\tau$ bins is obviously insufficient. There are only a few combinations (less than 1\%) that lead to $\chi_{\mathrm{C}}<10\%$ and $\chi_{\mathrm{H}}<20\%$ and there is not a single one that reduces $\chi_{\mathrm{C}}$ and $\chi_{\mathrm{H}}$ below 5\% and 20\% simultaneously. 
    
    How do the deviation measures vary with the location of the three separators? An illustrative example for the G2V star is shown in Fig.\  \ref{fig:chis_2x5_fixedTau0}, which shows the dependence of $\chi_{\mathrm{C}}$ and $\chi_{\mathrm{H}}$ on the location of the middle separator $\tau_1$ when the other two separators are fixed. To visualise the results, in each column of the figure the deepest separator is fixed at one of six values ($\log\tau_0 = 0.5, 0, -0.5, -1.2, -1.9, -2.9$). Each of the curves correspond to a different fixed value of the highest separator $\tau_2$. The position of the highest separator is indicated by the colour, the black curve corresponding to the smallest value of $\tau_2$ and the red curve to the largest one.
    
    The figure shows that $\chi_{\mathrm{C}}$ and $\chi_{\mathrm{H}}$ vary significantly with the location of $\tau_1$ if the location of the deepest separator $\tau_0$ is deeper than some critical depth (left-hand columns of Fig.\ \ref{fig:chis_2x5_fixedTau0}). That critical location of $\tau_0$ is located deeper for $\chi_{\mathrm{C}}$ than for $\chi_{\mathrm{H}}$ ($\log \tau_0 \simeq -0.5$ for $\chi_{\mathrm{C}}$ and $\log \tau_0 \simeq -1.2$ for $\chi_{\mathrm{H}}$). For $\chi_{\mathrm{C}}$ especially, if $\tau_0 \simeq -0.5$ the curves flatten out, meaning that the results are essentially insensitive to the location of the other two nodes.
    
    If $\tau_0$ is deeper than these critical values, the deviation measures are insensitive to the location of the other two separators only if $\tau_2$ is relatively close to $\tau_0$ (the red curve in the left-hand column of Fig.\ \ref{fig:chis_2x5_fixedTau0}).
    When $\tau_0$ is pushed higher up, the deviation measures are insensitive to the location of the remaining separators and their value increases (compare two right-hand columns of Fig.\ \ref{fig:chis_2x5_fixedTau0}).   
    
    When $\log \tau_0 \geq -0.5$, the shape of the curves for the $\chi_{\mathrm{C}}$ and $\chi_{\mathrm{H}}$ in Fig.\  \ref{fig:chis_2x5_fixedTau0} becomes more different. The cooling deviation shows one minimum around $\log\tau_1\approx -2.5$, while the heating deviation has two minima: one for the locations of  $\tau_1$ close to $\tau_0$ and another one that moves higher up with increasing height of $\tau_0$. When $\log\tau_0 = -0.7$, the minimum is at $ \log\tau_1 \approx -5$, and when $\log\tau_0 \ge -0.5$, the minimum is at $ \log\tau_1 \approx -2.5$.
    Since the location of one of the minima of $\chi_{\mathrm{H}}$ coincides with the minimum of $\chi_{\mathrm{C}}$, both can be minimised using the same set of separators even when $\tau_0$ is as deep as $\log\tau_0  = 0.5$.
    
    When the location of the separator $\tau_0$ is set above the critical values, the sensitivity of both $\chi_{\mathrm{C}}$ and $\chi_{\mathrm{H}}$ to the location of $\tau_1$ and $\tau_2$ mostly disappears (see the sequence from the left to the right panel of Fig.\  \ref{fig:chis_2x5_fixedTau0}). This happens for a higher up location of $\tau_0$ for $\chi_{\mathrm{H}}$ than for $\chi_{\mathrm{C}}$.

    Our analysis suggests that, in the case of the G2V star, there are two strategies for placing the separators so that $\chi_{\mathrm{C}}$ and $\chi_{\mathrm{H}}$ remain below $4\%$  and $15\%$ respectively. The first strategy is to set $\log\tau_0$ around the continuum formation height, [-0.5, 0] and $\log\tau_1\in[-3,-2.5]$. The second one is to place $\log\tau_0$ so that it includes most of the weak and intermediate photospheric lines ($\log\tau_0\in[-2.4,-1.9]$) and $\log\tau_1$ even higher in $[-5,-3.5]$. In both cases the location of  $\log\tau_2$ can be anywhere above $\log\tau_1$. The former case (see 2nd and 3rd column) contains the combinations that produce the minimal $\chi$ values in our experiment, although the curves are not entirely flat in such a way that the optimal solution is sensitive to the model atmosphere and to the ODF data. Alternatively, the latter case still guarantees small deviations, but it appears to be more robust as $\chi_\mathrm{C}$ and $\chi_\mathrm{H}$ essentially depend only on the location of the deepest bin. Since our model atmosphere does not have a chromospheric temperature rise, however, the strategy with all separators positioned relatively high may be limited to that type of models.
    
    For the spectral types F3V and K0V the results are similar to those for G2V. For $\tau_0$ located deeper than a certain critical value, $\chi_{\mathrm{C}}$ and $\chi_{\mathrm{H}}$ vary with $\tau_1$ similarly to the G star, showing two minima for $\chi_{\mathrm{H}}$, of which one coincides with the single minimum of $\chi_{\mathrm{C}}$. There is a little variation of the deviation curves with $\tau_2$ except when it is close to the $\tau_0$. For the F3V star, the optimal combinations that produce $\chi_{\mathrm{C}} \leq 8\%$ and $\chi_{\mathrm{H}} \leq 15\%$ are (a) $\log\tau_0\in[-0.2, 0.5]$ and $ \log\tau_1\in[-2.5, -2]$ and (b) $\log\tau_0\in[-2.2,-0.7]$ and $\log\tau_1\in[-5.5, -4.7]$, with $\log\tau_2$ again anywhere above $\log\tau_1$.

    For the K0V star, we identified three variants that produce $\chi_{\mathrm{C}} \leq 5\%$ and $\chi_{\mathrm{H}} \leq 15\%$: (a) $\log\tau_0\in[0, 0.3]$ and $\log\tau_1\in[-2, -1]$, (b) $\log\tau_0\in[-0.5, -0.2]$ and $\log\tau_1\in[-4, -3]$, and (c)$\log\tau_0\in[-1.9, -1.7]$ and $\log\tau_1\in[-5, -3]$, with $\log\tau_2 < \log\tau_1$.
    
    While the results for the M2V star presented in Fig.\ \ref{fig:chis_2x5_fixedTau0_M2V} show general similarity to the results for the other stars, there are two important differences.
    First, for the deep choice of $\log\tau_0$ there is only one minimum for the $\chi_{\mathrm{H}}$ varying with $\tau_1$ and it does not coincide anymore with the minimum for $\chi_{\mathrm{C}}$, but with its local maximum (cf.\ two left columns in Fig.\  \ref{fig:chis_2x5_fixedTau0_M2V}). 
    Secondly, once the $\tau_0$ separator is sufficiently high in the atmosphere so that the deviation measures are insensitive to changes in $\tau_1$, $\chi_{\mathrm{C}}$
    flattens out as for the other stars but its value quickly increases with the height of $\tau_0$ (see sequence from left to right in the figure),  from $\chi_{\mathrm{C}} \simeq 6\%$ when $\log\tau_0 = 0.2$ to $\chi_{\mathrm{C}} \simeq 15.5\%$ when $\log\tau_0 = -0.9$.
    Higher up, the value of $\chi_{\mathrm{C}}$ remains larger than 12\%, while $\chi_{\mathrm{H}}$ also starts to increase once it becomes flat (after $\tau_0$=-1.2). 
    Therefore, for the M2V star it is not possible to minimise both deviation measures simultaneously with four bins.      
    The best combination of parameters that we found gives $\chi_{\mathrm{C}} \leq 9\%$ and $\chi_{\mathrm{H}} \leq 27\%$ for $\log\tau_0\in[0.2, 0.5]$ and $\log\tau_1\in[-1.1, -0.6]$, with $\log\tau_2 < \log\tau_1$.

    Based on the analysis presented above, we selected optimal four-bins combinations as reference for other binning strategies that we described in the following sections: $\log\tau = \left\{ -1.9, -5, -7 \right\}$ for the F3V star; $\log\tau = \left\{ 0, -2.5, -5 \right\}$ for the G2V; $\log\tau = \left\{ -0.2, -3, -5 \right\}$ for the K0V; $\log\tau = \left\{ 0.3, -0.8, -2.5 \right\}$ for the M2V. 
    
    \subsection{Number of $\tau$-bins} \label{subsec:tau_bin_number}
    
    \begin{figure}
    \includegraphics[width=8.8cm]{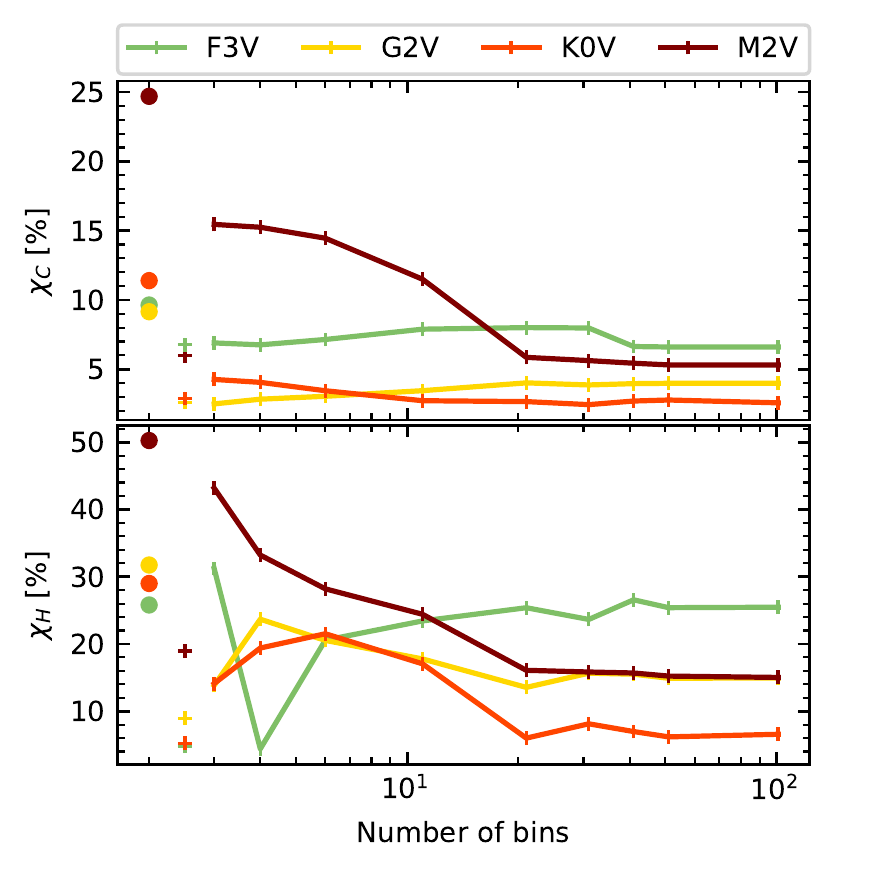}%png}
    \caption{Deviation measures $\chi_{\mathrm{C}}$ (\textit{upper panel}) and $\chi_{\mathrm{H}}$ (\textit{lower}) versus the number of bins for the four stars (\textit{colours} as in Fig.\ \ref{fig:1d_mod_atms}). \textit{Solid lines}: deviation measures for the $Q$ computed using 3, 4, 6, 11, 21, 31, 41, 51, and 101 $\tau$-bins distributed equidistantly. \textit{Circles}: deviations for $Q$ computed with the Rosseland opacity. \textit{Pluses}: the deviation of the optimal combinations of four $\tau$-bins (see Sect.\  \ref{subsec:tau_bin_location} and Table \ref{tab:values_for_error_measures}).}
             \label{fig:binning_number_of_separators}
    \end{figure}

    To understand the dependence of the $Q$ rate on the number of separators, we computed $Q$ for the four stars for different number of separators (2, 3, 5, 10, 20, 30, 40, 50, and 100)\footnote{In the present and following sections, there could be combinations of bins where one bin might end up empty. In this case, when computing the RTE solution, the empty bin is ignored. } that are uniformly distributed in the relevant range for each of the stars: $\log \tau^{\mathrm{ref}} \in [-9, 0.5]$ for F3V, $[-6.5, 0.5]$ for K0V, and G2V and $[-4.5, 0.5]$ for M2V. Again, the selected optical depth ranges ensure that all the stellar models are properly represented (see the vertical axis of Fig.\  \ref{fig:binning_indexes_example}).

    In Figure \ref{fig:binning_number_of_separators} $\chi_{\mathrm{C}}$ and $\chi_{\mathrm{H}}$ are shown as a function of the number of bins. 
    For F, G, and K -- type stars, there is very little variation in $\chi_{\mathrm{C}}$. The trend is slightly different for each of the stars individually. While $\chi_{\mathrm{C}}$ slowly decreases for the K star with an increasing number of bins, it increases for G2V, and it first increases and then decreases for the F3V -- type star. In all three cases the value of  $\chi_{\mathrm{C}}$ saturates at around 20 bins. 
    The M star shows a significant improvement in $\chi_{\mathrm{C}}$ when the number of bins increases from 4 to 21. For more bins it saturates at $\approx$$5\%$. 
    Although there is more variation in $\chi_{\mathrm{H}}$ than in $\chi_{\mathrm{C}}$, the overall trend is similar and, again, the results saturate for more than $\approx$$20$ bins.
    \citet{voegler_opacity_binning} tested the difference between the $Q$ rate computed with the binned opacity and with the ODF for an increasing number of bins (up to 60). Similarly to our results, they found that the solutions for $Q$ converge towards a limiting value that is different from Q$_{\mathrm{ODF}}$. 
    
    \citet{wray2006} investigated the performance of the OB and ODF methods 
    in the context of shock-generated radiation during simulated re-entry of Apollo AS-501 vehicle into the Earth's atmosphere. Although the physics of the experiment and implementation of the OB are different (e.g.\ planckian mean for $\overline{\varkappa}_l$, $\varkappa$-sorting instead of $\tau$-sorting), they also found that the results of the radiation computation quickly converge with increasing number of bins (reaching saturation at around 10 bins). However, note that their conclusions about the ODF solution are not directly translatable to the radiative transfer in the stellar atmospheres owing to different definitions of ODF.
    
    Therefore, the saturation values of $\chi_{\mathrm{C}}$ and $\chi_{\mathrm{H}}$ are the minimal deviations that cannot be eliminated by increasing the number of the $\tau$-separators. Depending on the stellar type, the minimal $\chi_{\mathrm{C}}$ values span between 2\% and 7\%, and the minimal $\chi_{\mathrm{H}}$ span between $\approx$$10\%$ and $\approx$$30\%$. These minimal deviations are the smallest for the K and the largest for the F star.
    However, it is possible to find certain combinations of the separators that minimise the deviations particularly well. For example, the $\chi_{\mathrm{H}}$ value for the F3V star is reduced to only a few per cent when three separators are set to $\log\tau = \left\{ -1.8, -4.3, -6.6 \right\}$.
    This is consistent with locations that produce simultaneous minimum $\chi_{\mathrm{C}}$ and $\chi_{\mathrm{H}}$ deviations, as discussed in Sect.\  \ref{subsec:tau_bin_location}.
    The deviations with the optimal combinations found using four bins for each of the stars in Sect.\  \ref{subsec:tau_bin_location} are shown as coloured plus markers in Fig.\  \ref{fig:binning_number_of_separators}. 
    These deviations are either similar to or smaller than the saturation values when the bins are distributed uniformly. We have also computed the $Q$ rates using the Rosseland opacity mean applied to the whole spectrum (Eq.\ \ref{eq:harmonic_mean}). The $\chi$ differences for this grey solution are shown as coloured circles in Fig.\  \ref{fig:binning_number_of_separators}. Their values are larger or similar to the the saturation ones, but always larger than the solutions with four bins from Sect.\  \ref{subsec:tau_bin_location}. 
    
    We experimented with several other strategies for distributing separators automatically.
    They all confirm the conclusion that careful distribution of the separators is more important than their number.
    
    \subsection{$\{\tau,\lambda\}$-binning} \label{subsec:tau_lambda_binning}
    
    \begin{figure*}
        \centering
        \includegraphics[width=17.6cm]{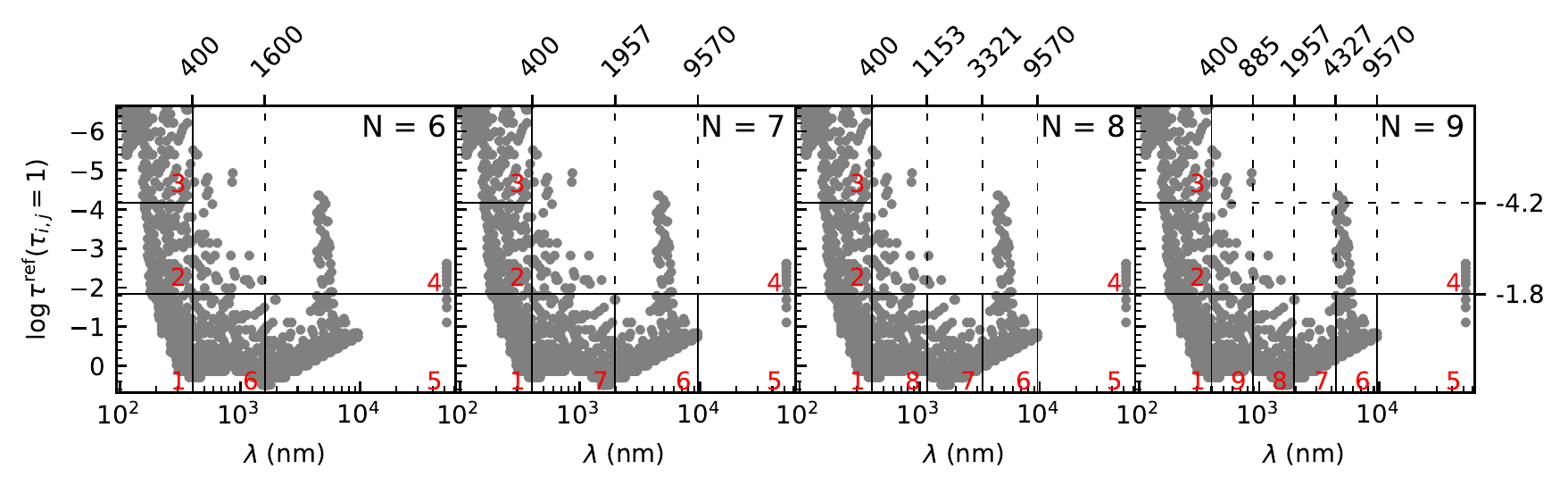}
        \caption{Example of the binning used to compute the $Q$ values for the G2V star, using different numbers of $\{\tau,\lambda\}$-bins. Similarly to Fig.\ \ref{fig:binning_indexes_example}, each \textit{grey dot} corresponds to one ODF substep. For this star, the $\tau$-separators in the UV in all the experiments are fixed at $\log\tau = \left\{ -1.8, -4.2 \right\}$.  \textit{Columns, from left to right}, show the cases for $N$=6,7,8,9 bins. The \textit{red numbers} show indices of the bins. The case for $N$=6 has two $\lambda$-separators at 400 and 1600 nm. For more than 6 bins, bins 1 to 5 are kept identical, but more $\lambda$ separators are added between 400 and 9750 nm. A similar binning set-up is used for the other stars with modified $\tau$-separators in the UV and the $\lambda$ separator between the short and long wavelengths for $N$>6.}
        \label{fig:indexes_tau-lambda_binning}
    \end{figure*}
    
    \begin{figure}
    \includegraphics[width=8.8cm]{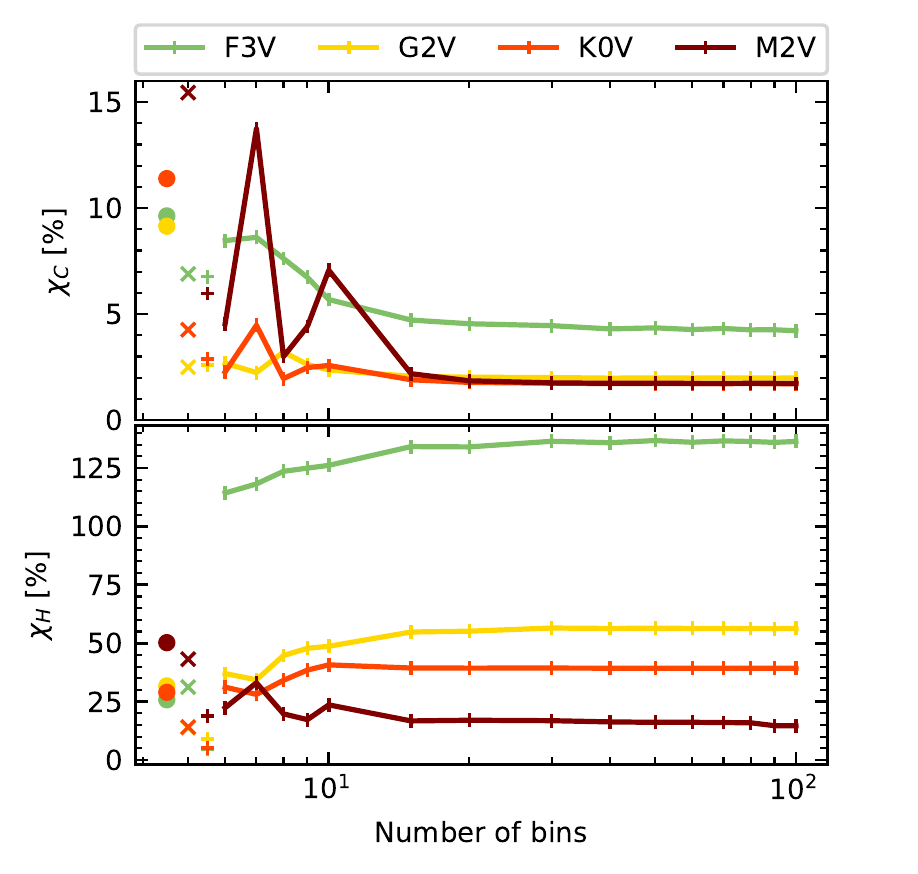}%ng}
    \caption{Deviation measures $\chi_{\mathrm{C}}$ (\textit{upper panel}) and $\chi_{\mathrm{H}}$ (\textit{lower}) versus the number of bins, for the four stars (\textit{colours} as in Fig.\ \ref{fig:1d_mod_atms}), for the case of $\{\tau,\lambda\}$-binning approach similar to \citet{2014MagicThesis}. 
    \textit{Solid lines}: deviation measures for the $Q$ computed using 6, 7, 8, 9, 10, 15, 20, 30, 40, 50, 60, 70, 80, 90, and 100 bins distributed as in Fig.\  \ref{fig:indexes_tau-lambda_binning}. \textit{Circles}: deviations for $Q$ computed with the Rosseland opacity. \textit{Crosses}: deviations for $Q$ computed with only two $\tau$-separators located at the heights of the bins in the UV (no $\lambda$-separators).  \textit{Pluses}: the deviation of the optimal combinations of four $\tau$-bins (see Sect.\  \ref{subsec:tau_bin_location} and Table \ref{tab:values_for_error_measures}).}
             \label{fig:tau-lambda_binning_with_Nseparators}
    \end{figure}

    \begin{figure*}
        \centering
        \includegraphics[width=17.6cm]{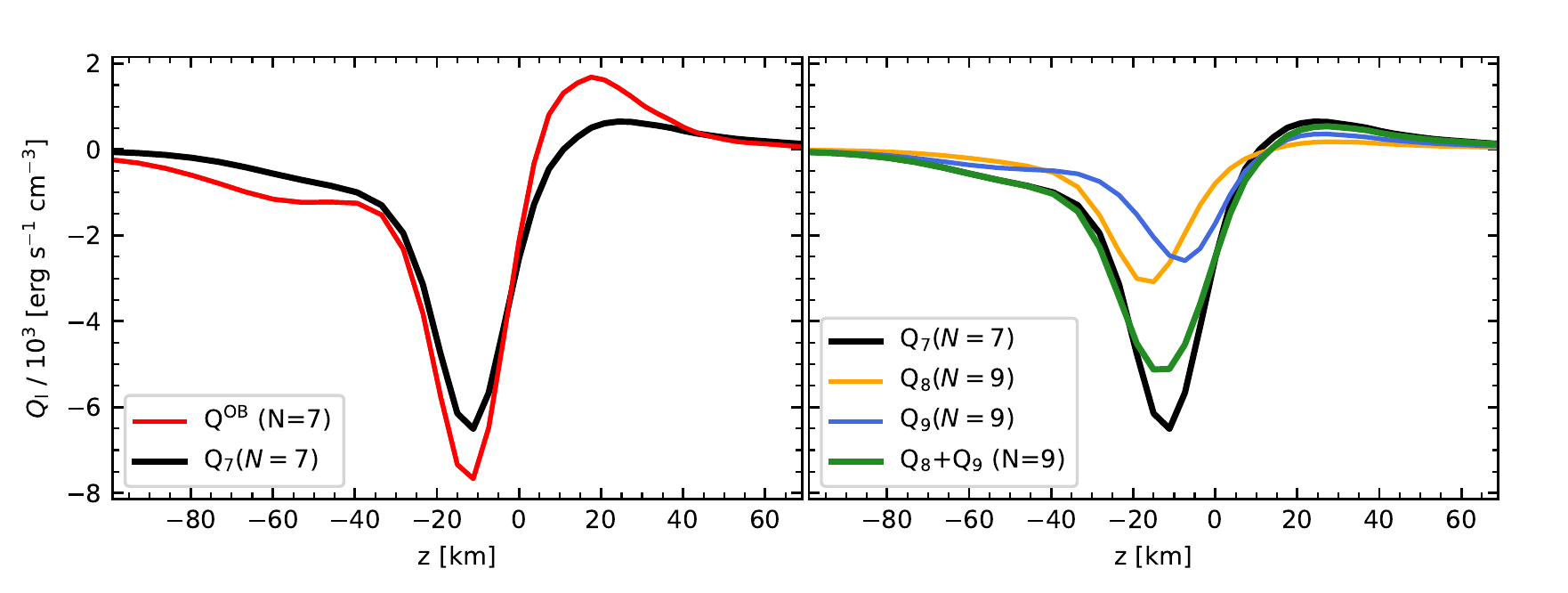}
        \caption{Comparison of the radiative energy exchange rate $Q$ from bin 7 for $N=7$ and bins 8 and 9 for $N=9$ corresponding to the cases shown in Fig.\  \ref{fig:tau-lambda_binning_with_Nseparators} for the M2V star. \textit{Left panel}: $Q_7$ dominates the total $Q^{\mathrm{OB}}=\sum_l Q_l$ for $N=7$. \textit{Right panel}: comparison of $Q_8$ and $Q_9$ for $N=9$ against $Q_7$ for $N=7$. }
        \label{fig:q7_q9_m2v}
    \end{figure*}
    
    \begin{figure}
    \includegraphics[width=8.8cm]{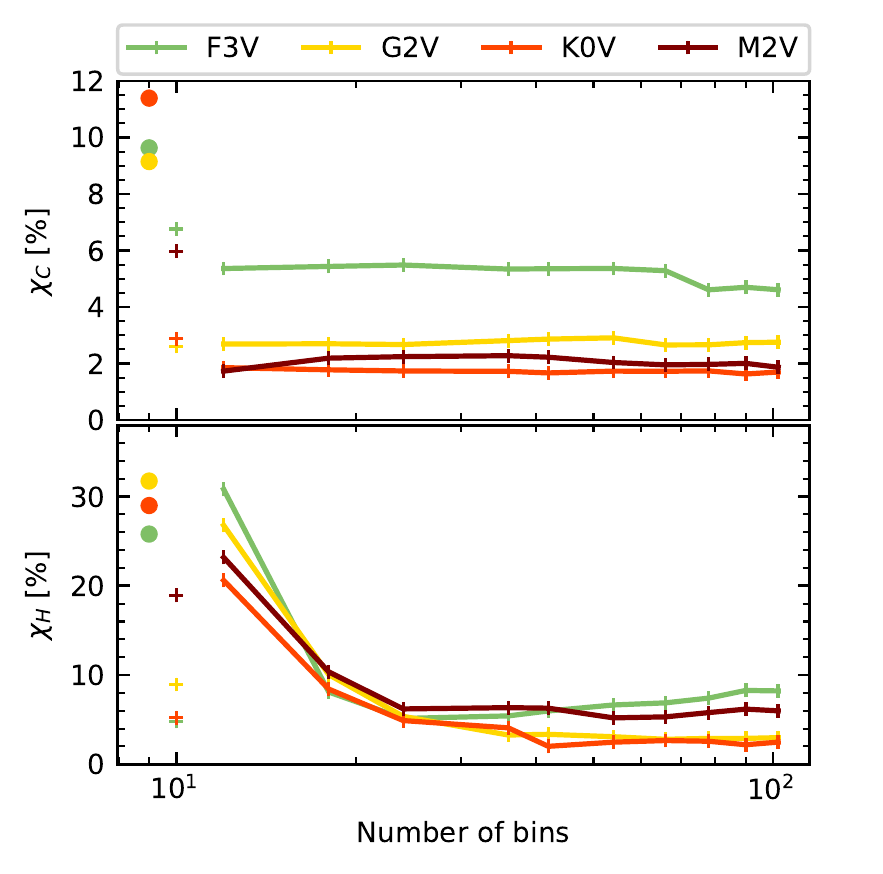}%ng}
    \caption{Deviation measures $\chi_{\mathrm{C}}$ (\textit{upper panel}) and $\chi_{\mathrm{H}}$ (\textit{lower}) against the number of bins for the four stars, computed  for the strategy with five fixed $\lambda$-separators at $\lambda\in(381,726,1383,2636, 5023)$ nm and with varying number of $\tau$-separators (1, 2, 3, 5, 6, 8, 10, 12, 14, and 16) and therefore varying total number of bins (12, 18, 24, 36, 42, 54, 66, 78, 90, and 102 bins). \textit{Solid lines}: deviation measures for $Q$ computed using a different number of $\{\tau,\lambda\}$-bins. \textit{Circles}: deviations for $Q$ computed with the Rosseland opacity. \textit{Pluses}: deviations of the optimal combinations of four $\tau$-bins (see Sect.\  \ref{subsec:tau_bin_location} and Table \ref{tab:values_for_error_measures}).}
             \label{fig:tau-lambda_binning_fixedLambdaSeps}
    \end{figure}
    The OB method using only $\tau$-separators does not reproduce the exact solution in the limit of infinite number of bins  because the contributions from different parts of the spectrum, each with its own height variation, are mixed in any given bin \citep[][see discussion in App.\ \ref{app:not_convergence}]{voegler_opacity_binning}. 
    One way to improve the approximation is to preserve some of the 
    wavelength dependence in the binning procedure by grouping the 
    opacity data points with respect to their spectral regions prior
    to the $\tau$-binning. 
    Therefore, $\lambda$-separators between these regions are introduced in addition to the $\tau$-separators. 
    Such bidimensional distribution of the opacities is non-uniform in $(\tau, \lambda)$ parameter space: in every wavelength interval the $\tau$-separators may be specified
    independently (both their number and location) to account for the height distribution of the opacities in that interval.
    \citet{ludwig_thesis} introduced the $\{\tau,\lambda\}$-binning, showing that splitting the continuum at the Balmer jump improved significantly the results for Vega, due to the systematically different behaviour of $Q$ for wavelengths short-ward and long-ward of the jump.
    \cite{voegler_thesis} explored the possibility of splitting the least opaque bin into two subgroups for long and short wavelengths and concluded that the improvement is probably not worth the extra computational effort.
    \cite{2018colletGiant} also attempted to improve the OB procedure by partly accounting for the lost wavelength dependence. \cite{2014MagicThesis} treated the UV and the visible + IR opacities separately. Each group is sorted with respect to its own set of $\tau$-separators. In addition, the least opaque bin of the visible + IR opacities is further split with a set $\lambda$-separators.
    
    We computed the $Q$ values using a $\{\tau,\lambda\}$-binning approach similar to \citet{2014MagicThesis} (see his figure  2.6) for different total number of bins, $N$ = 6, 7, 8, 9, 10, 15, 20, 30, 40, 50, 60, 70, 80, 90, and 100. The distribution of the bins for the G2V star for $N = 6, 7, 8$ and 9 is shown in Fig.\  \ref{fig:indexes_tau-lambda_binning}. 
    For any $N$, the spectrum is first divided into the short and long wavelength parts by placing one $\lambda$-separator at $\lambda_0$.
    The value of $\lambda_0$ is 400 nm for the F and G stars (effectively separating UV from visible), 500 nm for the K star, and 600 nm for the M star. 
    The short-wavelength opacities are then grouped into three $\tau$-bins, and the long-wavelength ones into two. The least opaque $\tau$-bin of the long-wavelength opacities is then split by one or more additional $\lambda$-separators. For $N=6$ the $\lambda$-separator is placed at 1600 nm. For $N>6$, the $\lambda$-separators of the least opaque $\tau$-bin are placed equidistantly in $\log \lambda$ in the range $[\lambda_0, 9570]$ nm. The $\tau$-separators in the short-wavelength region are customised for each stellar type: -2.6 and -5.8 for F3V,  -1.8 and -4.2 for G2V,  -1.8 and -4.0 for K0V, and -1.2 and -2.8 for M2V. The locations of the separators are chosen optimally to sample different features present for each of the stars when the ODF points are plotted in the $\tau$-$\lambda$ plane (cf.\ Figs.\  \ref{fig:binning_indexes_example} and \ref{fig:histogram_indexes}).
    
    We computed the $Q$ rate for every combination of $\{\tau,\lambda\}$-bins. The corresponding deviations $\chi_{\mathrm{C}}$ and $\chi_{\mathrm{H}}$ are shown in Fig.\  \ref{fig:tau-lambda_binning_with_Nseparators}. The behaviour of $\chi_{\mathrm{C}}$ with the number of bins is generally similar to that found in Fig.\  \ref{fig:binning_number_of_separators}. For four stars the $\chi_{\mathrm{C}}$ deviation decreases with increasing number of bins until it saturates after around 10--20 bins. The saturation values of $\chi_{\mathrm{C}}$ are lower than the ones found in Sect.\  \ref{subsec:tau_bin_number} (cf.\ Table \ref{tab:values_for_error_measures}). For less than 15 bins ($\Delta \log \lambda > 0.1$) $\chi_{\mathrm{C}}$ remains nearly constant for G2V and K0V, but increases significantly for F3V and M2V. The results for M2V show two peaks for $N = 7$ and 10, and a minimum between them for $N = 8$. 

    This strong variation in $\chi_{\mathrm{C}}$ is a good illustration of how sensitive the $Q$ rate is on how the ODF data points are distributed between the bins. 
    Figure \ref{fig:indexes_tau-lambda_binning} 
    shows the distribution of the bins with $N = 7$ and $N = 9$ (2nd and 4th panel). Bins 1 to 5 are identical in both cases. Bins 6 and 7 for $N=7$ are split into two bins each for $N=9$.  
    Bin 7 (for $N=7$) includes most of the visible spectrum, it coincides with the central part of the Planck function and thus it dominates the radiative losses. The rate $Q$ computed from that bin is compared to the total $Q$ in the left-hand panel of Fig.\ \ref{fig:q7_q9_m2v}.
    In the right-hand panel it is compared
    with $Q$ values computed from the bins 8 and 9 (for $N = 9$) and with their sum. Individual $Q$ profiles for the bins 8 and 9 (for $N = 9$) differ significantly in shape, amplitude and location of the maximal cooling. 
    Because of the non-linearity of RTE, their sum is not equal to $Q$ computed when the opacity of the two bins is merged in the bin 7 for $N=7$. 
    In terms of our deviation measure $\chi_{\mathrm{C}}$, this leads to a drop from $\approx$$14 \%$ for $N=7$ to $\approx$$3 \%$ for $N=9$ (top panel of Fig.\ \ref{fig:tau-lambda_binning_with_Nseparators}).
    The finer the distribution of the visible opacities in wavelength is, the more stable are the results leading to the saturation of $\chi_{\mathrm{C}}$. 
    
    For the M2V star the $\chi_{\mathrm{H}}$ saturation value is about the same as in Fig.\  \ref{fig:binning_number_of_separators}, but for the F, G, and K stars $\chi_{\mathrm{H}}$ saturates at significantly larger values owing to an excess of heating relative to the ODF solution. 
    
    In another experiment, we tested a variant of the previous method were bins are distributed in a checkerboard pattern in the $(\tau, \lambda)$ plane. The number of $\tau$-separators is varied between 1 and 16 and the number of $\lambda$-separators is fixed at 5 ($\lambda=381,726,1383,2636,5023$ nm). The $\tau$-separators for all the stars are distributed equidistantly at $\log\tau^{\mathrm{ref}} \in [-3, 0.5]$. The minimum number of bins in this experiment is 12, the maximum 102. 
    The results are shown in Fig.\  \ref{fig:tau-lambda_binning_fixedLambdaSeps}. For a small number of bins
    $\chi_{\mathrm{H}}$ is larger than the deviation measure we obtain with the optimised solutions from Sect.\  \ref{subsec:tau_bin_location}. 
    With 12 bins the values $\chi_{\mathrm{H}}$ lie between 20 and 30\%, similar to those in Fig.\  \ref{fig:binning_number_of_separators} with 11 bins distributed only in $\tau$. With finer sampling in $\tau$, the value of $\chi_{\mathrm{H}}$ is quickly reduced below 10\% and saturated at that level for all four stars.

    \begin{table}[]
    \caption{Deviation measures $\chi_{\mathrm{C}}$ and $\chi_{\mathrm{H}}$ for different binning strategies described in Sects.\ \ref{subsec:tau_bin_location}-\ref{subsec:tau_lambda_binning}, for the four stars F3V, G2V, K0V, and M2V (columns are named after the corresponding sections). 
    In column \ref{subsec:tau_bin_location} we show the deviations of the 4-bins case with the separators optimally distributed as described in Sect.\  \ref{subsec:tau_bin_location}.
    In column \ref{subsec:tau_bin_number} we show the deviations for the case with 21 bins from Fig.\  \ref{fig:binning_number_of_separators}. In column \ref{subsec:tau_lambda_binning}a we show the deviations for the case with 15 bins from Figure  \ref{fig:tau-lambda_binning_with_Nseparators}. Finally, in column \ref{subsec:tau_lambda_binning}b, we show the deviations for the case with 24 bins from Fig.\  \ref{fig:tau-lambda_binning_fixedLambdaSeps}.}
    \begin{tabular}{ccccccccc}
    \hline
      & \multicolumn{2}{c}{\ref{subsec:tau_bin_location}}      & \multicolumn{2}{c}{\ref{subsec:tau_bin_number}} & \multicolumn{2}{c}{\ref{subsec:tau_lambda_binning}a} &  \multicolumn{2}{c}{\ref{subsec:tau_lambda_binning}b} \\  \hline 
      & \multicolumn{1}{c}{$\chi_{\mathrm{C}}$} & \multicolumn{1}{c}{$\chi_{\mathrm{H}}$}  & \multicolumn{1}{c}{$\chi_{\mathrm{C}}$} & \multicolumn{1}{c}{$\chi_{\mathrm{H}}$} & \multicolumn{1}{c}{$\chi_{\mathrm{C}}$} & \multicolumn{1}{c}{$\chi_{\mathrm{H}}$} & \multicolumn{1}{c}{$\chi_{\mathrm{C}}$} & \multicolumn{1}{c}{$\chi_{\mathrm{H}}$}   \\ \hline% \hline
    F & \multicolumn{1}{c}{6.8}  & \multicolumn{1}{c}{4.8}  & \multicolumn{1}{c}{8.0}  & \multicolumn{1}{c}{25.4} & \multicolumn{1}{c}{4.7}  & \multicolumn{1}{c}{134.1} & \multicolumn{1}{c}{5.4}  & \multicolumn{1}{c}{6.0}   \\ \hline
    G & \multicolumn{1}{c}{2.6}  & \multicolumn{1}{c}{8.9}  & \multicolumn{1}{c}{4.0}  & \multicolumn{1}{c}{13.5} & \multicolumn{1}{c}{2.1}  & \multicolumn{1}{c}{54.8}  & \multicolumn{1}{c}{2.9}  & \multicolumn{1}{c}{3.4}   \\ \hline
    K & \multicolumn{1}{c}{2.9}  & \multicolumn{1}{c}{5.2}  & \multicolumn{1}{c}{2.7}  & \multicolumn{1}{c}{6.0}  & \multicolumn{1}{c}{1.9}  & \multicolumn{1}{c}{39.4}  & \multicolumn{1}{c}{1.7}  & \multicolumn{1}{c}{2.0}   \\ \hline
    M & \multicolumn{1}{c}{6.0}  & \multicolumn{1}{c}{18.9} & \multicolumn{1}{c}{5.9}  & \multicolumn{1}{c}{16.1} & \multicolumn{1}{c}{2.2}  & \multicolumn{1}{c}{16.7}  & \multicolumn{1}{c}{2.2}  & \multicolumn{1}{c}{6.3}   \\ \hline
    \end{tabular}
    \label{tab:values_for_error_measures}
    \end{table}    
    
    From this analysis it follows that introduction of the $\lambda$ separators may reduce the minimum deviation of the OB in 
    particular cases when it is combined with carefully selected 
    separators in $\tau$. However, this improvement is relatively small in comparison with the basic method where sorting is done only by optical depths (cf.\ Figs.\  \ref{fig:tau-lambda_binning_fixedLambdaSeps} and \ref{fig:binning_number_of_separators}). 
    Table \ref{tab:values_for_error_measures} summarises the deviation measures $\chi_{\mathrm{C}}$ and $\chi_{\mathrm{H}}$ computed in the experiments described in this section. 
    The values shown for the experiments described in Sect.\  \ref{subsec:tau_bin_number} and \ref{subsec:tau_lambda_binning} correspond to the minimum number of bins at which the deviation measures approach their saturation values. 
    It is clear that the 4-bin approach with carefully tuned locations of the $\tau$-separators (Sect.\  \ref{subsec:tau_bin_location}) performs similarly or better than the other approaches for all 4 stars. The only exception is the M star, where taking into account the wavelength distribution simultaneously with distribution in $\tau$ for all $\lambda$ intervals is an obvious advantage (Fig.\  \ref{fig:tau-lambda_binning_with_Nseparators}).
    The strategy proposed by \citet{magic2013} reproduces the cooling component better than other strategies, but it produces large errors in the heating component. 

\begin{figure*}
        \centering
        \includegraphics[width=16cm]{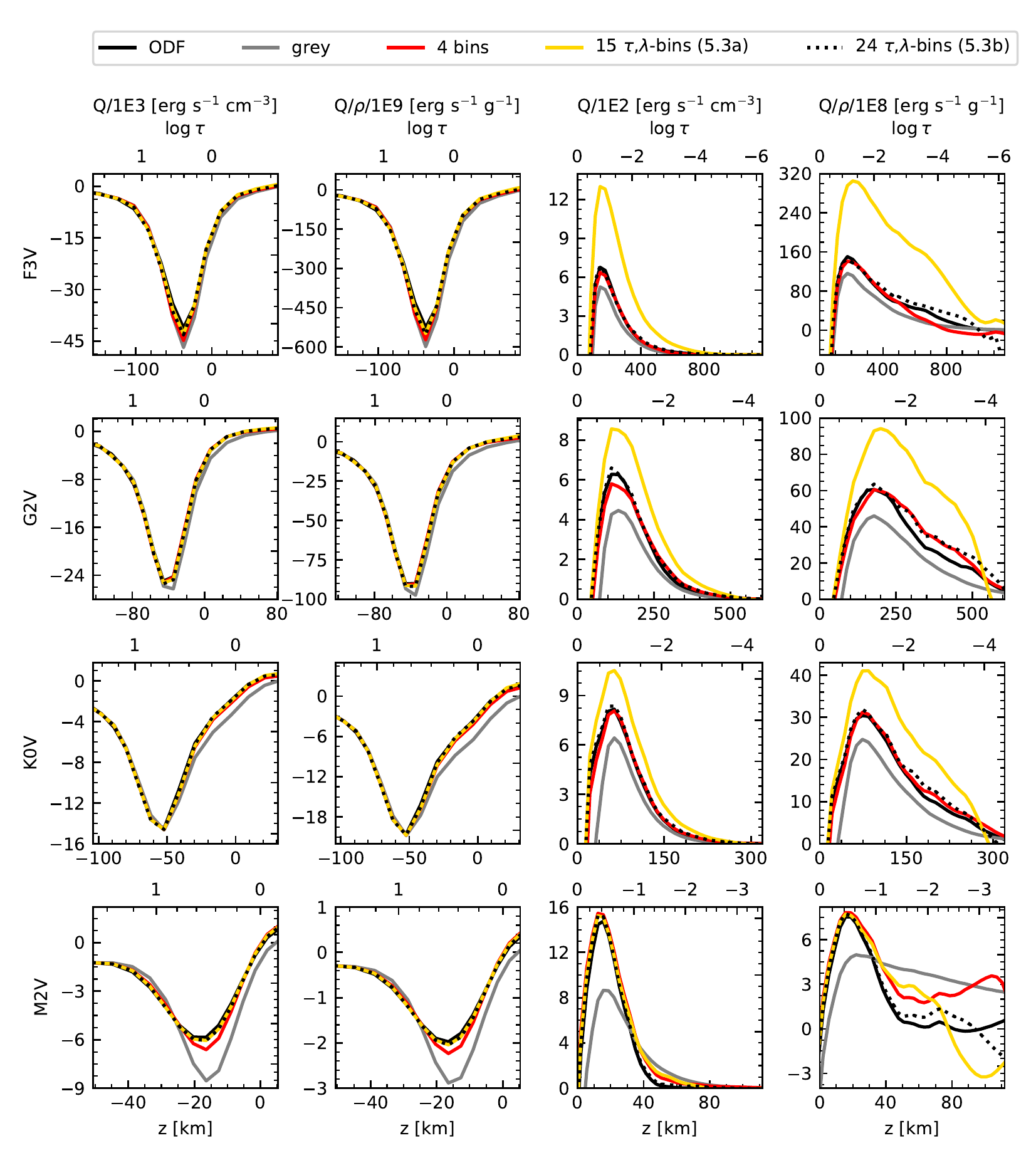}
        \caption{Comparison of $Q$ and $Q/\rho$ for the four stars (\textit{each row}) and five opacity approaches: ODF (\textit{black solid line}), grey opacity computed with Eq.\ \ref{eq:harmonic_mean} (\textit{grey solid line}), and the binned opacities for 4 bins (\textit{red solid line}), 15 bins (\textit{yellow solid line}), and 24 bins (\textit{black dotted line}) corresponding to columns \ref{subsec:tau_bin_location}, \ref{subsec:tau_lambda_binning}a and \ref{subsec:tau_lambda_binning}b in Table \ref{tab:values_for_error_measures}. \textit{Left column}: $Q$ for the deeper part of the atmosphere where mainly cooling occurs. \textit{Second column}: $Q/\rho$ for the same part of the atmosphere. \textit{Third column}: $Q$ for the higher layers where the heating occurs. \textit{Right column}: $Q/\rho$ for the same part of the atmosphere.}
        \label{fig:q_vs_qoverrho}
    \end{figure*}

    Finally, in Figure \ref{fig:q_vs_qoverrho} $Q$ and $Q/\rho$ are compared for the four stars and five opacity approaches (ODF, grey opacity, and three OB setups). The deeper part of the atmosphere where mainly cooling occurs is shown in the two left columns; the higher layers where the heating occurs are shown in the two right columns. 
    On one hand, the cooling can be fitted either using $Q$ or $Q/\rho$, since the shape of $Q$ is practically not modified by the relatively slowly changing density in sub-surface layers. 
    On the other hand, the density scale height diminishes faster in higher layers, thus, dividing by the rapidly decreasing density enhances the differences we see in $Q$. 
    For the part of the atmosphere studied in the present paper (top of the convection zone and lower photosphere), $Q$ and $Q/\rho$ show similar information regarding the quality of the computed $Q^{\mathrm{OB}}$ in respect to $Q^{\mathrm{ODF}}$. 
    Some differences are expected when choosing $Q/\rho$, but the general behaviours described in this work should be the same. If one is interested in layers closer to the chromosphere and beyond, $Q/\rho$ would be the quantity to optimise.
    
\section{Conclusions} \label{sec:conclusions}

     This work is based on detailed monochromatic opacity tables computed with the \texttt{SYNSPEC} \citep{SYNSPEC} code using complete up-to-date sets of atomic and molecular line lists. 
     The ODF is calculated using these opacities and compared with the ODF from the \texttt{ATLAS} code \citep{kurucz1993_odf}.
     Although there are some systematic differences between the two data sets, a good match is found for the F3V, G2V, and K0V stellar spectral types when the radiative energy exchange rate $Q$ is computed using each of the data sets. 
     For the M2V -- type star the comparison of \texttt{SYNSPEC} versus \texttt{ATLAS} ODFs reveals the importance of the molecules in the spectra of these stars. The lack of some significant opacity contributors such as VO and TiO in the \texttt{ATLAS} line lists produces less heating in the radiative energy exchange rate than in the \texttt{SYNSPEC} results. 
     
     In Sect.\  \ref{sec:opacity_binning} the OB method \citep{norlund1982} is applied to the \texttt{SYNSPEC} ODF. 
     This method enabled rapid progress in realistic 3D MHD simulations of the near-surface convection as it reduced the number of required radiative transfer computations by several orders of magnitude while preserving the essence of the non-grey nature of the radiative transfer solution. 
     However, the method involves several free parameters that cannot be intuitively guessed.
     We systematically tested several strategies for the opacity binning: positioning of the four bins (Sect.\  \ref{subsec:tau_bin_location}), the number of $\tau$ bins (Sect.\  \ref{subsec:tau_bin_number}) and simultaneously binning in optical depth and wavelength (Sect.\  \ref{subsec:tau_lambda_binning}).
    None of the strategies tested converge to the ODF solution in the limit of a large number of bins since the binning mixes parts of the spectrum with different height profiles of the opacity (App.\ \ref{app:not_convergence}). 
    It is possible to find combinations of small number of bins that significantly reduce the deviations. However, these combinations are model-dependent and with relatively narrow intervals in which the bins can be located. Strategies that include preservation of the wavelength dependence through combined $\tau$- and $\lambda$-binning may improve the deviation measures, but with a significant increase in computing overheads. For example, even in the most extreme case of the M star, to reduce the already small $\chi$ values by a factor of 3 (see the bottom row of Table \ref{tab:values_for_error_measures}), requires an increase in the number of bins, and thus the computing cost by a factor of 4 or 5. In most applications such accuracy is not needed as other approximations made in the RTE solution and the MHD modeling contribute more to the total error budget.

    The results presented in this study are limited to the solar chemical composition. In the stars with different metallicities both the atmospheric structure and the relative importance of the various opacity contributors will change. Therefore, our results should not be blindly applied to these stars. Nevertheless, the trends identified in our analysis may be used as guidelines for tuning the details of the OB procedure for the stars with non-solar metallicities.

    An open question left in this work is the importance of the measured deviations $\chi_\mathrm{C}$ and $\chi_\mathrm{H}$ in the final structure of the atmosphere in the context of simulations. To answer this question, one may start by extending the study for the Sun in section 3.4 of \citet{voegler_opacity_binning} for the case of other stellar spectral types in time dependent simulations.
    
    It should also be emphasised that the optimal binning approximation selected based on the 1D models will not necessarily remain optimal when implemented in 3D simulations. However, the tuning of the binning strategy is
    computationally too expensive to be routinely done in 3D. 
    In the follow-up of this study we shall explore how different
    binning strategies perform in realistic 3D atomospheres simulated with the MANCHA code \citep{2017khomenko, 2018khomenko}, especially in the case of the M -- type star. 

\begin{acknowledgements}
      This work was supported by the European Research Council through the Consolidator Grant ERC--2017--CoG--771310--PI2FA and by Spanish Ministry of Science through the grant PID2021--127487NB--I00.
      APG acknowledges support from the Agencia Estatal de Investigación (AEI) of the Ministerio de Ciencia, Innovación y Universidades (MCIU) and the European Social Fund (ESF) under grant with reference PRE2018--086567.
      CAP is thankful for funding from the Spanish goverment through grants AYA2014--56359--P, AYA2017--86389--P and PID2020--117493GB--100.
      We thank the referee for the detailed and careful reading of the manuscript and for asking the questions that lead to Appendix \ref{app:not_convergence} and Fig.\ \ref{fig:q_vs_qoverrho}.
      We thank Hans Günter Ludwig for reading the manuscript, the interesting discussions about the OB method and for suggesting the analysis in Fig.\ \ref{fig:OB_Ross_fashion}.
      We are thankful to Alexander Shapiro for reading the manuscript and giving his feedback.
      We are thankful to Terry Mahoney for the language editing.
      This research has made use of NASA's Astrophysics Data System Bibliographic Services. 
\end{acknowledgements}

%-----BIBILIOGRAPHY---------------------------------
% for the bibliography, at the end
\bibliographystyle{aa} % style aa.bst
\bibliography{references} % your references Yourfile.bib
%---------------------------------------------------

\begin{appendix} 
\section{Content of the \texttt{SYNSPEC} monochromatic opacity} \label{app:opac_sources} 
    
    In \texttt{SYNSPEC} each of the transitions taken into account to compute the opacity can be specified by the user. Only \textit{explicit} ions (in our computations, H, H$^{+}$, H$^{-}$, He, He$^{+}$, He$^{++}$, C, C$^{+}$, C$^{++}$, N, N$^{+}$, N$^{++}$, O, O$^{+}$, O$^{++}$, Na, Na$^{+}$, Na$^{++}$, Mg, Mg$^{+}$, Mg$^{++}$, Al, Al$^{+}$, Al$^{++}$, Si, Si$^{+}$, Si$^{++}$, Ca, Ca$^{+}$, Ca$^{++}$, Fe, Fe$^{+}$ and Fe$^{++}$) are considered in the continuum opacities solved in LTE (see sections 2.4 and 2.5 of \citealp{tlusty_guide_II} as well as \citealp{synspec_tlusty_guide_IV} for more details). We use the data included in the \texttt{synple} repository (version 1.2), as described in the following paragraphs. The contributions involving molecules and any of the Collision-Induced Absorptions are only taken into account below certain temperature, in this work chosen to be $8000 \mathrm{K}$ (i.e.\ when molecules are considered in the EOS).
    
    The free-free cross-sections for H and He$^{+}$ are computed hydrogenically with approximated free-free Gaunt factor \citep[see chapter 7.1 of ][]{Hubeny_Mihalas_2014}. For He, C, N, O, Na, Mg, Al, Si, Ca and Fe the cross-sections are computed again hydrogenically, but with Gaunt factor of 1. Other free-free contributions are also taken into account:
    \begin{itemize}
        \item H$^{-}$ from \citet{H-_bf_ff}.
        \item H$_2^+$ from \citet{h2+_bf_ff}.
        \item H$_2^-$ from \citet{h2-_ff}.
        \item He$^-$ from the polynomial fit of \citet{poly_fit_He-} to the data of \citet{he-_data}.
    \end{itemize}
    
    The bound-free transitions are mainly computed using hydrogenic cross-section, data from the Opacity Project 
    TOPbase \footnote{\url{http://cdsweb.u-strasbg.fr/topbase/topbase.html}} and the Iron Project \footnote{\url{https://cds.unistra.fr/topbase/TheIP.html}}: 
    \begin{itemize}
        \item H$^{-}$: the cross-section is computed hydrogenically with Gaunt factor of 1 \citep[see chapter 7.1 of ][]{Hubeny_Mihalas_2014}.
        \item H$_2^+$: the cross-section is computed using the expression from \citet{h2+_bf_ff}.
        \item H: the cross-section for the first nine transitions are computed hydrogenically with Gaunt factor of 1.
        \item He: the cross-sections for the first 14 transitions are computed using cubic fits from \citet{he_bf_fit_OPproject} from the Opacity Project data, taking into account the multiplicity of every transition (if these are triples or singlets).
        \item He$^{+}$ : the cross-section for 14 transitions is computed hydrogenically with exact Gaunt factors.
        \item C: the cross-sections for 104 transitions are interpolated from TOPbase data.
        \item C$^{+}$: the cross-sections for 40 transitions are interpolated from TOPbase data. 
        \item N: the cross-sections for 89 transitions are interpolated from TOPbase data.
        \item N$^{+}$: the cross-sections for the 51 transitions are interpolated from TOPbase data.
        \item O: the cross-sections for 54 transitions are interpolated from TOPbase data.
        \item O$^{+}$: the cross-sections for 74 transitions are interpolated from TOPbase data.
        \item Na: the cross-sections for 32 transitions are interpolated from TOPbase data.
        \item Na$^{+}$: the cross-sections for eight transitions are interpolated from TOPbase data.
        \item Mg: the cross-sections for 71 transitions are interpolated from TOPbase data. 
        \item Mg$^{+}$: the cross-sections for the 31 transitions are interpolated from TOPbase data.
        \item Al: the cross-sections for 33 transitions are interpolated from TOPbase data. 
        \item Al$^{+}$: the cross-sections for 81 transitions are interpolated from TOPbase data. 
        \item Si: the cross-sections for 57 transitions are interpolated from TOPbase data.
        \item Si$^{+}$: the cross-sections for 46 transitions are interpolated from TOPbase data. 
        \item Ca: the cross-sections for 79 transitions are interpolated from TOPbase data. 
        \item Ca$^{+}$:  the cross-sections for 32 transitions are interpolated from TOPbase data. 
        \item Fe: the cross-sections for 49 transitions are obtained from the Iron Project data. 
        \item Fe$^{+}$: the cross-sections for 41 transitions are obtained from the Iron Project data. 
    \end{itemize}
    
    Other contributions to the continuum and bands are also included:
    \begin{itemize}
        \item H, He and H$_2$ Rayleigh scattering from \citet{rayleigh_h_he_h2}.
        \item Thomson scattering: $\varkappa = 6.65\times 10^{-25} n_e/\rho \; \left(\mathrm{cm^2 g^{-1}}\right)$
        \item Collision-Induced Absorption of H$_2$-H$_2$ \citep{h2-h2_CIA}, H$_2$-He \citep{h2-he_CIA}, H$_2$-H \citep{h2_h_CIA} and H-He \citep{h-he_CIA}.
        \item CH, OH continuous absorption from \citet{OH_CH}. 
     \end{itemize}

    Bound-bound transitions are computed using the data from different line lists, from which the code selects which lines may actually contribute (e.g.\ with a threshold in the line-to-continuum opacity ratio). These are mainly from Kurucz line lists \footnote{\url{http://kurucz.harvard.edu/linelists.html}}, updated by data from the National Institute of Standards and Technology (NIST) when available. \texttt{SYNSPEC} also uses data from EXOMOL \footnote{\url{https://www.exomol.com}} for some of the molecular transitions.  In the case of the molecules, around a total of $2 \times 10^{7}$ transitions for H$_2$, CH, NH, OH, NaH, MgH, SiH, CaH, CrH, FeH, C$_2$, CN, CO, MgO, AlO, SiO, CaO, VO, H$_2$O and TiO are included (see Fig.\  \ref{fig:included_molecules}). In the case of the atoms, \texttt{SYNSPEC} includes a line list with around $2 \times 10^{6}$ transitions for:
    \begin{itemize}
        \item Neutral H, As, Se, Rb, Sb, Te, Cs, Pt, Au, Tl, and Bi.
        \item Neutral and first ion of He, Li, Ga, Ge, Sr, Y, Tc, Pd, Ag, Cd, In, Sn, Ba, La, Ce, Pr, Nd, Sm, Eu, Gd, Tb, Dy, Ho, Tm, Yb, Lu, Hf, Ta, W, Re, Os, Ir, Hg, Pb, Th, and U.
        \item Neutral and first 2 ions of Be, Nb, Rh, Zr, and Ru.
        \item Neutral and first 3 ions of B, C, and Mo.
        \item Neutral and first 5 ions of N, O, F, Ne, Na, Mg, Al, Si, P, S, Cl, Ar, and K.
        \item Neutral and first 7 ions of Zn.
        \item Neutral and first 8 ions of Ca, Sc, Ti, V, Cr, Mn, and Co.
        \item Neutral and first 9 ions of Fe, Ni, and Cu.
    \end{itemize}

\section{Details of the comparison of the opacity distribution function from \texttt{SYNSPEC} and \texttt{ATLAS}} \label{app:details_A_vs_S} 

    Figures in this appendix show examples of typical differences between the ODFs from \texttt{SYNSPEC} and \texttt{ATLAS} in the opacity and in the $Q$ rates for the four stellar models (see Sect.\ \ref{subsec:comparison_ODF_SYNS_vs_ATLAS}).

    \begin{figure*}
        \centering
        \includegraphics[width=17.6cm]{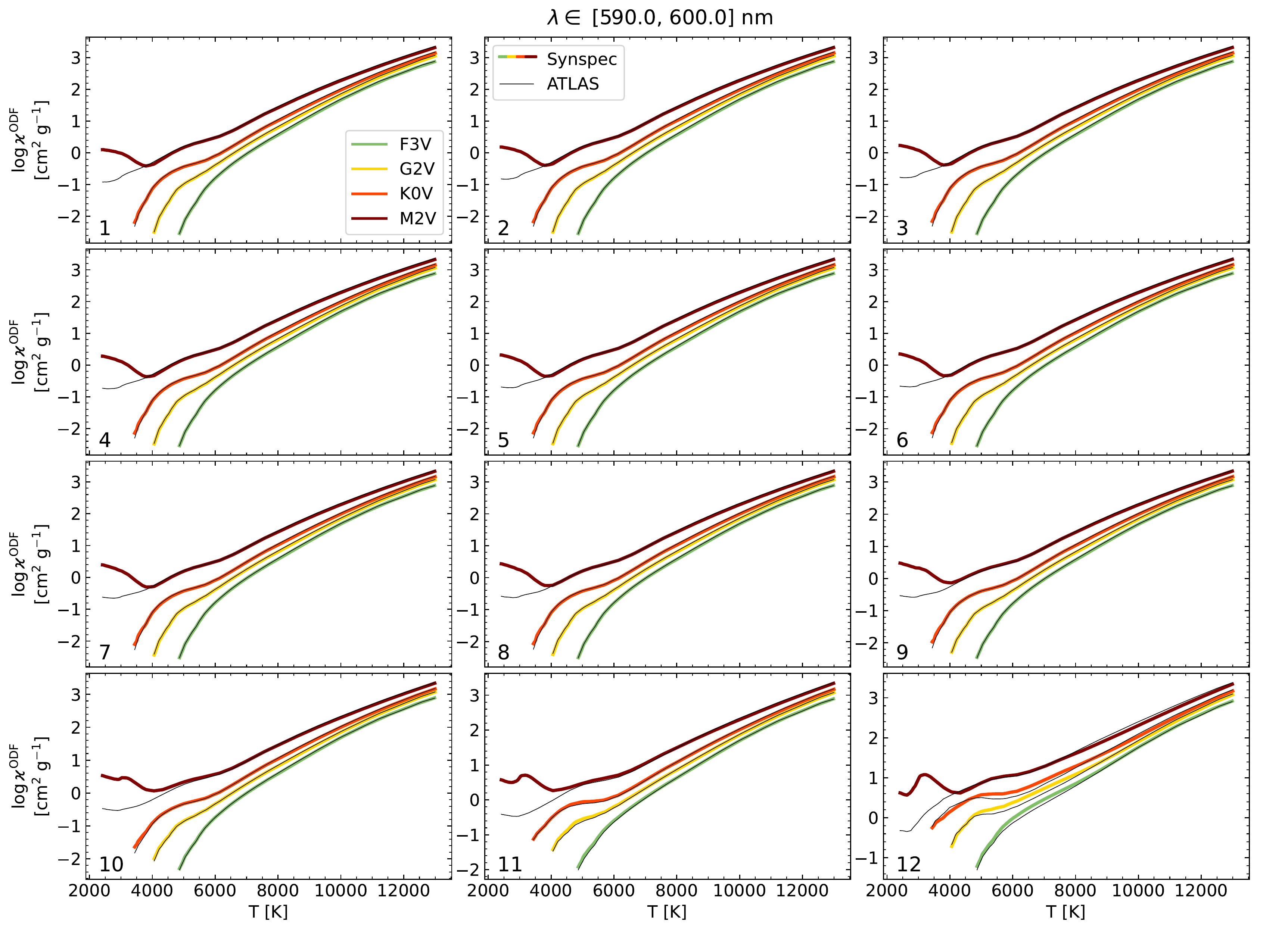}
        \caption{ODF opacity against the temperature of the four stellar atmospheres (different \textit{colour lines}, corresponding to a different spectral type as in Fig.\  \ref{fig:1d_mod_atms}), and for 12 substeps (\textit{different panels}) of the ODF step defined by the $\lambda \in [590,600]$ nm range: \texttt{SYNPEC} (\textit{thick coloured lines}) and \texttt{ATLAS} (\textit{thin black lines}). The \textit{number} at the left bottom corner of each panel indicates the substep in the order of increasing opacities.  
        This ODF step is an example of a generally good match between \texttt{SYNPEC} and \texttt{ATLAS} that is found in the majority of ODF steps in the visible and IR. The only systematic difference is present in all substeps at the top part of the M model.}  
        \label{fig:ODF_AvsS_590-600}
    \end{figure*}
    \begin{figure*}
        \centering
        \includegraphics[width=17.6cm]{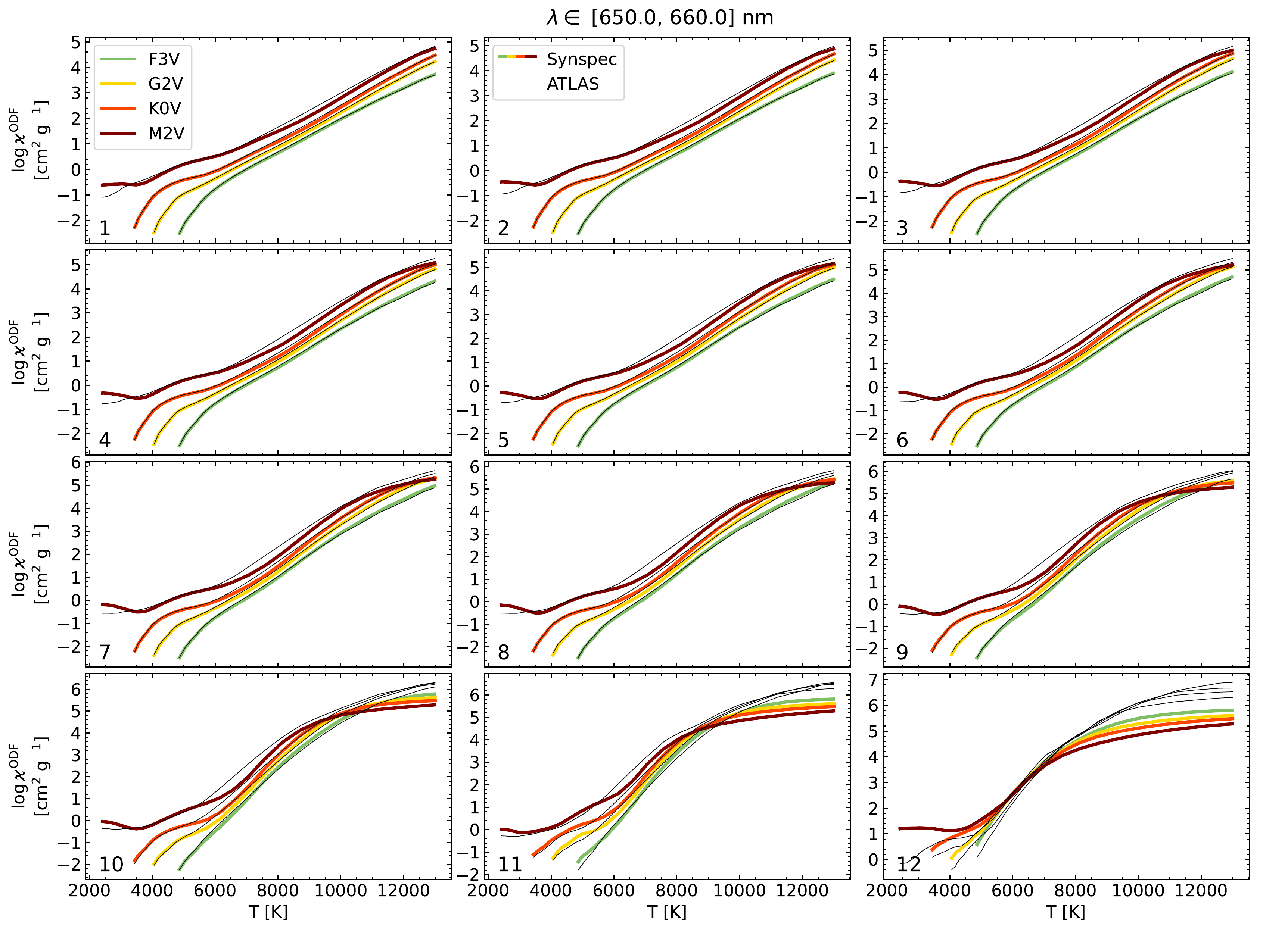}%.png}
        \caption{Same as Fig.\  \ref{fig:ODF_AvsS_590-600}, for the step defined by $\lambda \in [650,660]$ nm containing hydrogen H$\alpha$. 
        This example shows that there are significant differences between the two data sets in the strong hydrogen line. The differences are present in all substeps for the M star, but also in the more opaque substeps (\textit{rows 3 and 4}) for the three hotter stellar types.}
        \label{fig:ODF_AvsS_650-660}
    \end{figure*}
    \begin{figure*}
        \centering
        \includegraphics[width=17.6cm]{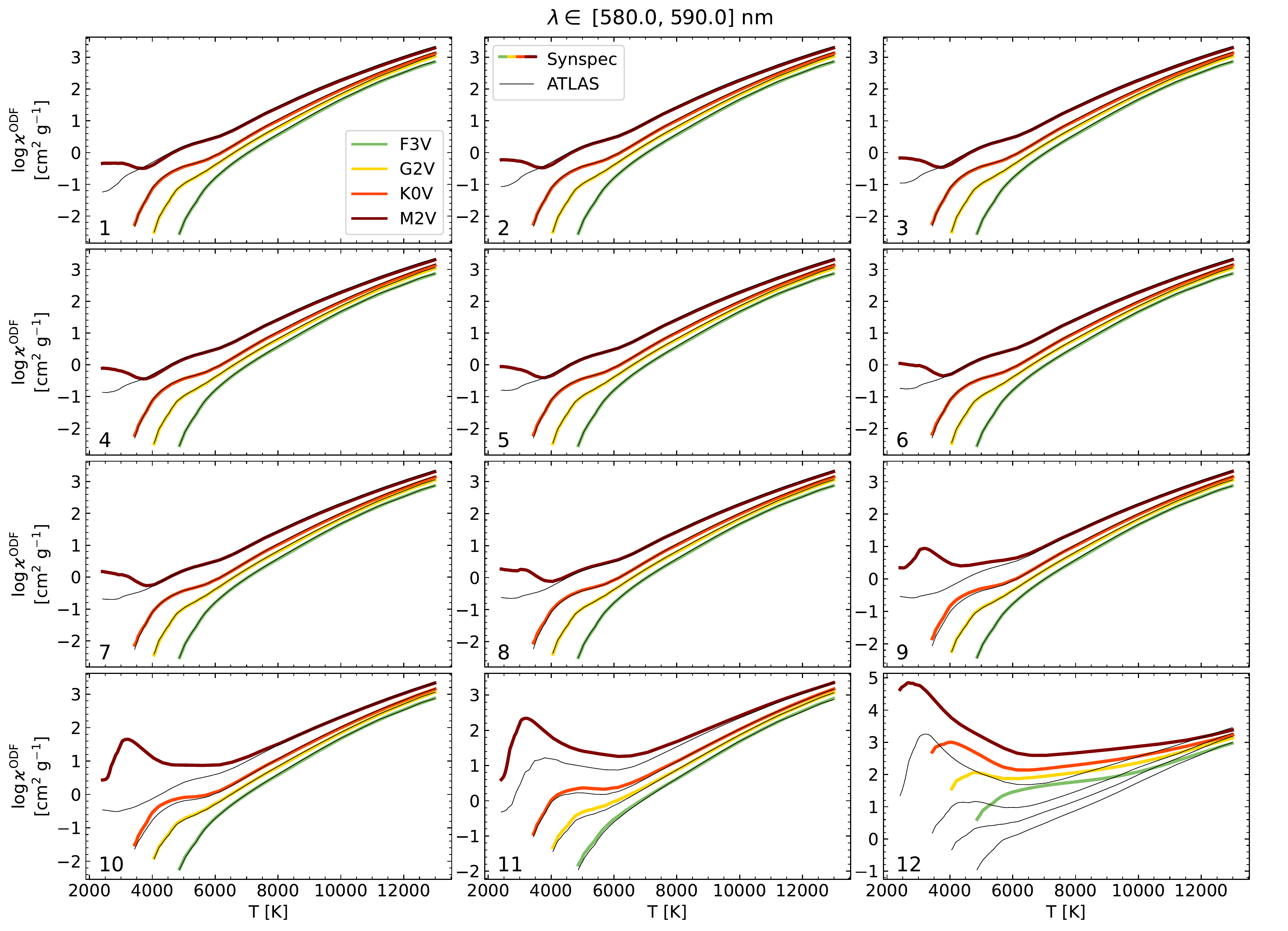} 
        \caption{Same as Fig.\  \ref{fig:ODF_AvsS_590-600} for the step defined by $\lambda \in [580,590]$ nm containing Na{\tiny I} D1 and D2 line.}
        \label{fig:ODF_AvsS_580-590} 
    \end{figure*}
    \begin{figure*}
        \centering
        \includegraphics[width=17.6cm]{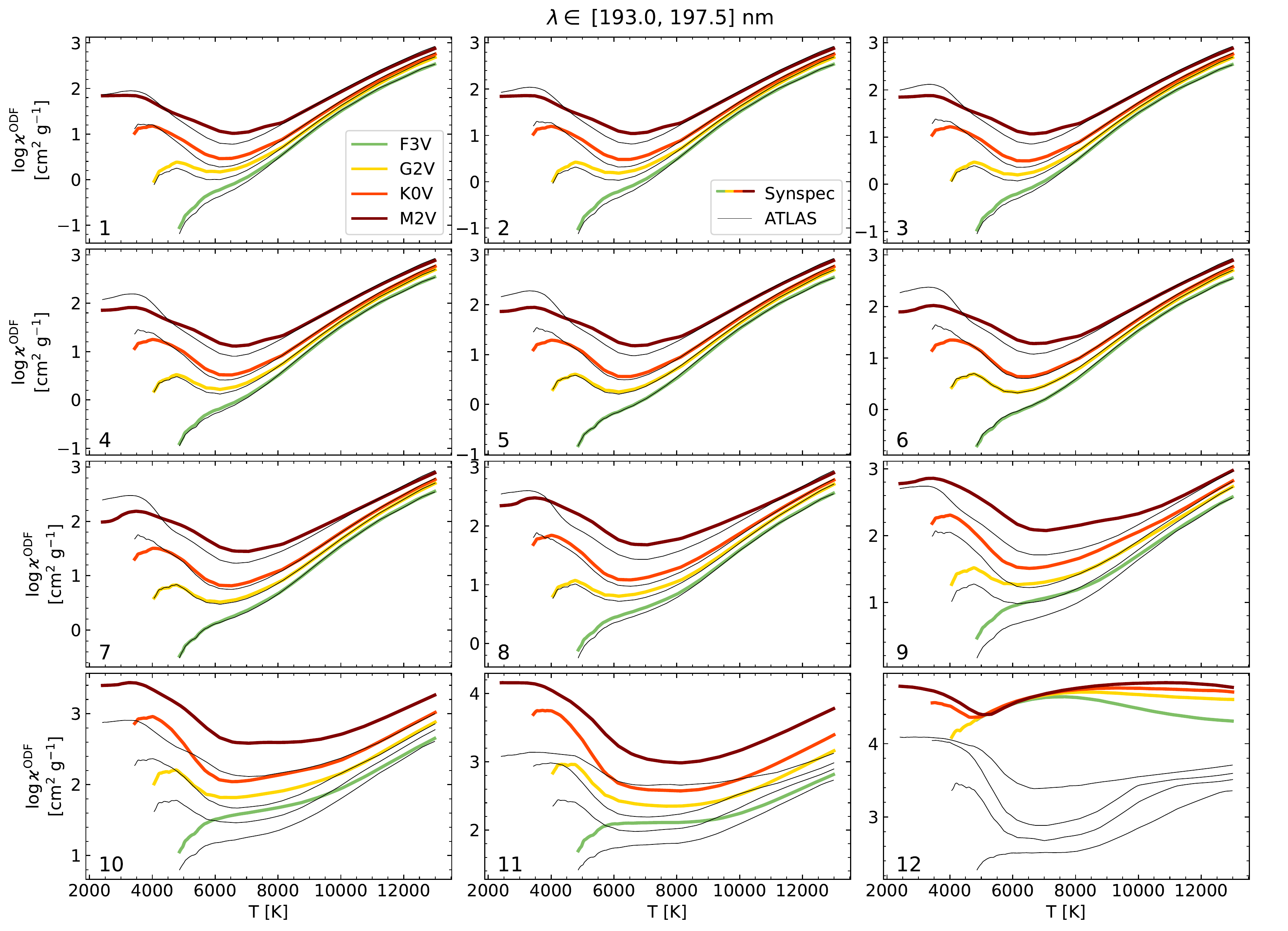}
        \caption{Same as Fig.\  \ref{fig:ODF_AvsS_590-600} for the step defined by $\lambda \in [193,197.5]$ nm. The differences in the UV between the two tables are large in all the substeps and for all stellar models.}
        \label{fig:ODF_AvsS_193-197}
    \end{figure*}

    \begin{figure}
        \includegraphics[width=8.8cm]{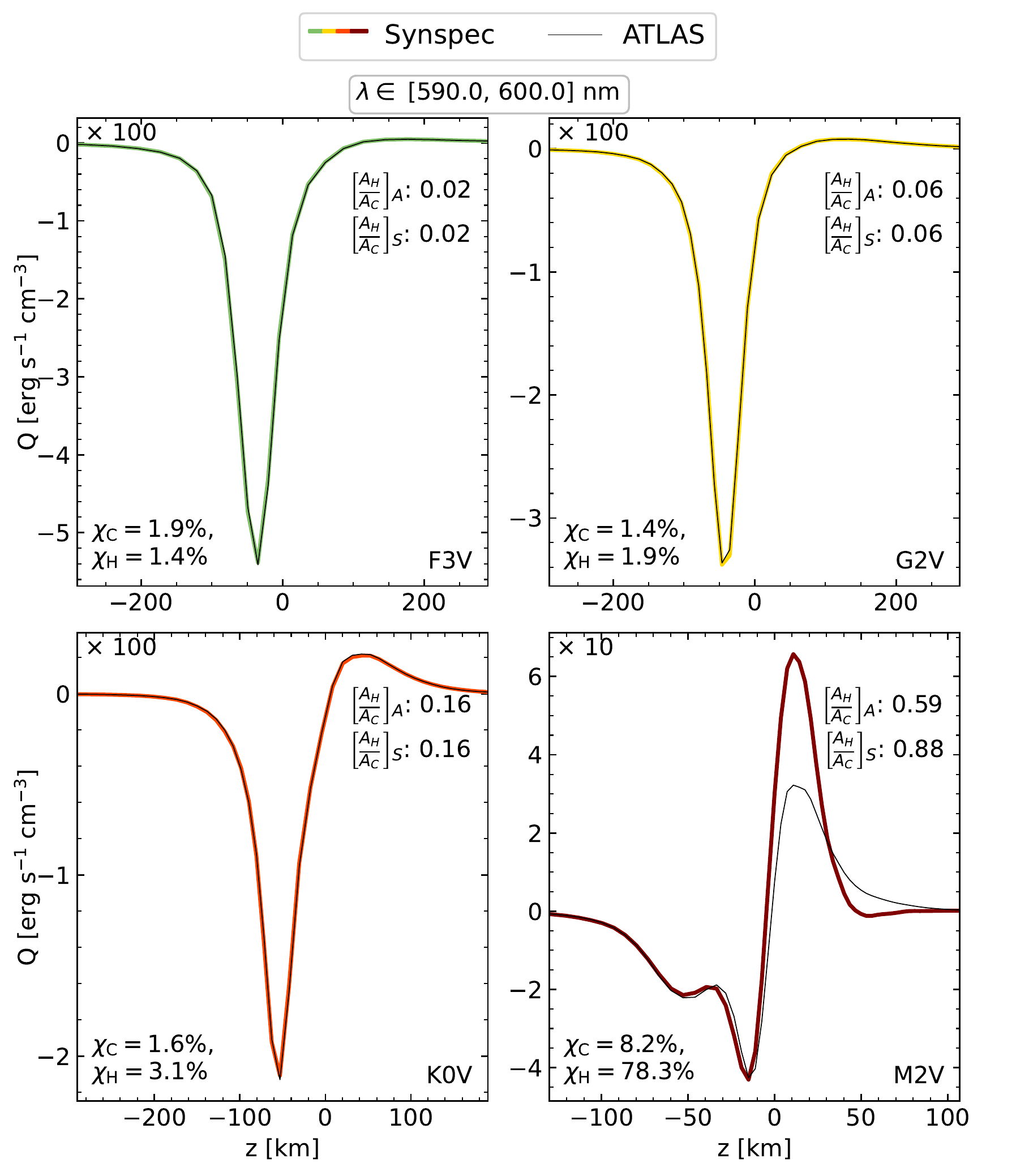}%.png}
        \caption{Radiative energy exchange rate $Q_i$ computed from the four model atmospheres using the opacity shown in Fig.\  \ref{fig:ODF_AvsS_590-600} and integrated over the interval $\lambda \in [590,600]$ nm. The values on the vertical axis are scaled with the factor indicated in the \textit{top left of each panel}. \textit{Bottom left of each panel}: deviation between the two ODFs computed using Eqs.\  \ref{eq:error_c}, \ref{eq:error_h}. \textit{Top right of each panel}: ratio of the heating area with respect to the cooling area for the $Q_i$ for both data sets (\textit{A} for \texttt{ATLAS} and \textit{S} for \texttt{SYNSPEC}).}
        \label{fig:Q_AvsS_590-600}
    \end{figure}
    \begin{figure}
        \includegraphics[width=8.8cm]{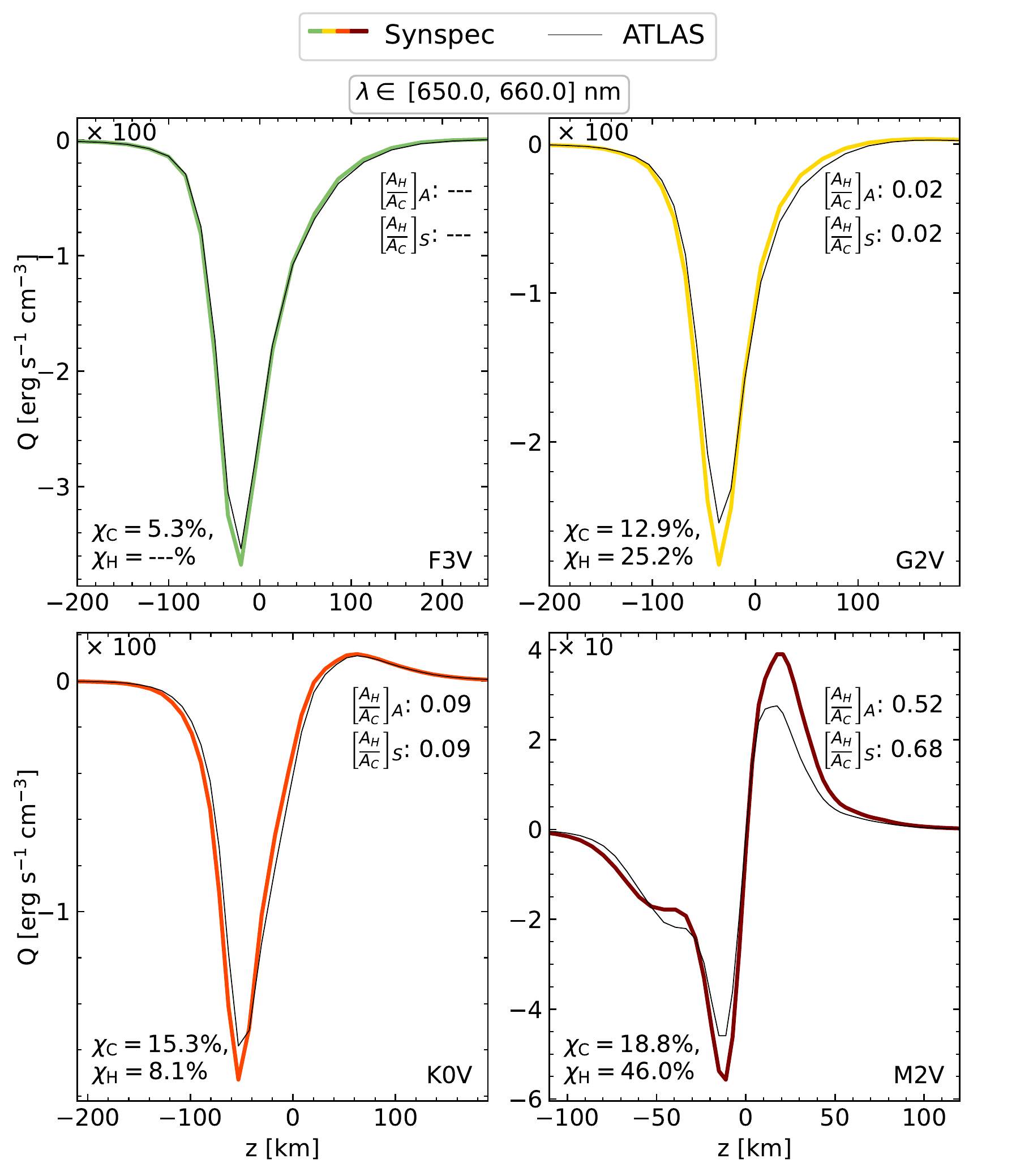}
        \caption{Same as Fig.\  \ref{fig:Q_AvsS_590-600}, but for the $\lambda \in [650,660]$ nm interval that contains the hydrogen H$\alpha$ line.
        Computed using the opacity shown in Fig.\  \ref{fig:ODF_AvsS_650-660}.}
        \label{fig:Q_AvsS_650-660}
    \end{figure}
    \begin{figure}
        \includegraphics[width=8.8cm]{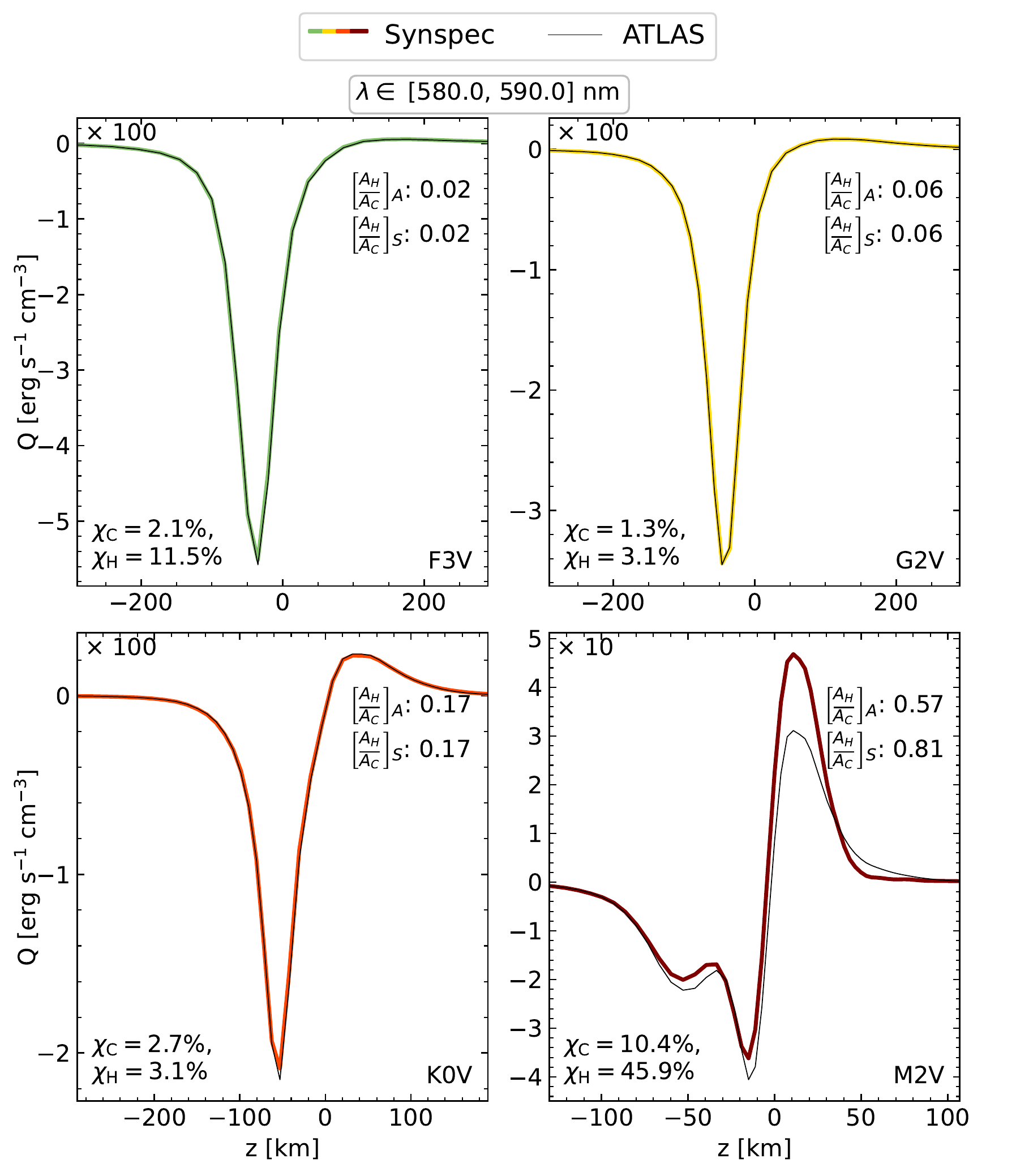}
        \caption{Same as Fig.\  \ref{fig:Q_AvsS_590-600}, but for $\lambda \in [580,590]$ nm, interval that contains the Na{\tiny I} D1 and D2 line. Computed using the opacity shown in Fig.\  \ref{fig:ODF_AvsS_580-590}.}
        \label{fig:Q_AvsS_580-590}
    \end{figure}
    \begin{figure}
        \includegraphics[width=8.8cm]{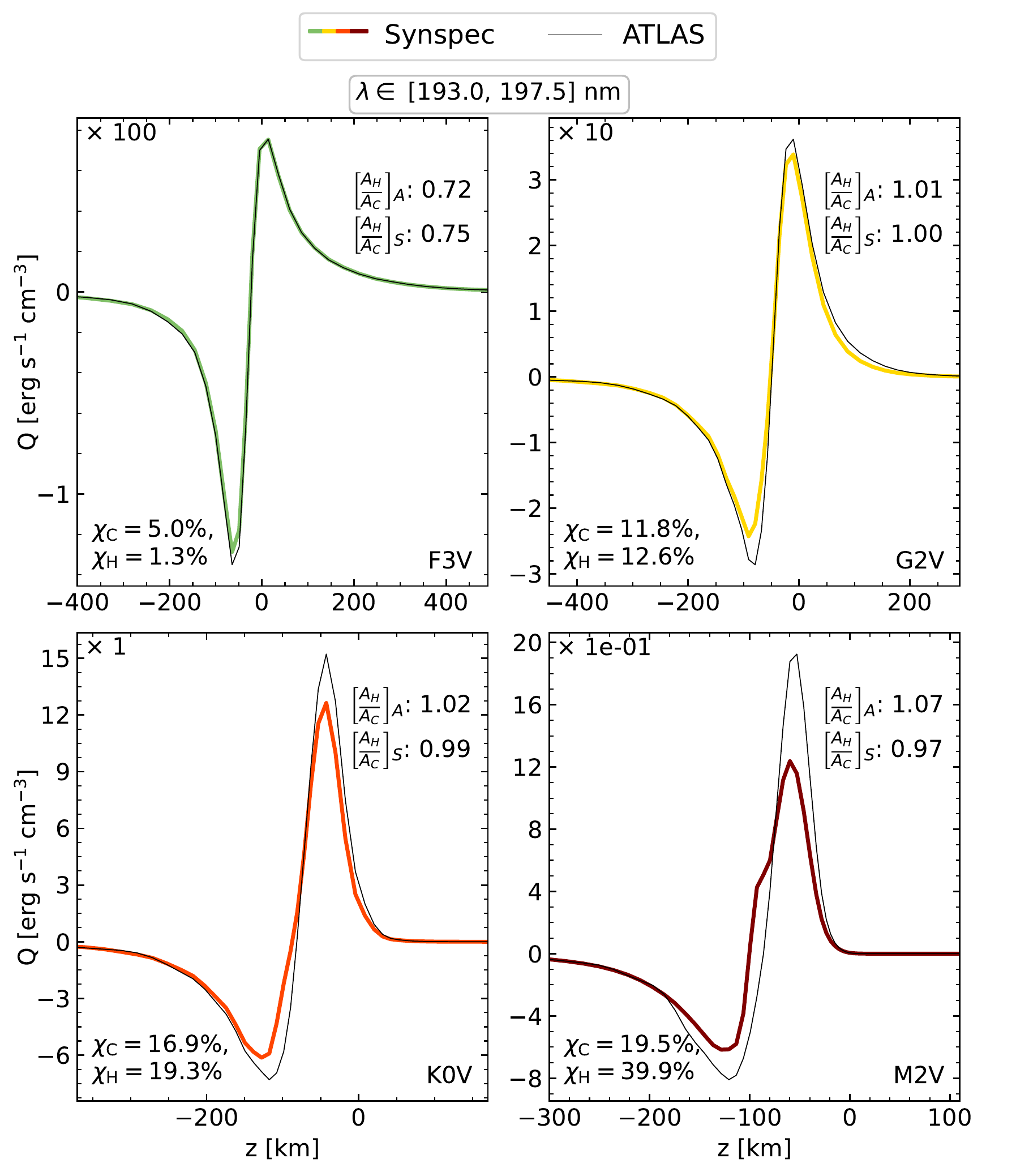}
        \caption{Same as Fig.\  \ref{fig:Q_AvsS_590-600}, but for $\lambda \in [193,197.5]$ nm, computed using the opacity shown in Fig.\  \ref{fig:ODF_AvsS_193-197}.}
        \label{fig:Q_AvsS_193-197}
    \end{figure}

\clearpage
\section{Saturation of the opacity binning method and comparison against grey opacity} \label{app:not_convergence}
\begin{figure}
        \includegraphics[width=8.8cm]{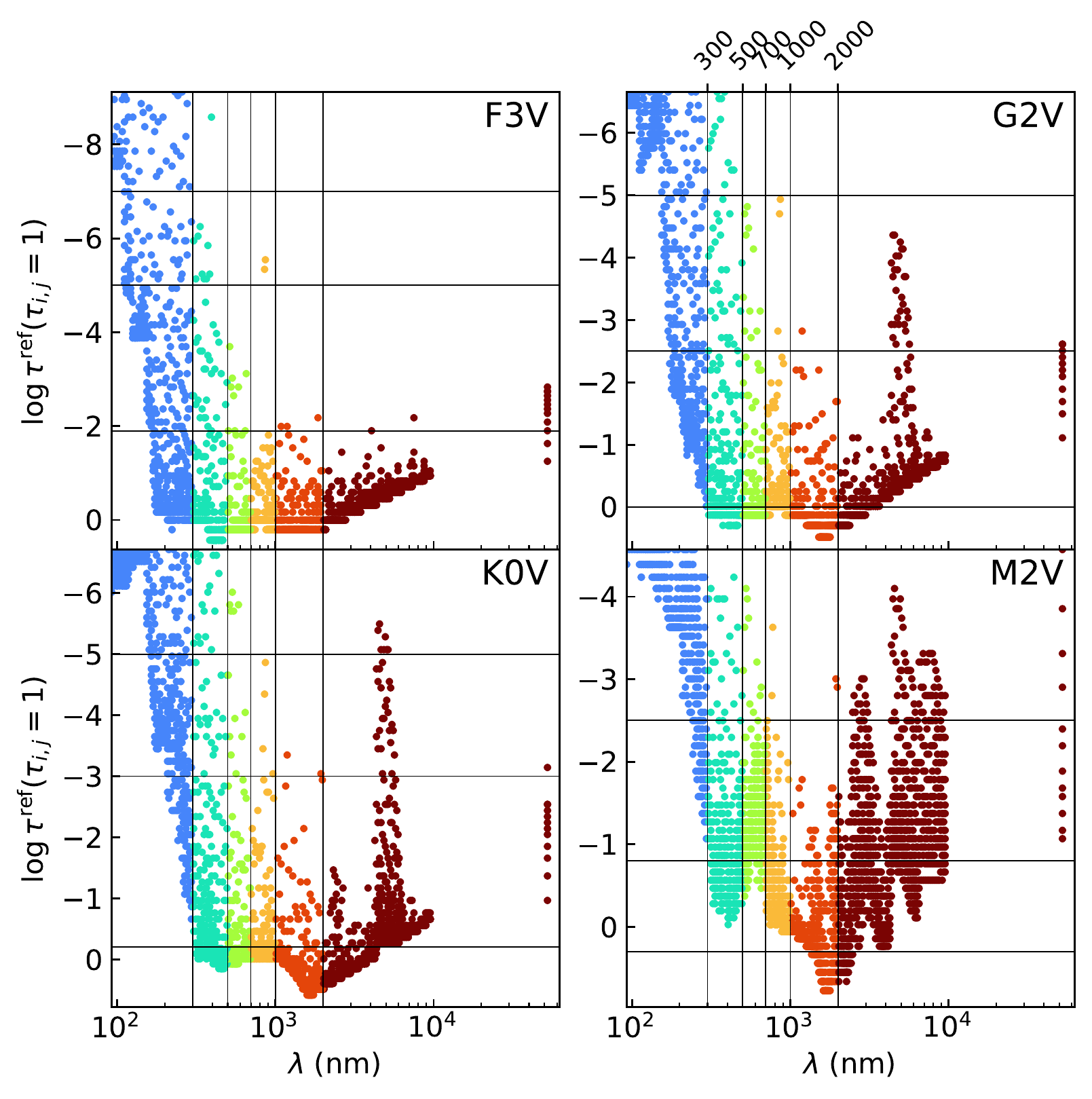}
        \caption{Distribution of the opacity bins in optical depth and wavelength for the four stars. Similarly to Fig.\ \ref{fig:binning_indexes_example}, each \textit{coloured dot} corresponds to one ODF substep. The wavelength separators are identical for all stars ($\lambda<300$, $\in [300,500]$, $\in [500,700]$, $\in [700,1000]$, $\in [1000-2000]$, and $>2000 \, \mathrm{nm}$).
        All values within one wavelength range are shown in the same \textit{colour}.
        The $\tau$-separators are selected from the optimal combinations found in Sect.\  \ref{subsec:tau_bin_location} for each star: $\log\tau = \left\{ -1.9, -5, -7 \right\}$ for the F3V star; $\log\tau = \left\{ 0, -2.5, -5 \right\}$ for the G2V; $\log\tau = \left\{ -0.2, -3, -5 \right\}$ for the K0V; $\log\tau = \left\{ 0.3, -0.8, -2.5 \right\}$ for the M2V.}
        \label{fig:index_not_convergence}
    \end{figure}

\begin{figure}
        \includegraphics[width=8.8cm]{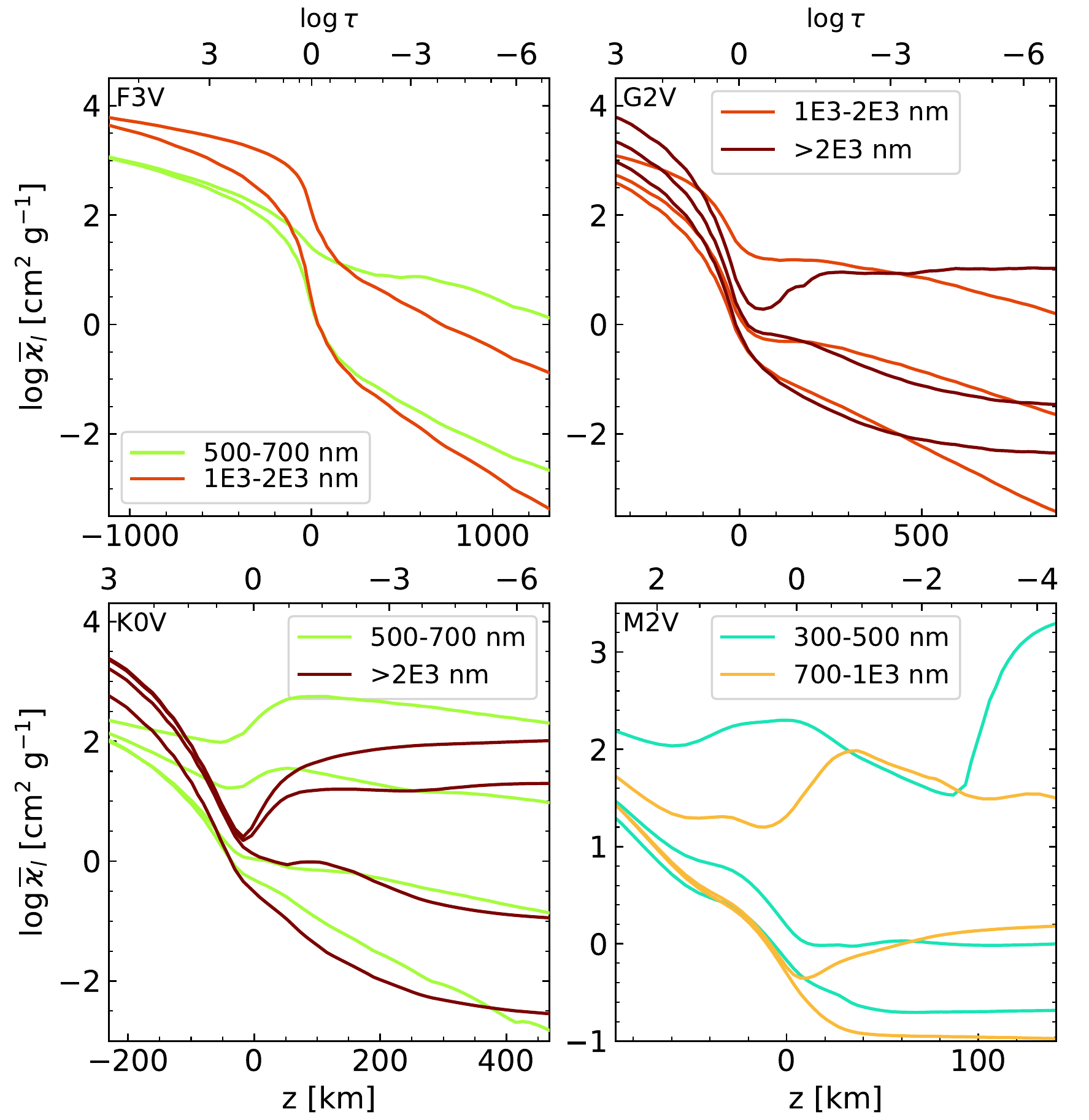}
        \caption{Opacity $\overline{\varkappa}_l$ of each $\tau$-bin in pairs of $\lambda$-ranges for the four stars (spectral type at the \textit{top left} of each panel) versus the geometrical hegiht (\textit{bottom axis} of each panel) and optical depth (\textit{top axis}). The $\lambda$-ranges are selected to illustrate the different variations of the opacity with height depending on the spectral region.
        The \textit{colours} of the curves correspond to the $\lambda$-ranges in Fig.\ \ref{fig:index_not_convergence}, with the values of each range in the legend. }
        \label{fig:not_convergence}
    \end{figure}

\FloatBarrier % To keep the figures after the title
    The OB method is not expected to converge to the ODF solution for a large number of $\tau$-bins \citep{voegler_opacity_binning}. 
    To explain why this is the case, we use the checkerboard distribution of the bins introduced in Fig.\ \ref{fig:index_not_convergence}, similar to that used in Sect.\ \ref{subsec:tau_lambda_binning}, but now with 4 bins in $\tau$ and 6 in $\lambda$. 
    The number of non-empty $\tau$-bins changes
    depending on the $\lambda$-bin. For example, in the case of the F3V star there are 4 $\tau$-bins for $\lambda < 300$ nm, while there are 2 $\tau$-bins for $\lambda \in [1000,2000]$ nm (Fig.\ \ref{fig:index_not_convergence}).
    For each star, in Fig.\ \ref{fig:not_convergence} we compare the binned opacity of pairs of wavelength ranges selected to better illustrate this behaviour. We compare $\lambda$-ranges that have the same number and location of non-empty $\tau$-bins. 
    For each star and any of the $\tau$-bins, the opacity from one of the two $\lambda$-ranges vary with height in a different way than the opacity from the other. 
    Even for arbitrarily increased number of bins,
    binning only in $\tau$ mixes wavelength regions which opacities have different height profiles 
    (this is illustrated in figure 4.13 from \citealp{voegler_thesis}, too). Although this effect may be reduced, it is present even when splitting in $\lambda$, since each contributor to the opacity may have its own variation with height.

\begin{figure}
        \includegraphics[width=8.8cm]{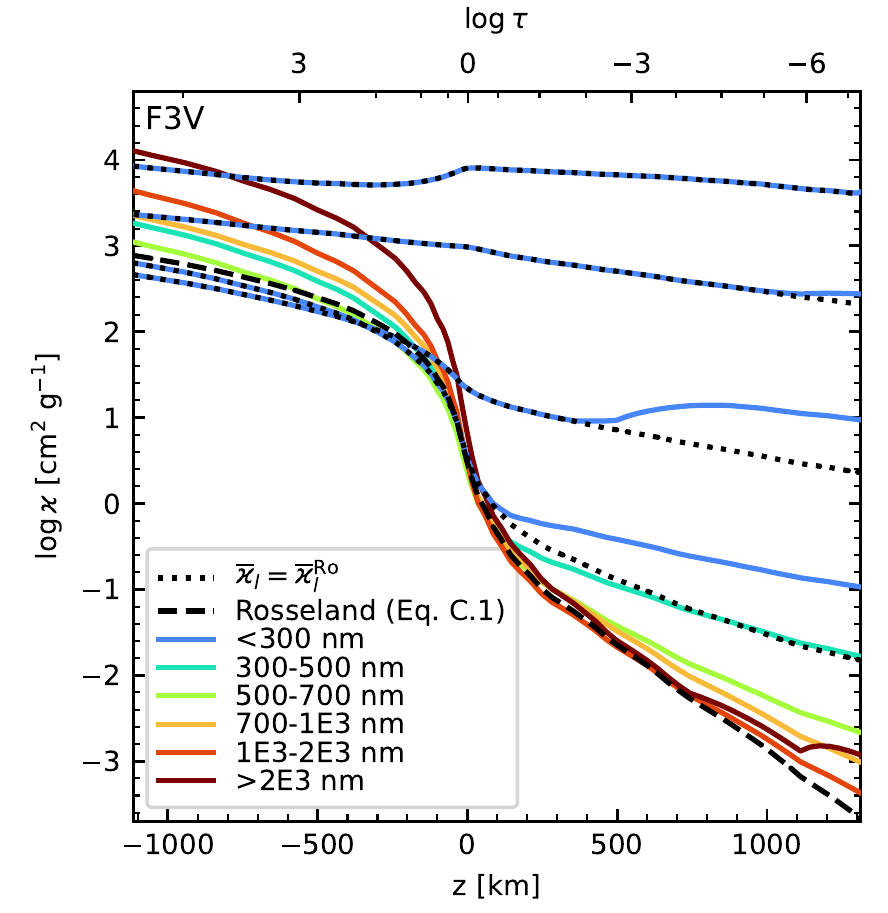}
        \caption{Opacity $\overline{\varkappa}_l$ for each $\tau$-bin in $\lambda<300 \, \mathrm{nm}$ computed with Eq.\ \ref{eq:mean_opac} (\textit{solid blue curves}) compared to the opacity $\overline{\varkappa}_l=\overline{\varkappa}^{\mathrm{Ro}}_l$ (\textit{dotted black curves}) for the F3V star. The rest of the \textit{coloured curves} correspond to the opacity for the least opaque bin for each wavelength range. \textit{Dashed black line}: Rosseland opacity mean computed with Eq.\ \ref{eq:harmonic_mean}. The \textit{bottom and top axis} show the geometrical height and optical depth, respectively. }
        \label{fig:OB_goes_grey}
    \end{figure}

    Figure \ref{fig:OB_goes_grey} 
    shows how the opacity computed with the OB method converges to the Rosseland opacity in each $\tau$-bin, $\overline{\varkappa}_l=\overline{\varkappa}^{\mathrm{Ro}}_l$ (dotted black lines), in the limit of large enough optical depths. We show only the case for the F3V star and the range for $\lambda<300 \, \mathrm{nm}$ to avoid overcrowding the plot.
    However, the conclusions are valid for
    any $\lambda$-range and star. Another example of this convergence can be found in figure 4.12 from \citet{voegler_thesis} for the case of the Sun. 
    Figure \ref{fig:OB_goes_grey} also shows the Rosseland opacity mean for the whole spectrum in black dashed line. The harmonic mean used to compute this averaged opacity,
    \begin{equation} \label{eq:harmonic_mean}
        \overline{\varkappa}_{\mathrm{Ro}} = \int \frac{\partial \mathrm{B}}{\partial \mathrm{T}} d\lambda \Bigg / \left( \int  \frac{\partial \mathrm{B}}{\partial \mathrm{T}} \frac{1}{\varkappa_\lambda} d\lambda \right),
    \end{equation}
    gives more weight to the continuum than to the line cores, 
    making that the Rosseland mean is close to the opacity of the least opaque bins.

\begin{figure}
        \includegraphics[width=8.8cm]{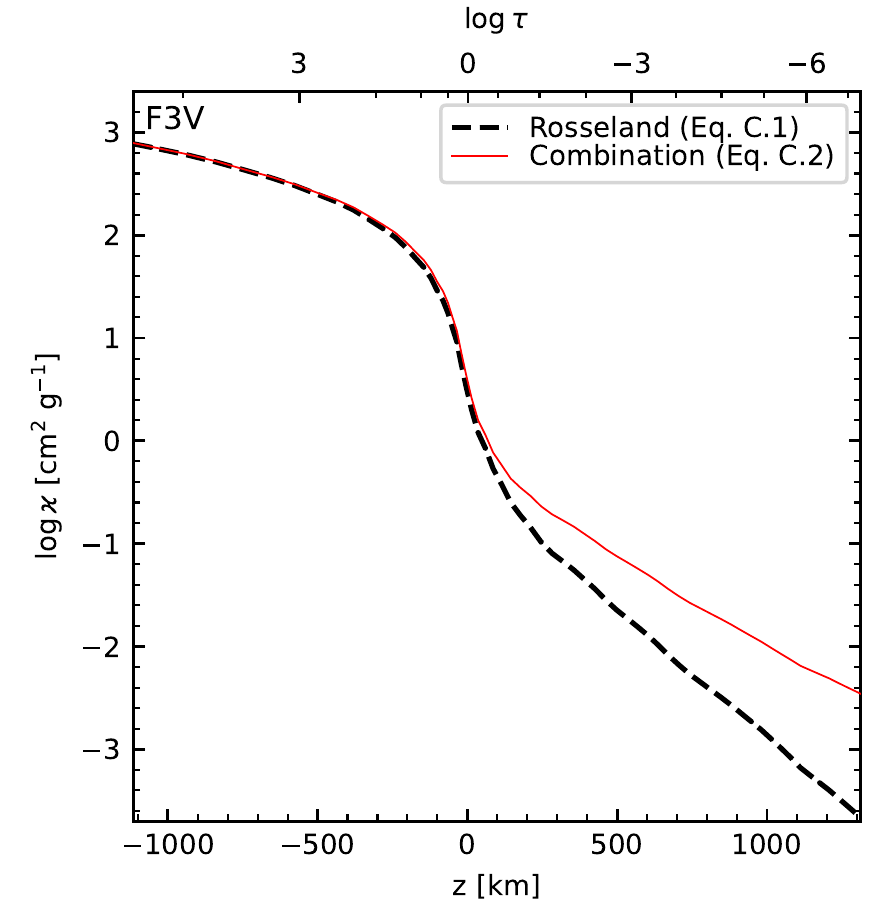}
        \caption{Rosseland opacity mean computed with Eq.\ \ref{eq:harmonic_mean} (\textit{black dashed line}) and Eq.\ \ref{eq:ross_fashion} (\textit{red thin solid line}) for the F3V star. The \textit{bottom and top axis} show the geometrical height and optical depth, respectively.}
        \label{fig:OB_Ross_fashion}
    \end{figure}

    Finally, Figure \ref{fig:OB_Ross_fashion} shows that the combination of the opacity bins in the Rosseland fashion, 
    \begin{equation} \label{eq:ross_fashion}
        \left. \sum_l \, \left. \frac{\partial B}{\partial T} \right|_l  \right/ \left( \sum_l \, \left. \frac{\partial B}{\partial T} \right|_l \, \frac{1}{\overline{\varkappa_l}} \right),
    \end{equation}
    yields the correct value for the Rosseland mean for deep enough optical depths. 
    For these optical depths (see Fig.\ \ref{fig:OB_goes_grey} and Eq.\ \ref{eq:mean_opac}) 
    \begin{equation}
        \overline{\varkappa}_l \approx \overline{\varkappa}^{\mathrm{Ro}}_l = \left. \frac{\partial \mathrm{B}}{\partial \mathrm{T}}\right|_l \Bigg / \left( \sum_{i(l)} \Delta \lambda_i \frac{\partial \mathrm{B}_{i}}{\partial \mathrm{T}} \sum_{j(i,l)} \frac{\omega_j}{\varkappa_{i,j}} \right),
    \end{equation}
    thus, the combination of the bins following Eq.\ \ref{eq:ross_fashion} gives
    \begin{multline}
        \left. \sum_l \, \left. \frac{\partial B}{\partial T} \right|_l  \right/ \left( \sum_l \, \left. \frac{\partial B}{\partial T} \right|_l \, \frac{1}{\overline{\varkappa_l}} \right) = \\
        \left. \sum_l \, \sum_{i(l)} \Delta \lambda_i \frac{\partial \mathrm{B}_{i}}{\partial \mathrm{T}}  \right/ \left( \sum_l \sum_{i(l)} \frac{\partial \mathrm{B}_{i}}{\partial \mathrm{T}} \sum_{j(i,l)} \frac{\omega_j}{\varkappa_{i,j}} \right).
    \end{multline}
    This equation is similar to Eq.\ \ref{eq:harmonic_mean}, but with the formalism of the ODF to compute the integrals and with the additional sums over the bins. It can be written in a more general way as 
    \begin{multline}
        \left. \sum_l \, \left. \frac{\partial B}{\partial T} \right|_l  \right/ \left( \sum_l \, \left. \frac{\partial B}{\partial T} \right|_l \, \frac{1}{\overline{\varkappa_l}} \right) \equiv \\
        \left. \sum_l \, \int_{\lambda(l)} \frac{\partial \mathrm{B}_{\lambda}}{\partial \mathrm{T}}  d \lambda \right/ \left( \sum_l \int_{\lambda(l)} \frac{\partial \mathrm{B}_{\lambda}}{\partial \mathrm{T}} \frac{1}{\varkappa_{\lambda(l)}} d \lambda \right),
    \end{multline}
    where $\lambda(l)$ corresponds to the wavelengths that enter the $l$th bin. Binning in $\tau$, $\lambda$, or both always is equivalent to group only separating in $\lambda$, since each opacity point corresponds to a different wavelength\footnote{In the case of the ODF, the substeps lose their wavelength dependence, but the ODF formalism takes care of this.}. This property as well as understanding the integrals of the opacity in terms of Riemann or Lebesque integration
    explains the result shown in Fig.\ \ref{fig:OB_Ross_fashion}.
    Similarly, the Planck mean should be recovered for higher layers.

\end{appendix}

\end{document}